\documentclass[sort&compress]{elsarticle}
\usepackage[utf8]{inputenc}

\usepackage[capposition=top]{floatrow}
\usepackage{setspace}
\usepackage[margin=1.0in]{geometry}
\usepackage{adjustbox}
\usepackage[colorlinks=true]{hyperref}
\hypersetup{linkcolor=blue}
\hypersetup{citecolor=blue}
\usepackage[small, justification=justified, singlelinecheck=false]{caption}
\usepackage{url}
\usepackage{booktabs}
\usepackage{multirow}
\usepackage{graphicx}
\usepackage{adjustbox}
\usepackage{amssymb}
\usepackage{amsmath}
\usepackage{dutchcal}
\usepackage{tabularx}
\usepackage[table,xcdraw,dvipsnames]{xcolor}
\usepackage{comment}
\usepackage{subcaption}
\usepackage{numprint}
\npthousandsep{\,}
\usepackage{tocbibind}
\usepackage{enumitem}
\usepackage{bigfoot}
\usepackage[toc,page]{appendix}
\usepackage{pifont}
\usepackage{seqsplit}
\usepackage{cancel}
\usepackage{pifont}
\usepackage{rotating}
\usepackage{natbib}
\usepackage[ruled,vlined]{algorithm2e}

\newcommand{\cmark}{\ding{51}}%
\newcommand{\xmark}{\ding{55}}

\makeatletter
 \def\ps@pprintTitle{%
  \let\@oddhead\@empty
  \let\@evenhead\@empty
  \def\@oddfoot{}%
  \let\@evenfoot\@oddfoot
 }
\makeatother

\begin{document}
	
\begin{frontmatter}
\title{HLOB -- Information Persistence and Structure \\ in Limit Order Books}
		
\author[myaddress2]{Antonio Briola \corref{mycorrespondingauthor}}
\cortext[mycorrespondingauthor]{Corresponding author}
\ead{antonio.briola.20@ucl.ac.uk}
\author[myaddress2]{Silvia Bartolucci}
\author[myaddress2,myaddress3]{Tomaso Aste}

\address[myaddress2]{Department of Computer Science, University College London, London, United Kingdom.}
\address[myaddress3]{Systemic Risk Centre, London School of Economics, London, United Kingdom.}
	
\begin{abstract}
We introduce a novel large-scale deep learning model for Limit Order Book mid-price changes forecasting, and we name it `HLOB'. This architecture (i) exploits the information encoded by an Information Filtering Network, namely the Triangulated Maximally Filtered Graph, to unveil deeper and non-trivial dependency structures among volume levels; and (ii) guarantees deterministic design choices to handle the complexity of the underlying system by drawing inspiration from the groundbreaking class of Homological Convolutional Neural Networks. We test our model against $9$ state-of-the-art deep learning alternatives on $3$ real-world Limit Order Book datasets, each including $15$ stocks traded on the NASDAQ exchange, and we systematically characterize the scenarios where HLOB outperforms state-of-the-art architectures. Our approach sheds new light on the spatial distribution of information in Limit Order Books and on its degradation over increasing prediction horizons, narrowing the gap between microstructural modeling and deep learning-based forecasting in high-frequency financial markets.


\hfill
	
\end{abstract}
	

\end{frontmatter}

\newpage

\section{Introduction}\label{sec:Introduction}
Financial markets are complex environments. Their complexity stems from two main factors: (i) the interaction of a large number of agents pursuing heterogeneous goals at different time scales through the implementation of trading strategies designed to leverage asymmetric information; (ii) the emergence of self-organizing collective behaviors that do not result from the existence of any central controller and are therefore difficult to anticipate. The concurrence of these aspects contributes to the sporadic and limited-in-time persistence of inefficiencies that make the trading practice profitable. The analysis of existing inefficiencies and the forecasting of new ones is made possible by the mathematical and statistical modeling of the time series reflecting the financial market's behavior. The granularity of these time series widely varies depending on the goal of the analysis, and, in the high-frequency case (i.e., the scenario we are mainly interested in), it can be order-driven with a resolution up to the nanosecond \cite{lobsterdata_what_is_lobster}. 

Indeed, the majority of modern financial exchanges store order-level updates in data structures known as Limit Order Books (LOBs). At each point in time, in a given automated exchange, these data structures contain a snapshot of the standing intentions of market participants to buy or sell different amounts (or volumes) of an asset at a given price. Such trading intentions, which are defined in jargon as `orders', can be of different types (i.e., market orders, limit orders, and cancellation orders) and their flux (i.e., incoming or outgoing) is generally managed by computerized systems exploiting a \textit{FIFO} (first-in, first-out) mechanism to establish execution's priority \cite{bouchaud2018trades, briola2024deep, briola2020deep, briola2021deep}. The timing of accessing information contained in LOBs guarantees asymmetric levels of information to market participants. At the finest-grained information's exploitation level, we refer to High-Frequency Trading (HFT) to indicate the strategies that gain an edge through speed, allowing certain traders to act on information not yet accessible to others \cite{lehalle2018market}. HFT strategies exploit market’s microstructure imperfections to the detriment of other traders, triggering a predator-prey dynamic with other actors \cite{farmer2013ecological}. HFT has been prominent in the financial landscape since 2005 \cite{isichenko2021quantitative}. Despite being object of criticism and regulatory scrutiny since its introduction, it has been demonstrated that this practice's reliance on various levels of market data, rather than external information, contributes to noise generation, thereby preserving unpredictability in stock price movements \cite{bouchaud2009markets}. 

The difficulty in handling the inherent complexity expressed by HFT systems and the availability of large amounts of data, has fostered the development of deep learning models as a solution to the related modeling and forecasting tasks. Over recent years, increasingly sophisticated solutions have emerged, with some of them evolving towards the creation of informed architectures that meticulously incorporate LOBs' components in the definition of derived features. Although many scientific investigations proved the potential of these approaches, there is still a noticeable disconnection between theoretical results and their real-world practicability \cite{prata2023lob}.
Furthermore, the recent research work by \citet{briola2024deep} highlights that the effectiveness of these methods considerably varies depending on the stocks' unique microstructural characteristics. More specifically, the authors show that microstructural properties of stocks exposed to higher trading risks (i.e., the so-called `small-tick stocks') impose sparser LOB structures, highly undermining the ability of deep learning architectures to model their hidden dynamics effectively. On the contrary, the microstructural properties of stocks exposed to lower trading risks (i.e., the so-called `large-tick stocks') impose more compact LOB structures, facilitating deep learning architectures in effectively processing the underlying information.

The contribution of our paper is threefold:

\begin{enumerate}
    \item We introduce `HLOB', a novel, large-scale deep learning architecture that exploits the class of the Homological Convolutional Neural Networks \cite{wang2023homological, briola2023homological} to superimpose a dependency structure among LOB volume levels and model deeper and non-trivial relationships among them\,\footnote{The code to reproduce all the experiments is available at \url{https://github.com/FinancialComputingUCL/LOBFrame/tree/main}.}.
    \item We show that the exploitability of the informational content encoded in the spatial structure imposed by our deep learning architecture is limited in time and the velocity of its degradation is highly dependent on the stocks' microstructural properties.
    \item We test our model against $9$ state-of-the-art deep learning alternatives on $3$ real-world LOB datasets, each including $15$ stocks traded on the NASDAQ exchange. Our findings highlight the difficulty in finding a model that consistently outperforms the others; hence, we provide the guidelines for selecting a model based on factors such as the desired level of interpretability, the specific forecasting horizon, and the available infrastructure.
\end{enumerate}

The rest of the paper is organized as follows. In Section \ref{sec:Related_Work}, we provide an overview of the essential scientific works describing (i) the functioning of LOB dynamics; (ii) the main architectures proposed in the past to solve LOB-related forecasting tasks; and (ii) the intuition behind Information Filtering Networks and the class of Homological Convolutional Neural Networks. In Section \ref{sec:Data}, we present an overview of the datasets used in our experiments. In Section \ref{sec:Methods}, we provide technical insights into the HLOB model and the framework used for its training and validation. In Section \ref{sec:Results}, we present the results of our experiments, while in Section \ref{sec:Conclusion}, we wrap up our findings, providing a comprehensive description of the power and weaknesses of our model compared to the existing ones, with an overview on open challenges in the field.

\section{Related Work}\label{sec:Related_Work}
In this Section, we provide the essential references to (i) understand the operational mechanics of the LOBs; (ii) become familiar with existing models designed to identify microstructural alphas; and (iii) grasp the theoretical foundations of the HLOB model. It is essential to notice that our investigation spans three distinct research fields: (i) market microstructure; (ii) deep learning; and (iii) network science. We do not claim to cover the entire related literature, but, for each domain, we selectively reference the works that are critically relevant to our research, equipping the reader with the foundational tools required to master the content of this paper.

\subsection{Limit Order Book}\label{sec:Limit_Order_Book}
Most modern financial exchanges utilize electronic systems to record and match the trading intentions of market participants. These systems are centered on a data structure called `Limit Order Book' (LOB), which is unique for each security traded on a given exchange and provides immediate access to real-time supply and demand in the visible market. Participants on the same side of the market (whether buying or selling) compete with each other while concurrently opposing those on the opposite side; the buyers want to buy cheaper, and the sellers want to sell at a higher price, but the two sides ultimately need each other to make trades happen. The LOB is, hence, subject to updates (or ticks) that occur at irregular time intervals. These events reflect changes in the market and are constrained by predefined adjustments: (i) the tick size ($\theta$) for price adjustments; and (ii) the lot size ($\psi$) for volume changes\,\footnote{The value of $\theta$ and $\psi$ depend on the exchange. In the NASDAQ exchange, which is the source of the data used in the current research work (see Section \ref{sec:Data}), $\theta=\$0.01$ and $\psi=1$.}. Updates are made possible through the submission of new orders. Based on their direction, they can be bid (buy) or ask (sell) orders; based on their aggressive or passive attitude, they can be market or limit orders. A market order expresses the necessity to buy or sell a certain amount of a given asset at the current best available price on the opposite side of the LOB; it is typically subject to higher transaction fees. A limit order expresses an intention to buy or sell a quantity of an asset at a price that is more advantageous to the one quoted on the best level of the LOB\,\footnote{A LOB is organized into price/volume levels. On the bid side, standing intentions to buy different quantities of a financial security are organized in a descending order (i.e., the first level contains the orders to be executed at the highest price among the quoted ones); on the ask side, standing intentions to sell different quantities of a financial security are organized in an ascending order (i.e., the first level contains the orders to be executed at the lowest price among the quoted ones).}; it populates a queue in one of the deeper levels of the LOB, it does not have any guarantee to be executed and is typically subject to lower transaction fees. Cancellations represent a third class of orders; they delete active limit orders and are typically not subject to transaction fees. 

Temporally, the LOB is structured as stacked snapshots reflecting the tick-by-tick evolution of the market, and takes the form of a multivariate time-series $\mathbb{L} \in \mathbb{R}^{T \times 4L}$, where $T$ is the history length, and $L$ is the number of levels\,\footnote{The dimensionality here $4L$ because, for each level, we register the corresponding ask price, ask volume, bid price, and  bid volume.}. Spatially, a LOB record can be represented as:

\begin{equation}\label{eq:LOB_equation}
    \mathbb{L}(\tau) = \{p_\ell^{\text{ask}}(\tau), v_\ell^{\text{ask}}(\tau), p_\ell^{\text{bid}}(\tau), v_\ell^{\text{bid}}(\tau)\}_{\ell=1}^{L} \ ,
\end{equation}

where $p_\ell^{\text{ask/bid}}(\tau)$ is the ask/bid price at level $\ell \in L$ and $v_\ell^{\text{ask/bid}}(\tau)$ is the volume on the same level $\ell \in L$. The mid-price $m_\tau$ of a stock at time $\tau$ is defined as the average between the best ask price (i.e., $p_1^{\text{ask}}(\tau)$) and the best bid price (i.e. $p_1^{\text{bid}}(\tau)$), $m_\tau = \frac{p_1^{\text{ask}}(\tau) + p_1^{\text{bid}}(\tau)}{2}$. The bid-ask spread $\sigma_\tau$ of the stock at time $\tau$ is defined as the difference between the best ask price and the best bid price, $\sigma_\tau = p_1^{\text{ask}}(\tau) - p_1^{\text{bid}}(\tau)$. 

The level-based representation in Equation \ref{eq:LOB_equation} is convenient from the perspective of human understanding of the functioning of a LOB. However, it suffers a significant drawback from an automated learning standpoint: indeed, there is no guarantee of homogeneous spatial separation between consecutive price levels. It is worth noticing that, when exacerbated by specific stock's microstructural properties (see the research works by \citet{sirignano2021universal,bouchaud2018trades, briola2024deep}), such heterogeneity in the spatial distribution of LOB data sensibly reduces the ability of specific classes of deep-learning models (e.g., Convolutional Neural Networks) in the micro alphas' discovering process \cite{wu2021towards}.

\subsection{Deep Learning for Limit Order Book Forecasting}
The difficulty in handling the complexity expressed by LOBs and the related data abundance has fostered the development of deep learning algorithms to solve related modeling and forecasting tasks. Among them, we are particularly interested in architectures designed to forecast the direction of mid-price changes at a high-frequency resolution. Foundational contributions in the field are offered by \citet{sirignano2019deep}, \citet{sirignano2021universal}, \citet{tsantekidis2017using, tsantekidis2017forecasting} and \citet{passalis2017time}. These studies introduce the use of Multilayer Perceptron (MLP), Long Short-Term Memory (LSTM) \cite{hochreiter1997long}, Convolutional Neural Network (CNN) \cite{lecun2015deep}, and Bag-of-Features (BoF) \cite{o2011introduction} architectures as viable approaches for the forecasting task. Subsequently, these modules served as core components in more complex architectures. This trend emerged in the works of \citet{zhang2019deep} and \citet{tsantekidis2020using}, where the authors utilize convolutional filters to capture the spatial structure of the LOB, as well as LSTM modules to capture long-term time dependencies, and in the works by \citet{passalis2020temporal} and \citet{tran2022attention}, where the authors enrich the BoF paradigm for LOB forecasting through the introduction of the attention mechanism \cite{vaswani2017attention}. Other relevant works are the ones by \citet{tran2018temporal, tran2021data} and \citet{shabani2022multi, shabani2023augmented}, where the authors propose an architecture that incorporates the idea of bi-linear projection as well as of attention to focus on crucial temporal and spatial information embedded in LOBs. 

Concerning the integration of attention mechanisms in LOB forecasting attempts, it is worth mentioning the work by \citet{guo2023forecasting}, where the authors introduce a dual-stage temporal attention mechanism to repeatedly highlight the most valuable time-dimension information, and the works by \citet{wallbridge2020transformers}, \citet{kisiel2022axial}, and \citet{zhang2021deep}, which use transformer-based architectures to accomplish similar forecasting tasks. Lastly, it is relevant to mention the research by \citet{briola2020deep, briola2024deep}, \citet{lucchese2022short} and \citet{kolm2023deep, kolm2024improving}, where the authors critically assess the efficacy of methodologies mentioned previously in this section to understand their effectiveness under different evaluation conditions.

\subsection{Information Filtering Networks \& Homological (Convolutional) Neural Networks}
One of the main contributions of this paper is the introduction of HLOB, a novel large-scale deep learning model for mid-price change forecasting at a high-frequency resolution. This architecture is designed to capture and exploit complex dependencies at deeper LOB levels. This approach overcomes traditional CNN-LSTM models (e.g., DeepLOB \cite{zhang2019deeplob}), which only capture dependencies between consecutive LOB levels, being inadequate to fully handle the inherent complexity of the underlying system. 

The key theoretical prior behind HLOB is represented by Information Filtering Networks (IFNs). \cite{mantegna1999hierarchical, aste2005complex, barfuss2016parsimonious, massara2016network, tumminello2005tool}. IFNs are an effective tool to represent and model dependency structures among variables characterizing complex systems through the instruments of network science, while imposing strict topological constraints (e.g., being a tree or a planar graph) and optimizing global properties (e.g., the model's likelihood) \cite{aste2022topological}. The filtering process can be performed in many different ways; historically, the three main examples of IFNs have been (i) the Minimum Spanning Tree (MST) \cite{west2001introduction}; (ii) the Planar Maximally Filtered Graph (PMFG) \cite{aste2006dynamical, tumminello2007correlation}; and (iii) the Triangulated Maximally Filtered Graph (TMFG) \cite{massara2017network}. In this paper, we are mainly interested in the latter. The TMFG captures higher-order relationships among up to four variables per clique being planar\,\footnote{A graph is said to be planar if it can be embedded in a sphere without edges crossing.} and chordal\,\footnote{A graph is said to be chordal if all cycles made of four or more vertices have a chord which reduces the cycle to a set of triangles. A chord is defined as an edge that is not part of the cycle but connects two vertices of the cycle itself.}, and maximizes the likelihood of the underlying system by deterministically joining in a recursive way covariates expressing the highest similarity \cite{massara2017network, briola2022dependency}. This class of IFNs also inspired the groundbreaking class of Deep Neural Networks at the core of HLOB: the Homological Convolutional Neural Networks (HCNNs) \cite{briola2023homological}. This architecture, which has an archetype in the simpler class of Homological Neural Network \cite{wang2023homological}, is entirely data-centric and leverages the power of convolutions to take advantage of the topological priors in the TMFG \cite{briola2023homological}. An in-depth description of the building process of a TMFG, of an HCNN, and of the HLOB originating from them, is provided in Sections \ref{sec:TMFG_building_process} and \ref{sec:HCNN_building_process}.

\section{Data}\label{sec:Data}
We analyze $15$ stocks from $6$ sectors and $13$ industries, all listed on the NASDAQ exchange. The chosen dataset was originally proposed by \citet{briola2024deep} and contains only assets maintaining a large- (i.e., $10$B-$200$B) to -mega (i.e., $\geq 200$B) capitalization on a $3$-year analysis period spanning from January $2017$ to December $2019$. Stock-related information are summarised in Table \ref{tab:stocks_introduction}, where the assets are organized into $3$ groups based on their tick size. The \textit{first group} (i.e., CHTR, GOOG, GS, IBM, MCD, NVDA) contains `small-tick stocks' (i.e., the stocks characterized by $\langle \sigma \rangle \geq 3\theta$, where $\langle \sigma \rangle$ indicates the average bid-ask spread). The \textit{second group} (i.e., AAPL, ABBV, PM) contains `medium-tick stocks' (i.e., the stocks characterized by $1.5\theta \lesssim \langle \sigma \rangle \lesssim 3\theta$). The \textit{third group} (i.e., BAC, CSCO, KO, ORCL, PFE, VZ) contains large-tick stocks (i.e., the stocks characterized by $\langle \sigma \rangle \lesssim 1.5\theta$). An in-depth description of the effectiveness of this classification in capturing stocks- and class-related microstructural effects is available in the original research work \cite{briola2024deep}.

{
\vspace{0.1cm}
\renewcommand{\arraystretch}{2.3}
\begin{table}[H]
    \centering
    \caption{Overview of the stocks used in the paper. For each asset, we report the ticker, the extended name, the sector, the industry and the capitalization during $2017$, $2018$ and $2019$. To determine stocks' sector and industry affiliation, we follow the taxonomy proposed by the NASDAQ exchange \cite{nasdaq_stock_screener}. To determine the stock's capitalization, we rely on the data provided by \href{https://companiesmarketcap.com}{companiesmarketcap.com} \cite{capitalization_provider}.}
    \label{tab:stocks_introduction}
    \resizebox{\columnwidth}{!}{%
        \begin{tabular}{@{}ccccccc@{}}
        \toprule
        \textbf{Stock Symbol} &
          \textbf{Stock Name} &
          \textbf{Sector} &
          \textbf{Industry} &
          \textbf{Capitalization (2017)} &
          \textbf{Capitalization (2018)} &
          \textbf{Capitalization (2019)} \\ \midrule
        CHTR & Charter Communications, Inc.                & Telecommunications     & Cable \& Other Pay Television Services          & \$83.94 B  & \$64.21 B  & \$101.85 B \\
        GOOG & Alphabet, Inc.                              & Technology             & Computer Software: Programming, Data Processing & \$729.45 B & \$723.55 B & \$921.13 B \\
        GS   & Goldman Sachs Group, Inc.                   & Finance                & Investment Bankers/Brokers/Service              & \$96.09 B  & \$61.43 B  & \$79.86 B  \\
        IBM  & International Business Machines Corporation & Technology             & Computer Manufacturing                          & \$142.03 B & \$101.44 B & \$118.90 B \\
        MCD  & McDonald's Corporation                      & Consumer Discretionary & Restaurants                                     & \$137.21 B & \$136.21 B & \$147.47 B \\
        NVDA & NVIDIA Corporation                          & Technology             & Semiconductors                                  & \$117.26 B & \$81.43 B  & \$144.00 B \\ \midrule
        AAPL & Apple, Inc.                                 & Technology             & Computer Manufacturing                          & \$860.88 B & \$746.07 B & \$1.287 T  \\
        ABBV & AbbVie, Inc.                                & Health Care            & Biotechnology: Pharmaceutical Preparations      & \$154.39 B & \$136.33 B & \$130.94 B \\
        PM   & Philip Morris International, Inc.           & Health Care            & Medicinal Chemicals and Botanical Products      & \$164.09 B & \$103.78 B & \$132.39 B \\ \midrule
        BAC  & Bank of America Corporation                 & Finance                & Major Banks                                     & \$307.91 B & \$238.25 B & \$311.20 B \\
        CSCO & Cisco Systems, Inc.                         & Telecommunications     & Computer Communications Equipment               & \$189.34 B & \$194.81 B & \$203.45 B \\
        KO   & Coca-Cola Company                           & Consumer Staples       & Beverages (Production/Distribution)             & \$195.47 B & \$202.08 B & \$236.89 B \\
        ORCL & Oracle Corporation                          & Technology             & Computer Software: Prepackaged Software         & \$195.72 B & \$162.03 B & 169.94 B   \\
        PFE  & Pfizer, Inc.                                & Health Care            & Biotechnology: Pharmaceutical Preparations      & \$215.89 B & \$249.54 B & \$216.82 B \\
        VZ   & Verizon Communications, Inc.                & Telecommunications     & Telecommunications Equipment                    & \$215.92 B & \$232.30 B & \$253.93 B \\ \bottomrule
        \end{tabular}
    }
\end{table}
\vspace{0.1cm}
}

For each stock, high-resolution, tick-by-tick LOB data obtained from the LOBSTER provider \cite{lobsterdata_what_is_lobster} are employed. For each trading day, we use a LOB characterized by $L=10$ price and volume levels for both the bid and ask sides (see Equation \ref{eq:LOB_equation}). As outlined in Table \ref{tab:training_validation_test_split}, for each year, we allocate $40$ days for training, $5$ days for validation, and $10$ consecutive days for testing. Notably, the training days are chosen to form a sequence where most of the entries are consecutive, with only few exceptions. Indeed, the $5$ days of validation are randomly selected from the same period characterizing the training set. This choice guarantees greater robustness in the validation step, and it is made possible by the $5$-days feature-wise rolling window \textit{z}-score standardization procedure, which prevents any data leakage \cite{briola2024deep}. The raw LOB data are processed in accordance with the rigorous pipeline initially proposed by \citet{lucchese2022short} and subsequently refined by \citet{briola2024deep}.

{
\vspace{0.1cm}
\renewcommand{\arraystretch}{2}
\begin{table}[H]
\centering
\caption{Basic structure of the datasets used during the training, validation and test stage. For each year, for the training and test set, we report the starting and the ending day (both included in the analysis), while, for the validation set, we report all the dates explicitly. It is worth noting that weekends and public holidays are not trading days and, consequently, do not belong to any of the datasets.}
\scriptsize
\label{tab:training_validation_test_split}
\begin{tabular}{c|cc|c|cc}
\hline
\textbf{year} & \multicolumn{2}{c|}{\textbf{training}} & \textbf{validation}                                                           & \multicolumn{2}{c}{\textbf{test}} \\ \cline{2-6} 
\textbf{}     & \textbf{from}       & \textbf{to}      & \textbf{days}                                                                 & \textbf{from}    & \textbf{to}    \\ \hline
2017          & 03-13               & 05-22            & \begin{tabular}[c]{@{}c@{}}03-23, 04-05,\\ 04-13, 04-18,\\ 05-02\end{tabular} & 05-23            & 06-06          \\ \hline
2018          & 08-09               & 10-18            & \begin{tabular}[c]{@{}c@{}}08-15, 08-16,\\ 09-19, 09-26\\ 10-03\end{tabular}  & 10-19            & 11-01          \\ \hline
2019          & 06-04               & 08-13            & \begin{tabular}[c]{@{}c@{}}06-14, 06-27,\\ 07-08, 07-10,\\ 07-24\end{tabular} & 08-14            & 08-27          \\ \hline
\end{tabular}
\end{table}
\vspace{0.1cm}
}

Consistently with the work by \citet{briola2024deep}, we study the predictability of mid-price changes' direction\,\footnote{We decide to use the simple difference in mid-prices to gain higher control over the amplitude of the change at different time horizons, preserving, at the same time, the stationarity property of the resulting time series. Many alternatives have been proposed as target variables in the literature (e.g., \cite{zhang2019deeplob, ntakaris2018benchmark, lucchese2022short, tsantekidis2017forecasting}). All of them are based on the usage of the log-return as fundamental quantity, and apply different smoothing methods to mitigate the strong fit between labels and actual prices. While these methods are academically acceptable, their practicability is questionable as they are more tailored towards tracking mid-price trends than immediate fluctuations, thereby offering limited control on tick-by-tick changes crucial for developing high-frequency trading strategies.} at $3$ different horizons (i.e., H$\Delta_\tau \in \{10, 50, 100\}$) when such a movement is larger than or equal to $\theta$. The labeling step is consequently defined as follows:

\vspace{0.1cm}
\begin{equation}\label{eq:labelling_step}
\left\{
    \begin{array}{ll}
        (m_{\tau+\Delta \tau} - m_\tau) \leq - \, \theta \; \rightarrow \; -1 \; \rightarrow \; \text{Down} \ , \\
        - \, \theta < (m_{\tau+\Delta \tau} - m_\tau) < + \, \theta \; \rightarrow \; 0 \; \rightarrow \; \text{Stable}\ , \\
        (m_{\tau+\Delta \tau} - m_\tau) \geq + \, \theta \; \rightarrow \; 1 \; \rightarrow \; \text{Up} \ ,
    \end{array}
\right.
\end{equation}
\vspace{0.1cm}

where $\theta=\$0.01$ is the tick size on the NASDAQ exchange and $m_\tau$ is the mid-price at tick time $\tau$. It is worth noticing that horizons are always defined in terms of LOB updates (which are unevenly spaced), while physical time is never used. 

\section{Methods}\label{sec:Methods}
The HLOB model is centered on two primary mechanisms: (i) exploiting the informational content of topological priors in an IFN as inputs for a tailored version of Homological Convolutional Neural Networks (HCNNs) \cite{briola2023homological}, which in turn handles the dependency structures among LOB's spatial components (i.e., volume levels); and (ii) employing an LSTM module to capture long-term temporal patterns. A complete description of this system requires (i) a preliminary discussion on the process to distillate the necessary information to build the TMFG; and (ii) a detailed description of the required modifications to the original HCNN architecture to make it suitable for processing LOB inputs.

\subsection{The TMFG's Building Process}\label{sec:TMFG_building_process}
The building block of the HCNN and, consequently, of the HLOB architecture, is represented by an arbitrary IFN encoding higher-order dependency structures among variables in the underlying system. In this study, in line with the work by \citet{briola2023homological}, we choose the Triangulated Maximally Filtered Graph (TMFG) \cite{massara2017network}. As a \textit{first step}, we process LOB data by removing price levels from the bid and ask side, focusing solely on volume-related data. Formally, we reduce the dimension of each LOB snapshot from $\mathbb{L}(\tau) \in \mathbb{R}^{4L}$ to $\mathbb{L}(\tau) \in \mathbb{R}^{2L}$. This choice is needed to ensure homogeneity in the information used to build the IFN. Volume levels are inherently discrete, with the minimum tradable quantity set by the exchange's $\psi$ parameter (see Section \ref{sec:Limit_Order_Book}), and minor variations from consecutive LOB updates can introduce a non-negligible level of noise. To mitigate this effect, we categorize volumes into equally spaced bins. The number of bins is optimized on the training and validation set, and remains constant across stocks characterized by different microstructural properties (i.e., small-, medium-, and large-tick stocks). The size of the bins is calculated for each stock individually, and across all volume levels for each training day. 

As a \textit{second step}, for each stock and for each day in the training set, we calculate the pairwise mutual information (MI) between volume levels, obtaining positive and symmetric $(2L \times 2L)$ similarity matrices (i.e., MI matrices). It is worth noticing that the reliability of the MI computation is strengthened through a bootstrapping process applied on a daily basis on LOB data. For each stock, the final MI matrix is derived by averaging daily MI matrices in the training set.

As a \textit{third step}, stock-related TMFGs are computed by using average MI matrices as similarity matrices. We remark that, given the multivariate system $\mathbb{L}$, our primary goal is to estimate the multivariate probability density function $\widetilde{f}(\mathbb{L}|\mathcal{G}^*)$ with representation structure $\mathcal{G}^*$ that best describes the true and unknown $f(\mathbb{L})$. From an information theoretic perspective, the learning of an optimal network representation $\mathcal{G}^*$ consists of minimizing the Kullback-Leibler divergence ($D_{KL}$) \citep{kullback1951information} between $f(\mathbb{L})$ and $\widetilde{f}(\mathbb{L}|\mathcal{G})$, and, consequently, the cross-entropy ($H$) of the underlying system:
            
\begin{equation}\label{eq:new_general_problem_formulation}
    \begin{aligned}
        \mathcal{G}^* &\Rightarrow \arg\min_{\mathcal{G}}{D_{KL}(f(\mathbb{L}) \;||\; \widetilde{f}(\mathbb{L} |\mathcal{G}})) \\
    &\Rightarrow \arg\min_{\mathcal{G}}{\cancel{\mathbb{E}_f (\log f(\mathbb{L}))} - \mathbb{E}_f (\log \widetilde{f}(\mathbb{L} | \mathcal{G}))} \\
    &\Rightarrow \arg\min_{\mathcal{G}}(H(\mathbb{L}|\mathcal{G}))
    \end{aligned}
\end{equation}

The term $\mathbb{E}_f (\log f(\mathbb{L}))$ in Equation \ref{eq:new_general_problem_formulation} is independent from $\mathcal{G}$ and therefore its value is irrelevant to the purpose of discovering the optimal representation network. The second term, $- \mathbb E(\log \widetilde{f}(\mathbb{L} | \mathcal G))$ (notice the minus), instead, depends on $\mathcal G$ and must be minimized. It is the estimate of the entropy of the multivariate system under analysis and corresponds to the so-called cross-entropy ($H$). This minimization problem can be incrementally solved by joining the system's disconnected parts sharing the largest MI, which is exactly what the TMFG algorithm does (see  \citet{massara2017network}).

\vspace{0.1cm}
\begin{figure}[H]
    \centering
    \includegraphics[scale=0.55]{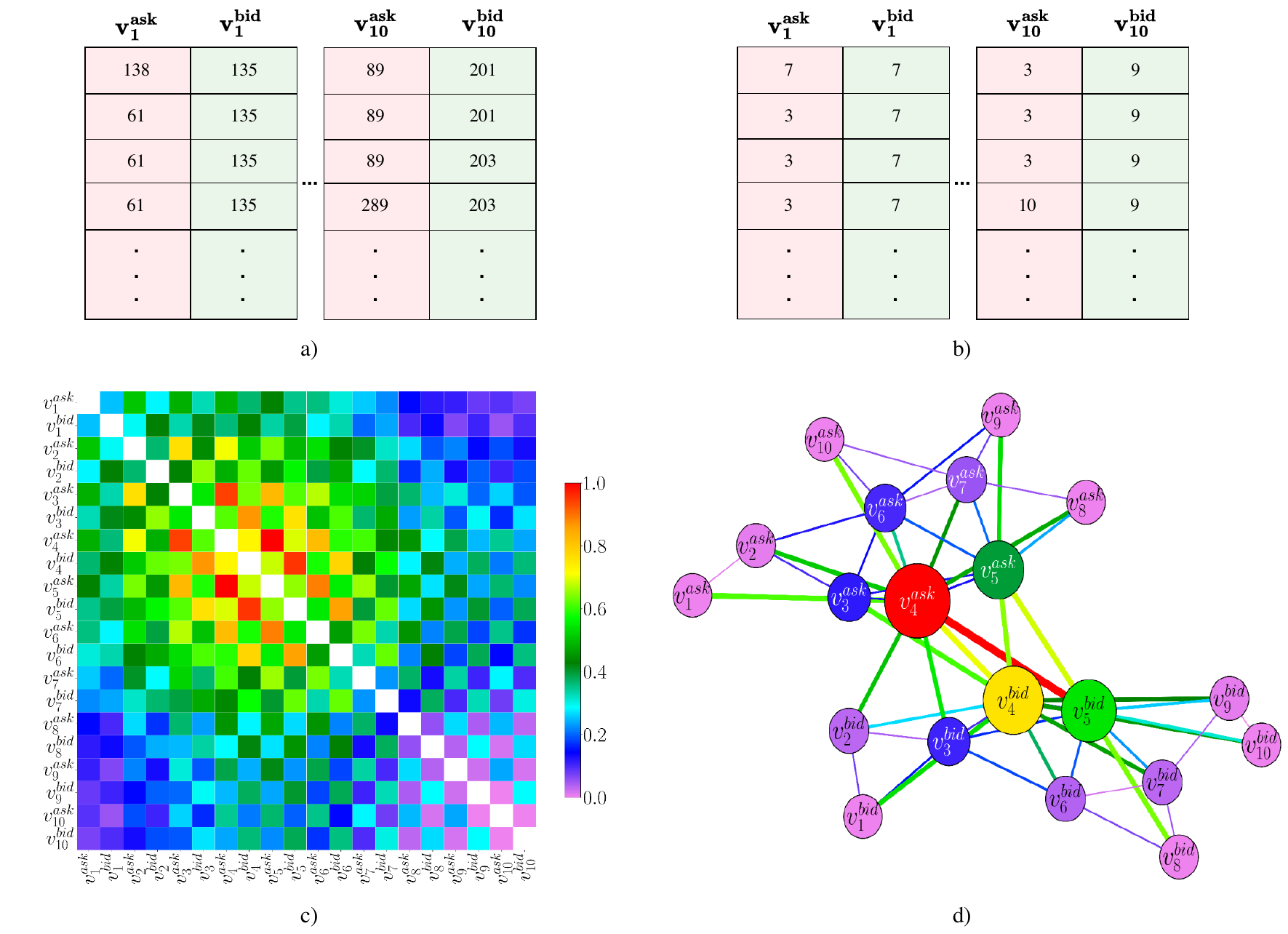}
    \caption{Schematic representation of the TMFG's building process: (a) we start from a simplified version of the LOB containing only volumes data; (b) we mitigate the noise affecting the LOB by categorizing volumes into bins of uniform size; (c) we compute the pairwise MI between volume levels; and (d) we build the TMFG using the MI matrix as input. We remark that in the proposed graph representation, both nodes' and edges' color/dimension depend on their betweenness centrality. The color bar remains consistent for both the MI matrix and the corresponding TMFG representation.}
    \label{fig:TMFG_pipeline}
\end{figure}
\vspace{0.1cm}

\subsection{From HCNN to HLOB}\label{sec:HCNN_building_process}
From each TMFG computed as described in Section \ref{sec:TMFG_building_process}, we isolate the realizations of $3$ simplicial families: (i) maximal cliques with size $4$ (i.e., $3$-dimensional simplices or \textit{tetrahedra}); (ii) maximal cliques with size $3$ (i.e., $2$-dimensional simplices or \textit{triangles}); (iii) maximal cliques with size $2$ (i.e., $1$-dimensional simplices or \textit{edges}). These three higher-order structures are sufficient to capture all the dependencies described by the chosen IFN. Given that the number of observed volume levels is constant across different stocks and trading days, we can deterministically compute (i) the shape of the vector of tetrahedra ($17 \times 4$); (ii) the shape of the vector of triangles ($52 \times 3$); and (iii) the shape of the vector of edges, ($54 \times 2$). All of them serve as input for the HLOB model, which, however, is designed to handle not only the \textit{spatial} dynamics captured by the TMFG, but also the \textit{temporal} dynamics of the LOB. In this sense, as model's input, consistently with the work by \citet{zhang2019deeplob}, we also use a history window of $100$ LOB's updates\,\footnote{We remark the existence of strong designing analogies between the HLOB model and the DeepLOB one \cite{zhang2019deeplob}, which, indeed, represents an archetype for the architecture introduced in this research paper. For this reason, in Section \ref{sec:Results}, we will systematically discuss the comparison between the forecasting performances of these two models.}.

\vspace{0.1cm}
\begin{figure}[H]
    \centering
    \begin{adjustbox}{center}
        \includegraphics[scale=0.35]{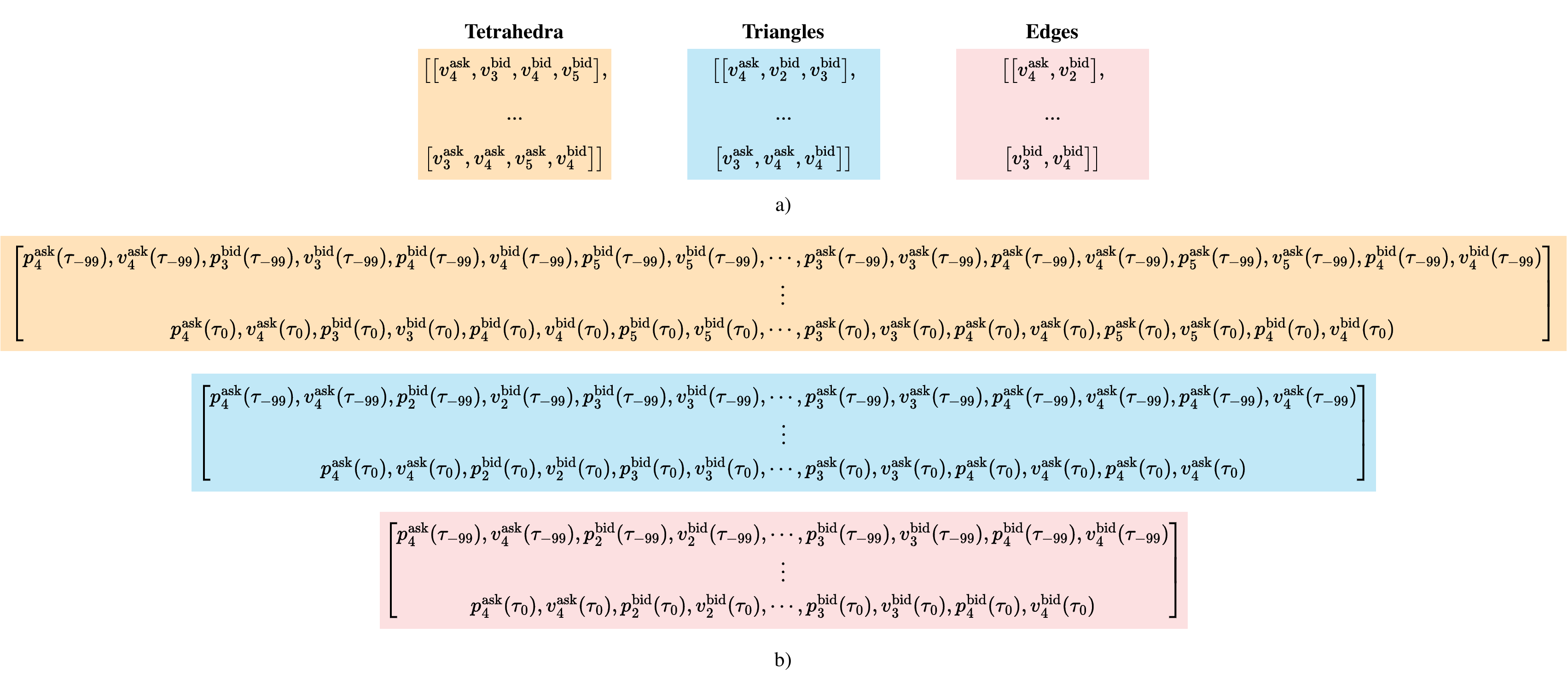}
    \end{adjustbox}
    \caption{This diagram illustrates the sequence of steps transitioning (a) from the output of the TMFG building process (b) to the input of the HLOB model. To construct the TMFG, we exclusively utilize volume levels from the LOB, forming a network characterized by three topological structures: tetrahedra, triangles, and edges. To prepare the inputs for the HLOB model, we perform two main tasks: (i) for each timestamp in the input's temporal dimension, we flatten each of the aforementioned sets; (ii) we incorporate the corresponding price levels' data into each representative of these three new input sets. Note that there is a  direct mapping between the colours used in this Figure and the ones used later to highlight the inputs of the HLOB model in Figure \ref{fig:HCNN_schema}.}
    \label{fig:input_definition}
\end{figure}
\vspace{0.1cm}

As described in Section \ref{sec:TMFG_building_process}, we use only the volume levels in the building process of the TMFG. However, price levels carry significant information that cannot be ignored. For this reason, we include them in the HLOB's building stage: for each timestamp constituting the input's historical dimension, we flat the vector of tetrahedra, triangles, and edges, and, for each volume level, we insert the corresponding price level. This transformation is schematically depicted in Figure \ref{fig:input_definition}, where we show the flattening step for the $3$ sets of simplicial families constituting the average TMFG, and the insertion of the price levels for each timestamp in the historical dimension of the HLOB's input. This operation produces $3$ conceptually new 2D input vectors: (i) one of size $(100 \times 136)$ (i.e., the re-shaped vector of tetrahedra); (ii) one of size $(100 \times 312)$ (i.e., the re-shaped vector of triangles); and (iii) one of size $(100 \times 216)$ (i.e., the re-shaped vector of edges). Each vector is separately passed as input to one head of the HLOB model. 

\vspace{0.1cm}
\begin{figure}[H]
    \centering
    \begin{adjustbox}{center}
        \includegraphics[scale=0.5]{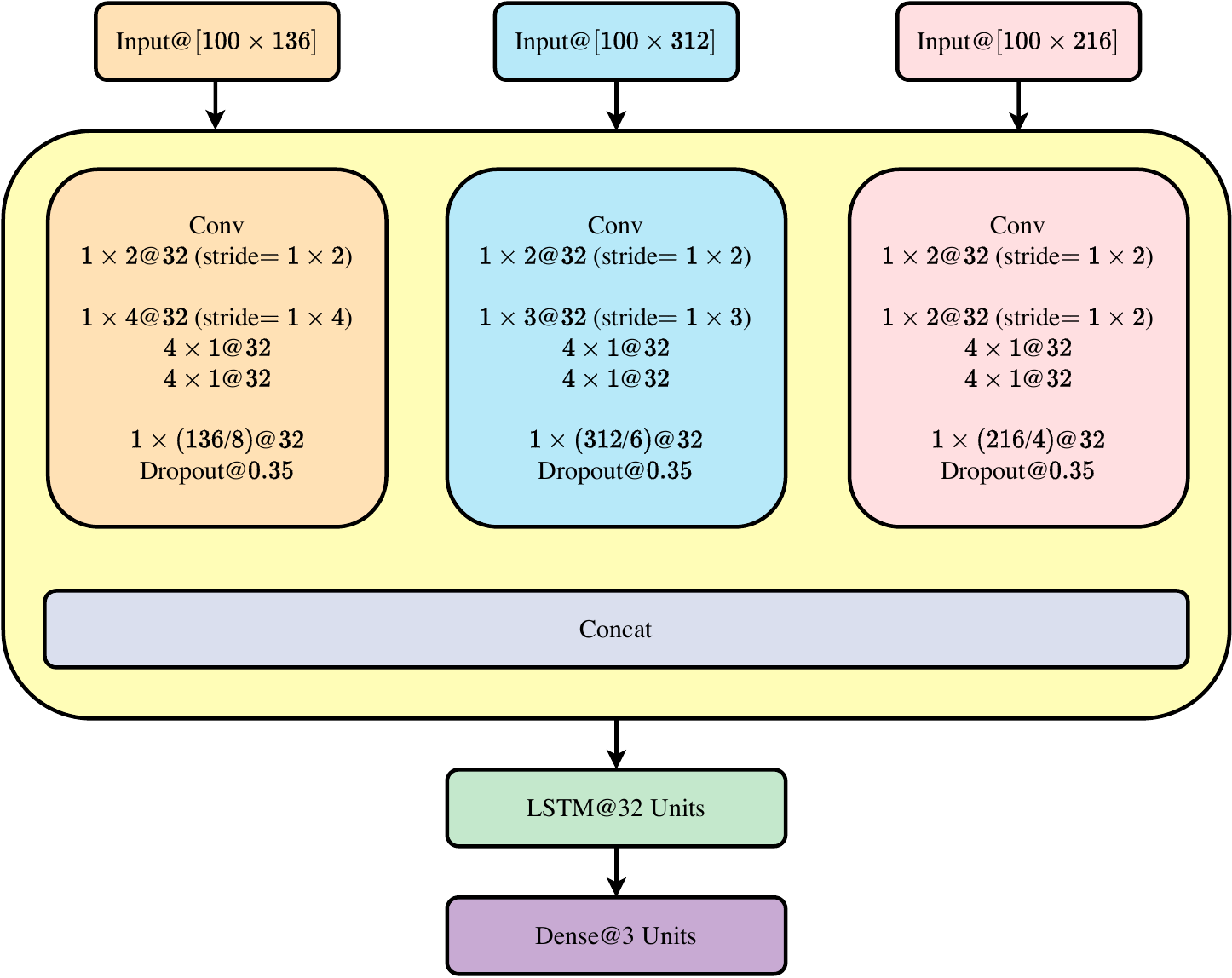}
    \end{adjustbox}
    \caption{Visual overview of the HLOB model's operational framework. Note that   there is a direct mapping between the colors used to denote the inputs of the HLOB model here and the colors used to represent the three categories of topological priors derived from a TMFG in Figure \ref{fig:input_definition}.}
    \label{fig:HCNN_schema}
\end{figure}
\vspace{0.1cm}

For each head, the size of the \textit{first convolutional filter} is $(1 \times 2)$ with a stride of $(1 \times 2)$. As described in the work by \citet{zhang2019deeplob}, this first layer summarises the information between the price and the volume $\{p_\ell^s, v_\ell^s\}_{\ell \in L}^{s \in \{\text{ask, bid}\}}$ at level $\ell$ and side $s$ of the LOB. At the same time, the stride prevents parameter sharing between geographically (but not logically) consecutive inputs. The number of parameters corresponding to this operation equals $96$ for each of the three heads of the architecture. The \textit{second convolutional layer} captures the relationships between components of a single realization of each simplicial family: (i) in the case of tetrahedra, between nodes composing each $3-$dimensional simplex (i.e., $4$-cliques); (ii) in the case of triangles, between nodes composing each $2-$dimensional simplex (i.e., $3$-cliques); (iii) in the case of edges, between nodes composing each $1-$dimensional simplex (i.e., $2$-cliques). A stride equal to $(1 \times 4)$ for tetrahedra, $(1 \times 3)$ for triangles, and $(1 \times 2)$ for edges, one more time, prevents parameter sharing between components of the same simplicial family. Here, the number of parameters for the convolutional operation involving tetrahedra is equal to $12\,384$, the number of parameters for the convolutional operation involving triangles is equal to $11\,360$, while the number of parameters involving edges is equal to $10\,336$. The \textit{third convolutional layer} captures the relationships between components of each simplicial family. The size of the convolutional filter is $(1 \times \Omega)$, where $\Omega$ is the cardinality of each original set of simplexes: $136/8 = 17$ in the case of tetrahedra, $312/6 = 52$ in the case of triangles, and $216/4 = 54$ in the case of edges. This further level of convolution is proved to be effective (see the work by \citet{briola2023homological}) in capturing information that is not necessarily related in the original network representation, but that can positively affect the characterization of the true but unknown $f(\mathbb{L})$ (see Equation \ref{eq:new_general_problem_formulation}). Since relationships modeled in this layer do not directly stem from the structure of the underlying TMFG, for each head of the HLOB, we apply a dropout with a rate of $0.35$. The number of parameters for this convolutional layer equals $17\,440$ in the case of tetrahedra, $53\,280$ in the case of triangles, and $55\,328$ in the case of edges. After these three layers of convolution, the dimension of each head's feature map is $(100 \times 1)$. These outputs are concatenated and passed through an \textit{LSTM module} to capture long-term temporal dependencies. The activation of an LSTM unit is fed back to itself, and the memory of past activations is kept with a separate set of weights, so the temporal dynamics of input features can be effectively modeled. The number of parameters of this additional layer is equal to $16\,640$. Lastly, the \textit{output layer} consists of a linear layer with a number of outputs equal to the number of classes. The model returns the logits for increased numerical stability while associated probabilities are computed in a separate stage.

\subsection{Experimental Settings}\label{sec:Experimental_Settings}
We test the HLOB architecture against $6$ state-of-the-art (SOTA) models in LOB mid-price changes forecasting: (i) CNN1 \cite{tsantekidis2017forecasting}; (ii) CNN2 \cite{tsantekidis2020using}; (iii) DLA \cite{guo2023forecasting}; (iv) BinBTabl \cite{tran2021data}; (v) BinCTabl \cite{tran2021data}; (vi) DeepLOB \cite{zhang2019deeplob}. All these models were proposed in the scientific literature between $2017$ and $2022$, and later systematically organized in the review paper by \citet{prata2023lob}. We also test our model against $2$ pure transformer-based architectures for time-series forecasting that we adapt for LOB mid-price changes forecasting: (i) Transformer \cite{vaswani2017attention}; (ii) iTransformer \cite{liu2023itransformer}. Finally, as an additional benchmark model, we combine the power of Transformers and CNNs in the LobTransformer architecture, which takes inspiration from the work of \citet{wallbridge2020transformers}, and is proposed here in a revised version. As pointed out in the work by \citet{briola2024deep}, the majority of these architectures suffer from a fundamental drawback in the science domain: the original code is not provided, severely compromising the results' reproducibility. For the first set of models described above, results discussed in the current paper are obtained by exploiting the code provided by \citet{prata2023lob}. All the other architectures are implemented from scratch. All models are included in the `LOBFrame' \cite{briola2024deep} pipeline to simplify their execution, while guaranteeing the highest reproducibility standards. A summary of the benchmark models is reported in Table \ref{tab:models_summary}.

{
\vspace{0.1cm}
\renewcommand{\arraystretch}{3}
\centering
\begin{table}[H]
\centering
\resizebox{\columnwidth}{!}{%
\begin{tabular}{c|c|c|c|c|c|c|c|c|c|c}
\toprule
 & \rotatebox{90}{\shortstack{\textbf{Tsantekidis et al. \cite{tsantekidis2017forecasting}} \\ \textbf{CNN1 (2017)}}} &
  \rotatebox{90}{\shortstack{\textbf{Tsantekidis et al. \cite{tsantekidis2020using}} \\ \textbf{CNN2 (2020)}}} &
  \rotatebox{90}{\shortstack{\textbf{Guo et al. \cite{guo2023forecasting}} \\ \textbf{DLA (2022)}}} &
  \rotatebox{90}{\shortstack{\textbf{Vaswani et al. \cite{vaswani2017attention}} \\ \textbf{Transformer (2017)}}} &
  \rotatebox{90}{\shortstack{\textbf{Liu et al. \cite{liu2023itransformer}} \\ \textbf{iTransformer (2023)}}} &
  \rotatebox{90}{\shortstack{\textbf{Briola et al.} \\ \textbf{LobTransformer (2024)}}} &
  \rotatebox{90}{\shortstack{\textbf{Tran et al. \cite{tran2021data}} \\ \textbf{BinBTabl (2021)}}} &
  \rotatebox{90}{\shortstack{\textbf{Tran et al. \cite{tran2021data}} \\ \textbf{BinCTabl (2021)}}} &
  \rotatebox{90}{\shortstack{\textbf{Zhang et al. \cite{zhang2019deeplob}} \\ \textbf{DeepLOB (2019)}}} &
  \rotatebox{90}{\shortstack{\textbf{Briola et al.} \\ \textbf{HLOB (2024)}}} \\ \midrule
\shortstack{\textbf{original code} \\ \textbf{availability}} & \xmark & \xmark & \xmark & - & - & - & \xmark & \xmark & \cmark & - \\ \hline
\shortstack{\textbf{n. trainable} \\ \textbf{parameters}} & $3.5\times10^4$ & $2.8\times10^5$ & $2.2\times10^5$ & $1.1\times10^5$ & $1.1\times10^5$ & $2.0\times10^6$ & $6.6\times10^3$ & $2.2\times10^4$ & $1.4\times10^5$ & $1.8\times10^5$ \\ \hline
\shortstack{\textbf{inference} \\ \textbf{time (\textit{ms})}} & 0.07 & 0.14 & 0.15 & 0.16 & 0.15 & 0.29 & 0.19 & 0.13 & 0.16 & 0.16 \\ \bottomrule
\end{tabular}
}
\caption{We report a summary of three main characteristics of benchmark models: (i) original code availability; (ii) model's number of trainable parameters; and (iii) model's inference time in milliseconds. The original code is not provided for $5$ out of $6$ of the models having a direct reference in the literature. BinBTabl is the most parsimonious among benchmark models with a number of trainable parameters equal to $6.6 \times 10^3$, while LobTransformer is the less parsimonious one with a number of trainable parameters equal to $2.0 \times 10^6$. The model with the lowest inference time is CNN1 (i.e., $0.07$\textit{ms}, while the model with the highest inference time is LobTransformer (i.e., $0.29$\textit{ms}).}
\label{tab:models_summary}
\end{table}
\vspace{0.1cm}
}

When possible, model-specific hyper-parameters are inherited from the work by \citet{prata2023lob}, while optimal weights are learned by minimizing the categorical cross-entropy loss using mini-batches of size $32$ \cite{zhang2019deeplob}. The mini-batches sampling procedure differs for the training, validation, and test sets. During training, the (sub)-sampling is random and balanced. From each trading day (see Table \ref{tab:training_validation_test_split}), we detect the number of samples for the least represented class, and (i) if this value is $\geq 5000$, then we sample $5000$ random representatives for each of the three classes (see Eq. \ref{eq:labelling_step}), otherwise, (ii) if this value is $< 5000$, we sample a number of random representatives for each class equal to the number of samples for the least represented class. During validation and test stages, we still sample mini-batches with a size of $32$, but they are always sequential and cover the totality of data in the two sets. In line with the related literature \cite{zhang2019deeplob}, all models are trained for a maximum number of epochs equal to $100$. Training halts if the validation loss fails to drop by at least $0.003$ units over a span of $15$ consecutive epochs. We use a modified version of the Adam optimizer \cite{kingma2014adam} with decoupled weight decay \cite{loshchilov2017decoupled}, commonly known as `AdamW'. Following the latest applied research findings \cite{nanoGPT, brown2020language}, we use a learning rate equal to $6 \times 10^{-5}$, a $\beta_1$ decay rate equal to $0.90$, and a $\beta_2$ decay rate equal to $0.95$. As described in the work by \citet{briola2024deep}, the choice of values for these parameters is determined by the training pipeline described above.

All the models considered in this paper are coded in Python using the PyTorch deep learning library \cite{paszke2019pytorch}. Experiments are run on the University College London Computer Science Department’s High-Performance Computing Cluster \cite{ucl2023cluster}. Given the $15$ stocks in Table \ref{tab:stocks_introduction}, and knowing that for each of the $3$ years we challenge the $10$ models described in Table \ref{tab:models_summary} on $3$ prediction horizons, we obtain that the number of year-wise experiments is equal to $450$. Consequently, the total number of executed experiments is equal to $1\,350$ for a cumulative GPU runtime of $7\,192$ hours, $20$ minutes, and $31$ seconds. To accomplish the task, we used $10$ different GPU models: (i) NVIDIA A$100$ $80$GB PCIe ($21$ experiments); (ii) NVIDIA A100-PCIE-40GB ($6$ experiments); (iii) NVIDIA GeForce GTX $1080$ Ti ($362$ experiments); (iv) NVIDIA GeForce RTX $2080$ Ti ($439$ experiments); (v) NVIDIA GeForce RTX $4090$ ($187$ experiments); (vi) NVIDIA RTX $6000$ Ada Generation ($139$ experiments); (vii) NVIDIA TITAN X (Pascal) ($96$ experiments); (viii) NVIDIA TITAN Xp ($60$ experiments); (ix) Tesla V100-PCIE-$16$GB ($10$ experiments); (x) Tesla V100-PCIE-$32$GB ($30$ experiments).

\section{Results}\label{sec:Results}
We present the results of our analysis (i) evaluating the effectiveness of the models introduced in Section \ref{sec:Experimental_Settings} in predicting the direction of mid-price changes; and (ii) examining the HLOB behavior to unveil intricate patterns into the LOB levels' structural dependencies. In all the experiments, we assess the behavior for the three classes of stocks (i.e., small-, medium- and large-tick stocks) at $\text{H}\Delta_\tau \in \{10, 50, 100\}$. This approach enables us to examine the models' performances across different scenarios, thereby linking their effectiveness to the microstructural characteristics of the stocks.

\subsection{Comparison of Model Performances}\label{sec:Models_Performance_Comparison}
We investigate models' effectiveness in predicting mid-price change direction through $3$ key metrics: (i) the F1 score; (ii) the Matthews Correlation Coefficient (MCC) \cite{gorodkin2004comparing}; and (iii) the probability of correctly executing a round-trip transaction ($p_{\text{T}}$) \cite{briola2024deep}. We report the results of this analysis in Tables \ref{tab:comprehensive_comparison_models_h10}, \ref{tab:comprehensive_comparison_models_h50}, and \ref{tab:comprehensive_comparison_models_h100}, highlighting the best performing model (green), the second-best performing model (blue) and the worst performing alternative (red). For each stock, a model is considered superior to the others if the sum of the $3$ performance metrics is maximal. Year-wise metrics are computed, and, for each horizon $\text{H}\Delta_\tau \in \{10, 50, 100\}$, only the average value is provided. 

{
\centering
\renewcommand{\arraystretch}{2}
\begin{table}[H]
\centering
\begin{adjustbox}{center,max width=\linewidth}
\caption{Models' performances at $\text{H}\Delta_\tau = 10$. For each deep learning architecture we report three key metrics: (i) the F1 score; (ii) the MCC; and (iii) the $p_{\text{T}}$. For each stock, we highlight the best performing model (green), the second-best performing model (blue) and the worst performing alternative (red); a model is considered superior to the others if the sum of the $3$ performance metrics is maximal.}
\label{tab:comprehensive_comparison_models_h10}
\tiny
\setlength{\tabcolsep}{1pt}
\begin{tabular}{c|cccccccccccccccccccccccccccccc}
\hline
 &
  \multicolumn{30}{c}{\textbf{H10}} \\ \cline{2-31} 
\multirow{-2}{*}{} &
  \multicolumn{3}{c|}{\textbf{cnn1}} &
  \multicolumn{3}{c|}{\textbf{cnn2}} &
  \multicolumn{3}{c|}{\textbf{dla}} &
  \multicolumn{3}{c|}{\textbf{transformer}} &
  \multicolumn{3}{c|}{\textbf{itransformer}} &
  \multicolumn{3}{c|}{\textbf{lobtransformer}} &
  \multicolumn{3}{c|}{\textbf{binbtabl}} &
  \multicolumn{3}{c|}{\textbf{binctabl}} &
  \multicolumn{3}{c|}{\textbf{deeplob}} &
  \multicolumn{3}{c}{\textbf{hlob}} \\ \cline{2-31} 
 &
  \textbf{F1} &
  \textbf{MCC} &
  \multicolumn{1}{c|}{\textbf{$p_T$}} &
  \textbf{F1} &
  \textbf{MCC} &
  \multicolumn{1}{c|}{\textbf{$p_T$}} &
  \textbf{F1} &
  \textbf{MCC} &
  \multicolumn{1}{c|}{\textbf{$p_T$}} &
  \textbf{F1} &
  \textbf{MCC} &
  \multicolumn{1}{c|}{\textbf{$p_T$}} &
  \textbf{F1} &
  \textbf{MCC} &
  \multicolumn{1}{c|}{\textbf{$p_T$}} &
  \textbf{F1} &
  \textbf{MCC} &
  \multicolumn{1}{c|}{\textbf{$p_T$}} &
  \textbf{F1} &
  \textbf{MCC} &
  \multicolumn{1}{c|}{\textbf{$p_T$}} &
  \textbf{F1} &
  \textbf{MCC} &
  \multicolumn{1}{c|}{\textbf{$p_T$}} &
  \textbf{F1} &
  \textbf{MCC} &
  \multicolumn{1}{c|}{\textbf{$p_T$}} &
  \textbf{F1} &
  \textbf{MCC} &
  \textbf{$p_T$} \\ \hline
CHTR &
  0.39 &
  0.11 &
  \multicolumn{1}{c|}{0.06} &
  0.38 &
  0.09 &
  \multicolumn{1}{c|}{0.05} &
  0.39 &
  0.11 &
  \multicolumn{1}{c|}{0.04} &
  0.40 &
  0.12 &
  \multicolumn{1}{c|}{0.06} &
  0.35 &
  0.05 &
  \multicolumn{1}{c|}{0.04} &
  {\color[HTML]{FE0000} \textbf{0.31}} &
  {\color[HTML]{FE0000} \textbf{0.07}} &
  \multicolumn{1}{c|}{{\color[HTML]{FE0000} \textbf{0.04}}} &
  0.42 &
  0.15 &
  \multicolumn{1}{c|}{0.06} &
  {\color[HTML]{3531FF} \textbf{0.43}} &
  {\color[HTML]{3531FF} \textbf{0.16}} &
  \multicolumn{1}{c|}{{\color[HTML]{3531FF} \textbf{0.06}}} &
  0.39 &
  0.10 &
  \multicolumn{1}{c|}{0.05} &
  {\color[HTML]{32CB00} \textbf{0.43}} &
  {\color[HTML]{32CB00} \textbf{0.17}} &
  {\color[HTML]{32CB00} \textbf{0.06}} \\
GOOG &
  0.42 &
  0.16 &
  \multicolumn{1}{c|}{0.04} &
  0.42 &
  0.16 &
  \multicolumn{1}{c|}{0.04} &
  0.39 &
  0.13 &
  \multicolumn{1}{c|}{0.03} &
  0.41 &
  0.15 &
  \multicolumn{1}{c|}{0.04} &
  {\color[HTML]{FE0000} \textbf{0.27}} &
  {\color[HTML]{FE0000} \textbf{0.04}} &
  \multicolumn{1}{c|}{{\color[HTML]{FE0000} \textbf{0.04}}} &
  0.44 &
  0.18 &
  \multicolumn{1}{c|}{0.05} &
  0.45 &
  0.18 &
  \multicolumn{1}{c|}{0.08} &
  {\color[HTML]{32CB00} \textbf{0.46}} &
  {\color[HTML]{32CB00} \textbf{0.20}} &
  \multicolumn{1}{c|}{{\color[HTML]{32CB00} \textbf{0.08}}} &
  0.45 &
  0.19 &
  \multicolumn{1}{c|}{0.04} &
  {\color[HTML]{3531FF} \textbf{0.46}} &
  {\color[HTML]{3531FF} \textbf{0.21}} &
  {\color[HTML]{3531FF} \textbf{0.05}} \\
GS &
  0.36 &
  0.10 &
  \multicolumn{1}{c|}{0.09} &
  0.29 &
  0.06 &
  \multicolumn{1}{c|}{0.06} &
  0.38 &
  0.09 &
  \multicolumn{1}{c|}{0.08} &
  0.38 &
  0.12 &
  \multicolumn{1}{c|}{0.10} &
  0.31 &
  0.05 &
  \multicolumn{1}{c|}{0.02} &
  {\color[HTML]{FE0000} \textbf{0.16}} &
  {\color[HTML]{FE0000} \textbf{0.00}} &
  \multicolumn{1}{c|}{{\color[HTML]{FE0000} \textbf{0.00}}} &
  0.40 &
  0.15 &
  \multicolumn{1}{c|}{0.09} &
  {\color[HTML]{3531FF} \textbf{0.41}} &
  {\color[HTML]{3531FF} \textbf{0.15}} &
  \multicolumn{1}{c|}{{\color[HTML]{3531FF} \textbf{0.10}}} &
  0.34 &
  0.10 &
  \multicolumn{1}{c|}{0.08} &
  {\color[HTML]{32CB00} \textbf{0.41}} &
  {\color[HTML]{32CB00} \textbf{0.17}} &
  {\color[HTML]{32CB00} \textbf{0.12}} \\
IBM &
  0.36 &
  0.09 &
  \multicolumn{1}{c|}{0.12} &
  0.36 &
  0.08 &
  \multicolumn{1}{c|}{0.08} &
  0.35 &
  0.08 &
  \multicolumn{1}{c|}{0.11} &
  0.35 &
  0.11 &
  \multicolumn{1}{c|}{0.11} &
  {\color[HTML]{FE0000} \textbf{0.30}} &
  {\color[HTML]{FE0000} \textbf{0.06}} &
  \multicolumn{1}{c|}{{\color[HTML]{FE0000} \textbf{0.02}}} &
  0.30 &
  0.05 &
  \multicolumn{1}{c|}{0.05} &
  0.36 &
  0.10 &
  \multicolumn{1}{c|}{0.14} &
  0.37 &
  0.10 &
  \multicolumn{1}{c|}{0.14} &
  {\color[HTML]{3531FF} \textbf{0.38}} &
  {\color[HTML]{3531FF} \textbf{0.11}} &
  \multicolumn{1}{c|}{{\color[HTML]{3531FF} \textbf{0.13}}} &
  {\color[HTML]{32CB00} \textbf{0.40}} &
  {\color[HTML]{32CB00} \textbf{0.13}} &
  {\color[HTML]{32CB00} \textbf{0.14}} \\
MCD &
  0.37 &
  0.08 &
  \multicolumn{1}{c|}{0.10} &
  0.35 &
  0.08 &
  \multicolumn{1}{c|}{0.08} &
  0.38 &
  0.09 &
  \multicolumn{1}{c|}{0.10} &
  0.38 &
  0.10 &
  \multicolumn{1}{c|}{0.11} &
  0.31 &
  0.04 &
  \multicolumn{1}{c|}{0.02} &
  {\color[HTML]{FE0000} \textbf{0.28}} &
  {\color[HTML]{FE0000} \textbf{0.04}} &
  \multicolumn{1}{c|}{{\color[HTML]{FE0000} \textbf{0.01}}} &
  0.39 &
  0.11 &
  \multicolumn{1}{c|}{0.12} &
  0.40 &
  0.11 &
  \multicolumn{1}{c|}{0.12} &
  {\color[HTML]{3531FF} \textbf{0.41}} &
  {\color[HTML]{3531FF} \textbf{0.12}} &
  \multicolumn{1}{c|}{{\color[HTML]{3531FF} \textbf{0.11}}} &
  {\color[HTML]{32CB00} \textbf{0.41}} &
  {\color[HTML]{32CB00} \textbf{0.13}} &
  {\color[HTML]{32CB00} \textbf{0.13}} \\
NVDA &
  0.31 &
  0.06 &
  \multicolumn{1}{c|}{0.07} &
  {\color[HTML]{FE0000} \textbf{0.24}} &
  {\color[HTML]{FE0000} \textbf{0.00}} &
  \multicolumn{1}{c|}{{\color[HTML]{FE0000} \textbf{0.00}}} &
  0.33 &
  0.07 &
  \multicolumn{1}{c|}{0.06} &
  0.36 &
  0.08 &
  \multicolumn{1}{c|}{0.10} &
  0.24 &
  0.02 &
  \multicolumn{1}{c|}{0.03} &
  {\color[HTML]{FE0000} \textbf{0.22}} &
  {\color[HTML]{FE0000} \textbf{0.02}} &
  \multicolumn{1}{c|}{{\color[HTML]{FE0000} \textbf{0.00}}} &
  {\color[HTML]{3531FF} \textbf{0.41}} &
  {\color[HTML]{3531FF} \textbf{0.13}} &
  \multicolumn{1}{c|}{{\color[HTML]{3531FF} \textbf{0.13}}} &
  {\color[HTML]{32CB00} \textbf{0.41}} &
  {\color[HTML]{32CB00} \textbf{0.13}} &
  \multicolumn{1}{c|}{{\color[HTML]{32CB00} \textbf{0.14}}} &
  0.34 &
  0.08 &
  \multicolumn{1}{c|}{0.09} &
  0.40 &
  0.12 &
  0.14 \\ \hline
AAPL &
  0.42 &
  0.16 &
  \multicolumn{1}{c|}{0.13} &
  0.39 &
  0.14 &
  \multicolumn{1}{c|}{0.09} &
  0.41 &
  0.14 &
  \multicolumn{1}{c|}{0.12} &
  0.41 &
  0.16 &
  \multicolumn{1}{c|}{0.13} &
  {\color[HTML]{FE0000} \textbf{0.35}} &
  {\color[HTML]{FE0000} \textbf{0.09}} &
  \multicolumn{1}{c|}{{\color[HTML]{FE0000} \textbf{0.07}}} &
  0.39 &
  0.17 &
  \multicolumn{1}{c|}{0.10} &
  0.41 &
  0.15 &
  \multicolumn{1}{c|}{0.15} &
  {\color[HTML]{3531FF} \textbf{0.42}} &
  {\color[HTML]{3531FF} \textbf{0.16}} &
  \multicolumn{1}{c|}{{\color[HTML]{3531FF} \textbf{0.16}}} &
  {\color[HTML]{32CB00} \textbf{0.43}} &
  {\color[HTML]{32CB00} \textbf{0.17}} &
  \multicolumn{1}{c|}{{\color[HTML]{32CB00} \textbf{0.15}}} &
  {\color[HTML]{32CB00} \textbf{0.42}} &
  {\color[HTML]{32CB00} \textbf{0.18}} &
  {\color[HTML]{32CB00} \textbf{0.15}} \\
ABBV &
  0.38 &
  0.13 &
  \multicolumn{1}{c|}{0.12} &
  0.39 &
  0.12 &
  \multicolumn{1}{c|}{0.10} &
  0.39 &
  0.13 &
  \multicolumn{1}{c|}{0.11} &
  {\color[HTML]{3531FF} \textbf{0.40}} &
  {\color[HTML]{3531FF} \textbf{0.15}} &
  \multicolumn{1}{c|}{{\color[HTML]{3531FF} \textbf{0.13}}} &
  {\color[HTML]{FE0000} \textbf{0.31}} &
  {\color[HTML]{FE0000} \textbf{0.06}} &
  \multicolumn{1}{c|}{{\color[HTML]{FE0000} \textbf{0.03}}} &
  0.33 &
  0.10 &
  \multicolumn{1}{c|}{0.04} &
  0.36 &
  0.13 &
  \multicolumn{1}{c|}{0.13} &
  0.37 &
  0.14 &
  \multicolumn{1}{c|}{0.13} &
  0.39 &
  0.13 &
  \multicolumn{1}{c|}{0.13} &
  {\color[HTML]{32CB00} \textbf{0.42}} &
  {\color[HTML]{32CB00} \textbf{0.18}} &
  {\color[HTML]{32CB00} \textbf{0.14}} \\
PM &
  0.35 &
  0.08 &
  \multicolumn{1}{c|}{0.09} &
  0.39 &
  0.10 &
  \multicolumn{1}{c|}{0.09} &
  0.37 &
  0.08 &
  \multicolumn{1}{c|}{0.08} &
  0.36 &
  0.09 &
  \multicolumn{1}{c|}{0.09} &
  {\color[HTML]{FE0000} \textbf{0.29}} &
  {\color[HTML]{FE0000} \textbf{0.02}} &
  \multicolumn{1}{c|}{{\color[HTML]{FE0000} \textbf{0.02}}} &
  0.28 &
  0.04 &
  \multicolumn{1}{c|}{0.04} &
  0.36 &
  0.10 &
  \multicolumn{1}{c|}{0.12} &
  0.36 &
  0.11 &
  \multicolumn{1}{c|}{0.13} &
  {\color[HTML]{3531FF} \textbf{0.36}} &
  {\color[HTML]{3531FF} \textbf{0.12}} &
  \multicolumn{1}{c|}{{\color[HTML]{3531FF} \textbf{0.13}}} &
  {\color[HTML]{32CB00} \textbf{0.39}} &
  {\color[HTML]{32CB00} \textbf{0.13}} &
  {\color[HTML]{32CB00} \textbf{0.13}} \\ \hline
BAC &
  0.43 &
  0.23 &
  \multicolumn{1}{c|}{0.04} &
  0.38 &
  0.21 &
  \multicolumn{1}{c|}{0.06} &
  0.38 &
  0.23 &
  \multicolumn{1}{c|}{0.04} &
  0.44 &
  0.28 &
  \multicolumn{1}{c|}{0.05} &
  {\color[HTML]{FE0000} \textbf{0.36}} &
  {\color[HTML]{FE0000} \textbf{0.18}} &
  \multicolumn{1}{c|}{{\color[HTML]{FE0000} \textbf{0.09}}} &
  0.46 &
  0.29 &
  \multicolumn{1}{c|}{0.05} &
  0.45 &
  0.27 &
  \multicolumn{1}{c|}{0.06} &
  0.45 &
  0.28 &
  \multicolumn{1}{c|}{0.07} &
  {\color[HTML]{3531FF} \textbf{0.46}} &
  {\color[HTML]{3531FF} \textbf{0.30}} &
  \multicolumn{1}{c|}{{\color[HTML]{3531FF} \textbf{0.07}}} &
  {\color[HTML]{32CB00} \textbf{0.47}} &
  {\color[HTML]{32CB00} \textbf{0.32}} &
  {\color[HTML]{32CB00} \textbf{0.06}} \\
CSCO &
  0.47 &
  0.29 &
  \multicolumn{1}{c|}{0.08} &
  0.50 &
  0.29 &
  \multicolumn{1}{c|}{0.09} &
  0.47 &
  0.28 &
  \multicolumn{1}{c|}{0.07} &
  0.48 &
  0.29 &
  \multicolumn{1}{c|}{0.08} &
  {\color[HTML]{FE0000} \textbf{0.41}} &
  {\color[HTML]{FE0000} \textbf{0.19}} &
  \multicolumn{1}{c|}{{\color[HTML]{FE0000} \textbf{0.11}}} &
  0.47 &
  0.27 &
  \multicolumn{1}{c|}{0.08} &
  0.45 &
  0.28 &
  \multicolumn{1}{c|}{0.08} &
  0.44 &
  0.27 &
  \multicolumn{1}{c|}{0.07} &
  {\color[HTML]{3531FF} \textbf{0.49}} &
  {\color[HTML]{3531FF} \textbf{0.30}} &
  \multicolumn{1}{c|}{{\color[HTML]{3531FF} \textbf{0.08}}} &
  {\color[HTML]{32CB00} \textbf{0.50}} &
  {\color[HTML]{32CB00} \textbf{0.33}} &
  {\color[HTML]{32CB00} \textbf{0.08}} \\
KO &
  0.47 &
  0.26 &
  \multicolumn{1}{c|}{0.08} &
  0.47 &
  0.27 &
  \multicolumn{1}{c|}{0.10} &
  0.46 &
  0.28 &
  \multicolumn{1}{c|}{0.09} &
  0.47 &
  0.28 &
  \multicolumn{1}{c|}{0.09} &
  {\color[HTML]{FE0000} \textbf{0.39}} &
  {\color[HTML]{FE0000} \textbf{0.17}} &
  \multicolumn{1}{c|}{{\color[HTML]{FE0000} \textbf{0.10}}} &
  {\color[HTML]{3531FF} \textbf{0.48}} &
  {\color[HTML]{3531FF} \textbf{0.30}} &
  \multicolumn{1}{c|}{{\color[HTML]{3531FF} \textbf{0.10}}} &
  0.45 &
  0.28 &
  \multicolumn{1}{c|}{0.10} &
  0.43 &
  0.27 &
  \multicolumn{1}{c|}{0.10} &
  0.48 &
  0.28 &
  \multicolumn{1}{c|}{0.10} &
  {\color[HTML]{32CB00} \textbf{0.49}} &
  {\color[HTML]{32CB00} \textbf{0.31}} &
  {\color[HTML]{32CB00} \textbf{0.10}} \\
ORCL &
  0.47 &
  0.27 &
  \multicolumn{1}{c|}{0.10} &
  0.45 &
  0.26 &
  \multicolumn{1}{c|}{0.09} &
  0.45 &
  0.26 &
  \multicolumn{1}{c|}{0.10} &
  0.48 &
  0.30 &
  \multicolumn{1}{c|}{0.11} &
  {\color[HTML]{FE0000} \textbf{0.38}} &
  {\color[HTML]{FE0000} \textbf{0.16}} &
  \multicolumn{1}{c|}{{\color[HTML]{FE0000} \textbf{0.07}}} &
  0.48 &
  0.31 &
  \multicolumn{1}{c|}{0.11} &
  0.46 &
  0.27 &
  \multicolumn{1}{c|}{0.10} &
  0.44 &
  0.26 &
  \multicolumn{1}{c|}{0.10} &
  {\color[HTML]{32CB00} \textbf{0.49}} &
  {\color[HTML]{32CB00} \textbf{0.32}} &
  \multicolumn{1}{c|}{{\color[HTML]{32CB00} \textbf{0.11}}} &
  {\color[HTML]{3531FF} \textbf{0.48}} &
  {\color[HTML]{3531FF} \textbf{0.32}} &
  {\color[HTML]{3531FF} \textbf{0.11}} \\
PFE &
  0.43 &
  0.24 &
  \multicolumn{1}{c|}{0.09} &
  0.42 &
  0.24 &
  \multicolumn{1}{c|}{0.09} &
  0.43 &
  0.25 &
  \multicolumn{1}{c|}{0.09} &
  0.44 &
  0.25 &
  \multicolumn{1}{c|}{0.09} &
  {\color[HTML]{FE0000} \textbf{0.36}} &
  {\color[HTML]{FE0000} \textbf{0.17}} &
  \multicolumn{1}{c|}{{\color[HTML]{FE0000} \textbf{0.11}}} &
  0.45 &
  0.27 &
  \multicolumn{1}{c|}{0.09} &
  {\color[HTML]{3531FF} \textbf{0.47}} &
  {\color[HTML]{3531FF} \textbf{0.29}} &
  \multicolumn{1}{c|}{{\color[HTML]{3531FF} \textbf{0.09}}} &
  0.46 &
  0.28 &
  \multicolumn{1}{c|}{0.10} &
  0.46 &
  0.27 &
  \multicolumn{1}{c|}{0.09} &
  {\color[HTML]{32CB00} \textbf{0.49}} &
  {\color[HTML]{32CB00} \textbf{0.32}} &
  {\color[HTML]{32CB00} \textbf{0.10}} \\
VZ &
  0.47 &
  0.23 &
  \multicolumn{1}{c|}{0.08} &
  0.42 &
  0.17 &
  \multicolumn{1}{c|}{0.07} &
  0.47 &
  0.26 &
  \multicolumn{1}{c|}{0.09} &
  0.47 &
  0.27 &
  \multicolumn{1}{c|}{0.10} &
  {\color[HTML]{FE0000} \textbf{0.39}} &
  {\color[HTML]{FE0000} \textbf{0.16}} &
  \multicolumn{1}{c|}{{\color[HTML]{FE0000} \textbf{0.10}}} &
  0.46 &
  0.25 &
  \multicolumn{1}{c|}{0.10} &
  0.45 &
  0.26 &
  \multicolumn{1}{c|}{0.10} &
  0.41 &
  0.24 &
  \multicolumn{1}{c|}{0.10} &
  {\color[HTML]{32CB00} \textbf{0.49}} &
  {\color[HTML]{32CB00} \textbf{0.28}} &
  \multicolumn{1}{c|}{{\color[HTML]{32CB00} \textbf{0.11}}} &
  {\color[HTML]{3531FF} \textbf{0.46}} &
  {\color[HTML]{3531FF} \textbf{0.28}} &
  {\color[HTML]{3531FF} \textbf{0.10}} \\ \hline
\end{tabular}
\end{adjustbox}
\end{table}
}

Looking at Table \ref{tab:comprehensive_comparison_models_h10}, we notice that, at $\text{H}\Delta_\tau = 10$, HLOB outperforms SOTA alternatives in the $73.3$\% of cases. For small-tick stocks, it is the best-performing model in $4/6$ scenarios (i.e., CHTR, GS, IBM, MCD); in the case of GOOG, it is the second-best alternative, while in the case of NVDA, it is the third-best alternative. For medium-tick stocks, HLOB is the best-performing model in $3/3$ scenarios (i.e., AAPL, ABBV, PM), while, for large-tick stocks, it is the best-performing option in $4/6$ cases (i.e., BAC, CSCO, KO, PFE) and the second-best alternative in the remaining $2$ scenarios (i.e., ORCL and VZ). The HLOB average F1 score is equal to $0.42$ for small-tick stocks, $0.41$ for medium-tick stocks, and $0.48$ for large-tick stocks. The average MCC is equal to $0.16$ for small- and medium-tick stocks, and to $0.33$ for large-tick stocks. The average $p_{\text{T}}$ is equal to $0.11$ for small-tick stocks, $0.14$ for medium-tick stocks, and $0.09$ for large-tick stocks. Focusing on inter-models' dynamics, we observe that, for small- to medium-tick stocks, performances are very similar for all the $3$ evaluation metrics except for iTransformer and LobTransformer (which are the worst-performing alternatives). For large-tick stocks, instead, we observe that also the worst-performing models, even showing a considerable distance from the best-performing alternative in traditional machine-learning metrics' realizations (i.e., F1 score and MCC), present competitive realizations in the case of $p_{\text{T}}$. Comparing HLOB performances with DeepLOB ones, we observe that (i) the average gain in F1 score is equal to $0.03$ for small-tick stocks, $0.02$ for medium-tick stocks, and  $0.003$ for large-tick stocks; (ii) the average gain in MCC is equal to $0.04$ for small-tick stocks, $0.02$ for medium-tick stocks, and $0.02$ for large-tick stocks; (iii) the average gain in $p_{\text{T}}$ is equal to $0.02$ for large-tick stocks, $0.01$ for medium-tick stocks, and $0.00$ for large-tick stocks.

Looking at Table \ref{tab:comprehensive_comparison_models_h50}, we notice that, at $\text{H}\Delta_\tau = 50$, HLOB model outperforms SOTA alternatives in the $60$\% of cases ($10\%$ less than what happens at $\text{H}\Delta_\tau = 10$). For small-tick stocks, it is the best-performing model in $1/6$ scenarios (i.e., IBM); in the case of GS and MCD, it is the second-best alternative, while in all the other cases (i.e., CHTR, GOOG, and NVDA), it is the third-best alternative. For medium-tick stocks, HLOB is the best-performing model in $3/3$ scenarios (i.e., AAPL, ABBV, PM), while, for large-tick stocks, it is the best-performing model in $5/6$ cases (i.e., BAC, CSCO, KO, ORCL, VZ), being the second-best alternative in the case of PFE. The HLOB average F1 score is equal to $0.36$ for small-tick stocks (with a percentage decrease of $16.66$\% compared to the realization at $\text{H}\Delta_\tau = 10$), $0.40$ for medium-tick stocks (with a percentage decrease of $2.50$\% compared to the realization at $\text{H}\Delta_\tau = 10$), and $0.58$ for large-tick stocks (with a percentage increase of $17.24$\% compared to the realization at $\text{H}\Delta_\tau = 10$). The average MCC is equal to $0.09$ for small-tick stocks (with a percentage decrease of $77.78$\% compared to the realization at $\text{H}\Delta_\tau = 10$), $0.11$ for medium-tick stocks (with a percentage decrease of $45.45$\% compared to the realization at $\text{H}\Delta_\tau = 10$), and to $0.38$ for large-tick stocks (with a percentage increase of $13.15$\% compared to the realization at $\text{H}\Delta_\tau = 10$). The average $p_{\text{T}}$ is equal to $0.07$ for small-tick stocks (with a percentage decrease of $57.14$\% compared to the realization at $\text{H}\Delta_\tau = 10$), $0.10$ for medium-tick stocks (with a percentage decrease of $40.00$\% compared to the realization at $\text{H}\Delta_\tau = 10$), and $0.14$ for large-tick stocks (with a percentage decrease of $35.71$\% compared to the realization at $\text{H}\Delta_\tau = 10$). Focusing on inter-models' dynamics, we observe that, also in this case, for small- to medium-tick stocks, performances are very similar for all the $3$ evaluation metrics, except for iTransformer and LobTransformer architectures; however, differently from what observed at $\text{H}\Delta_\tau = 10$, for the iTransformer model, this observation remains true also for large-tick stocks. Comparing HLOB performances with DeepLOB ones, we observe that (i) the average gain in F1 score is equal to $0.05$ for small-tick stocks (with a percentage increase of $40.00$\% compared to what observed at $\text{H}\Delta_\tau = 10$), $0.05$ for medium-tick stocks (with a percentage increase of $60.00$\% compared to what observed at $\text{H}\Delta_\tau = 10$), and $0.01$ for large-tick stocks (with a percentage increase of $70.00$\% compared to what observed at $\text{H}\Delta_\tau = 10$); (ii) the average gain in MCC is equal to $0.05$ for small-tick stocks (with a percentage increase of $20.00$\% compared to what observed at $\text{H}\Delta_\tau = 10$), $0.02$ for medium-tick stocks (with no increase compared to what observed at $\text{H}\Delta_\tau = 10$), and $0.03$ for large-tick stocks (with a percentage increase of $33.33$\% compared to what observed at $\text{H}\Delta_\tau = 10$); (iii) the average gain in $p_{\text{T}}$ is equal to $0.03$ for large-tick stocks (with a percentage increase of $33.33$\% compared to what observed at $\text{H}\Delta_\tau = 10$), $0.03$ for medium-tick stocks (with a percentage increase of $66.00$\% compared to what observed at $\text{H}\Delta_\tau = 10$), and $0.01$ for large-tick
stocks (with a percentage increase of $100.00$\% compared to what observed at $\text{H}\Delta_\tau = 10$).

{
\centering
\renewcommand{\arraystretch}{2}
\begin{table}[h!]
\centering
\begin{adjustbox}{center,max width=\linewidth}
\caption{Models' performances at $\text{H}\Delta_\tau = 50$. For each deep learning architecture we report three key metrics: (i) the F1 score; (ii) the MCC; and (iii) the $p_{\text{T}}$. For each stock, we highlight the best performing model (green), the second-best performing model (blue) and the worst performing alternative (red); a model is considered superior to the others if the sum of the $3$ performance metrics is maximal.}
\label{tab:comprehensive_comparison_models_h50}
\tiny
\setlength{\tabcolsep}{1pt}
\begin{tabular}{c|cccccccccccccccccccccccccccccc}
\hline
 &
  \multicolumn{30}{c}{\textbf{H50}} \\ \cline{2-31} 
\multirow{-2}{*}{} &
  \multicolumn{3}{c|}{\textbf{cnn1}} &
  \multicolumn{3}{c|}{\textbf{cnn2}} &
  \multicolumn{3}{c|}{\textbf{dla}} &
  \multicolumn{3}{c|}{\textbf{transformer}} &
  \multicolumn{3}{c|}{\textbf{itransformer}} &
  \multicolumn{3}{c|}{\textbf{lobtransformer}} &
  \multicolumn{3}{c|}{\textbf{binbtabl}} &
  \multicolumn{3}{c|}{\textbf{binctabl}} &
  \multicolumn{3}{c|}{\textbf{deeplob}} &
  \multicolumn{3}{c}{\textbf{hlob}} \\ \cline{2-31} 
 &
  \textbf{F1} &
  \textbf{MCC} &
  \multicolumn{1}{c|}{\textbf{$p_T$}} &
  \textbf{F1} &
  \textbf{MCC} &
  \multicolumn{1}{c|}{\textbf{$p_T$}} &
  \textbf{F1} &
  \textbf{MCC} &
  \multicolumn{1}{c|}{\textbf{$p_T$}} &
  \textbf{F1} &
  \textbf{MCC} &
  \multicolumn{1}{c|}{\textbf{$p_T$}} &
  \textbf{F1} &
  \textbf{MCC} &
  \multicolumn{1}{c|}{\textbf{$p_T$}} &
  \textbf{F1} &
  \textbf{MCC} &
  \multicolumn{1}{c|}{\textbf{$p_T$}} &
  \textbf{F1} &
  \textbf{MCC} &
  \multicolumn{1}{c|}{\textbf{$p_T$}} &
  \textbf{F1} &
  \textbf{MCC} &
  \multicolumn{1}{c|}{\textbf{$p_T$}} &
  \textbf{F1} &
  \textbf{MCC} &
  \multicolumn{1}{c|}{\textbf{$p_T$}} &
  \textbf{F1} &
  \textbf{MCC} &
  \textbf{$p_T$} \\ \hline
CHTR &
  0.34 &
  0.07 &
  \multicolumn{1}{c|}{0.03} &
  0.33 &
  0.03 &
  \multicolumn{1}{c|}{0.02} &
  0.36 &
  0.09 &
  \multicolumn{1}{c|}{0.04} &
  0.36 &
  0.09 &
  \multicolumn{1}{c|}{0.05} &
  0.29 &
  0.03 &
  \multicolumn{1}{c|}{0.04} &
  {\color[HTML]{FE0000} \textbf{0.22}} &
  {\color[HTML]{FE0000} \textbf{0.01}} &
  \multicolumn{1}{c|}{{\color[HTML]{FE0000} \textbf{0.00}}} &
  {\color[HTML]{32CB00} \textbf{0.38}} &
  {\color[HTML]{32CB00} \textbf{0.15}} &
  \multicolumn{1}{c|}{{\color[HTML]{32CB00} \textbf{0.05}}} &
  {\color[HTML]{3531FF} \textbf{0.38}} &
  {\color[HTML]{3531FF} \textbf{0.13}} &
  \multicolumn{1}{c|}{{\color[HTML]{3531FF} \textbf{0.05}}} &
  0.36 &
  0.06 &
  \multicolumn{1}{c|}{0.03} &
  0.36 &
  0.09 &
  0.05 \\
GOOG &
  0.31 &
  0.06 &
  \multicolumn{1}{c|}{0.03} &
  0.35 &
  0.06 &
  \multicolumn{1}{c|}{0.03} &
  0.30 &
  0.06 &
  \multicolumn{1}{c|}{0.05} &
  0.37 &
  0.12 &
  \multicolumn{1}{c|}{0.05} &
  {\color[HTML]{FE0000} \textbf{0.29}} &
  {\color[HTML]{FE0000} \textbf{0.03}} &
  \multicolumn{1}{c|}{{\color[HTML]{FE0000} \textbf{0.04}}} &
  0.31 &
  0.07 &
  \multicolumn{1}{c|}{0.03} &
  {\color[HTML]{3531FF} \textbf{0.42}} &
  {\color[HTML]{3531FF} \textbf{0.17}} &
  \multicolumn{1}{c|}{{\color[HTML]{3531FF} \textbf{0.07}}} &
  {\color[HTML]{32CB00} \textbf{0.43}} &
  {\color[HTML]{32CB00} \textbf{0.17}} &
  \multicolumn{1}{c|}{{\color[HTML]{32CB00} \textbf{0.07}}} &
  0.37 &
  0.09 &
  \multicolumn{1}{c|}{0.05} &
  0.42 &
  0.16 &
  0.06 \\
GS &
  0.30 &
  0.03 &
  \multicolumn{1}{c|}{0.04} &
  0.25 &
  0.00 &
  \multicolumn{1}{c|}{0.03} &
  0.29 &
  0.04 &
  \multicolumn{1}{c|}{0.05} &
  0.32 &
  0.04 &
  \multicolumn{1}{c|}{0.06} &
  0.26 &
  0.01 &
  \multicolumn{1}{c|}{0.03} &
  {\color[HTML]{FE0000} \textbf{0.22}} &
  {\color[HTML]{FE0000} \textbf{0.00}} &
  \multicolumn{1}{c|}{{\color[HTML]{FE0000} \textbf{0.00}}} &
  {\color[HTML]{3531FF} \textbf{0.32}} &
  {\color[HTML]{3531FF} \textbf{0.11}} &
  \multicolumn{1}{c|}{{\color[HTML]{3531FF} \textbf{0.08}}} &
  {\color[HTML]{32CB00} \textbf{0.35}} &
  {\color[HTML]{32CB00} \textbf{0.11}} &
  \multicolumn{1}{c|}{{\color[HTML]{32CB00} \textbf{0.08}}} &
  0.29 &
  0.03 &
  \multicolumn{1}{c|}{0.06} &
  {\color[HTML]{3531FF} \textbf{0.34}} &
  {\color[HTML]{3531FF} \textbf{0.09}} &
  {\color[HTML]{3531FF} \textbf{0.08}} \\
IBM &
  0.30 &
  0.04 &
  \multicolumn{1}{c|}{0.06} &
  0.29 &
  0.01 &
  \multicolumn{1}{c|}{0.04} &
  0.33 &
  0.05 &
  \multicolumn{1}{c|}{0.07} &
  0.34 &
  0.07 &
  \multicolumn{1}{c|}{0.07} &
  0.25 &
  0.02 &
  \multicolumn{1}{c|}{0.02} &
  {\color[HTML]{FE0000} \textbf{0.23}} &
  {\color[HTML]{FE0000} \textbf{0.03}} &
  \multicolumn{1}{c|}{{\color[HTML]{FE0000} \textbf{0.00}}} &
  {\color[HTML]{3531FF} \textbf{0.36}} &
  {\color[HTML]{3531FF} \textbf{0.07}} &
  \multicolumn{1}{c|}{{\color[HTML]{3531FF} \textbf{0.10}}} &
  0.34 &
  0.07 &
  \multicolumn{1}{c|}{0.10} &
  0.28 &
  0.03 &
  \multicolumn{1}{c|}{0.04} &
  {\color[HTML]{32CB00} \textbf{0.37}} &
  {\color[HTML]{32CB00} \textbf{0.08}} &
  {\color[HTML]{32CB00} \textbf{0.09}} \\
MCD &
  0.32 &
  0.05 &
  \multicolumn{1}{c|}{0.06} &
  0.29 &
  0.04 &
  \multicolumn{1}{c|}{0.04} &
  0.31 &
  0.06 &
  \multicolumn{1}{c|}{0.07} &
  0.32 &
  0.06 &
  \multicolumn{1}{c|}{0.06} &
  0.28 &
  0.02 &
  \multicolumn{1}{c|}{0.03} &
  {\color[HTML]{FE0000} \textbf{0.23}} &
  {\color[HTML]{FE0000} \textbf{0.01}} &
  \multicolumn{1}{c|}{{\color[HTML]{FE0000} \textbf{0.00}}} &
  {\color[HTML]{32CB00} \textbf{0.36}} &
  {\color[HTML]{32CB00} \textbf{0.08}} &
  \multicolumn{1}{c|}{{\color[HTML]{32CB00} \textbf{0.09}}} &
  {\color[HTML]{32CB00} \textbf{0.36}} &
  {\color[HTML]{32CB00} \textbf{0.08}} &
  \multicolumn{1}{c|}{{\color[HTML]{32CB00} \textbf{0.09}}} &
  0.32 &
  0.04 &
  \multicolumn{1}{c|}{0.05} &
  {\color[HTML]{3531FF} \textbf{0.36}} &
  {\color[HTML]{3531FF} \textbf{0.07}} &
  {\color[HTML]{3531FF} \textbf{0.08}} \\
NVDA &
  0.26 &
  0.02 &
  \multicolumn{1}{c|}{0.03} &
  0.17 &
  0.00 &
  \multicolumn{1}{c|}{0.00} &
  {\color[HTML]{3531FF} \textbf{0.34}} &
  {\color[HTML]{3531FF} \textbf{0.04}} &
  \multicolumn{1}{c|}{{\color[HTML]{3531FF} \textbf{0.06}}} &
  0.28 &
  0.05 &
  \multicolumn{1}{c|}{0.05} &
  {\color[HTML]{FE0000} \textbf{0.17}} &
  {\color[HTML]{FE0000} \textbf{0.01}} &
  \multicolumn{1}{c|}{{\color[HTML]{FE0000} \textbf{0.01}}} &
  0.23 &
  0.01 &
  \multicolumn{1}{c|}{0.00} &
  {\color[HTML]{32CB00} \textbf{0.37}} &
  {\color[HTML]{32CB00} \textbf{0.12}} &
  \multicolumn{1}{c|}{{\color[HTML]{32CB00} \textbf{0.09}}} &
  {\color[HTML]{32CB00} \textbf{0.37}} &
  {\color[HTML]{32CB00} \textbf{0.12}} &
  \multicolumn{1}{c|}{{\color[HTML]{32CB00} \textbf{0.09}}} &
  0.20 &
  0.01 &
  \multicolumn{1}{c|}{0.01} &
  0.30 &
  0.06 &
  0.07 \\ \hline
AAPL &
  0.35 &
  0.09 &
  \multicolumn{1}{c|}{0.11} &
  0.37 &
  0.09 &
  \multicolumn{1}{c|}{0.06} &
  0.38 &
  0.09 &
  \multicolumn{1}{c|}{0.11} &
  {\color[HTML]{3531FF} \textbf{0.39}} &
  {\color[HTML]{3531FF} \textbf{0.11}} &
  \multicolumn{1}{c|}{{\color[HTML]{3531FF} \textbf{0.10}}} &
  {\color[HTML]{FE0000} \textbf{0.26}} &
  {\color[HTML]{FE0000} \textbf{0.04}} &
  \multicolumn{1}{c|}{{\color[HTML]{FE0000} \textbf{0.02}}} &
  0.32 &
  0.09 &
  \multicolumn{1}{c|}{0.04} &
  {\color[HTML]{32CB00} \textbf{0.40}} &
  {\color[HTML]{32CB00} \textbf{0.11}} &
  \multicolumn{1}{c|}{{\color[HTML]{32CB00} \textbf{0.12}}} &
  {\color[HTML]{32CB00} \textbf{0.40}} &
  {\color[HTML]{32CB00} \textbf{0.11}} &
  \multicolumn{1}{c|}{{\color[HTML]{32CB00} \textbf{0.12}}} &
  0.36 &
  0.11 &
  \multicolumn{1}{c|}{0.07} &
  {\color[HTML]{32CB00} \textbf{0.40}} &
  {\color[HTML]{32CB00} \textbf{0.12}} &
  {\color[HTML]{32CB00} \textbf{0.11}} \\
ABBV &
  0.35 &
  0.09 &
  \multicolumn{1}{c|}{0.08} &
  0.34 &
  0.08 &
  \multicolumn{1}{c|}{0.05} &
  0.36 &
  0.10 &
  \multicolumn{1}{c|}{0.11} &
  0.36 &
  0.10 &
  \multicolumn{1}{c|}{0.09} &
  {\color[HTML]{FE0000} \textbf{0.24}} &
  {\color[HTML]{FE0000} \textbf{0.03}} &
  \multicolumn{1}{c|}{{\color[HTML]{FE0000} \textbf{0.01}}} &
  0.25 &
  0.04 &
  \multicolumn{1}{c|}{0.01} &
  0.37 &
  0.11 &
  \multicolumn{1}{c|}{0.11} &
  {\color[HTML]{3531FF} \textbf{0.38}} &
  {\color[HTML]{3531FF} \textbf{0.11}} &
  \multicolumn{1}{c|}{{\color[HTML]{3531FF} \textbf{0.11}}} &
  0.32 &
  0.08 &
  \multicolumn{1}{c|}{0.06} &
  {\color[HTML]{32CB00} \textbf{0.40}} &
  {\color[HTML]{32CB00} \textbf{0.12}} &
  {\color[HTML]{32CB00} \textbf{0.10}} \\
PM &
  0.35 &
  0.07 &
  \multicolumn{1}{c|}{0.09} &
  0.33 &
  0.06 &
  \multicolumn{1}{c|}{0.05} &
  0.35 &
  0.06 &
  \multicolumn{1}{c|}{0.09} &
  0.37 &
  0.07 &
  \multicolumn{1}{c|}{0.09} &
  {\color[HTML]{FE0000} \textbf{0.26}} &
  {\color[HTML]{FE0000} \textbf{0.02}} &
  \multicolumn{1}{c|}{{\color[HTML]{FE0000} \textbf{0.02}}} &
  {\color[HTML]{FE0000} \textbf{0.27}} &
  {\color[HTML]{FE0000} \textbf{0.02}} &
  \multicolumn{1}{c|}{{\color[HTML]{FE0000} \textbf{0.01}}} &
  {\color[HTML]{3531FF} \textbf{0.37}} &
  {\color[HTML]{3531FF} \textbf{0.09}} &
  \multicolumn{1}{c|}{{\color[HTML]{3531FF} \textbf{0.11}}} &
  0.37 &
  0.09 &
  \multicolumn{1}{c|}{0.10} &
  0.36 &
  0.07 &
  \multicolumn{1}{c|}{0.10} &
  {\color[HTML]{32CB00} \textbf{0.39}} &
  {\color[HTML]{32CB00} \textbf{0.09}} &
  {\color[HTML]{32CB00} \textbf{0.10}} \\ \hline
BAC &
  0.59 &
  0.39 &
  \multicolumn{1}{c|}{0.09} &
  0.55 &
  0.39 &
  \multicolumn{1}{c|}{0.08} &
  0.53 &
  0.36 &
  \multicolumn{1}{c|}{0.08} &
  0.58 &
  0.41 &
  \multicolumn{1}{c|}{0.09} &
  {\color[HTML]{FE0000} \textbf{0.41}} &
  {\color[HTML]{FE0000} \textbf{0.24}} &
  \multicolumn{1}{c|}{{\color[HTML]{FE0000} \textbf{0.08}}} &
  0.55 &
  0.33 &
  \multicolumn{1}{c|}{0.07} &
  0.59 &
  0.40 &
  \multicolumn{1}{c|}{0.11} &
  0.59 &
  0.40 &
  \multicolumn{1}{c|}{0.09} &
  {\color[HTML]{3531FF} \textbf{0.61}} &
  {\color[HTML]{3531FF} \textbf{0.44}} &
  \multicolumn{1}{c|}{{\color[HTML]{3531FF} \textbf{0.10}}} &
  {\color[HTML]{32CB00} \textbf{0.62}} &
  {\color[HTML]{32CB00} \textbf{0.47}} &
  {\color[HTML]{32CB00} \textbf{0.09}} \\
CSCO &
  0.55 &
  0.34 &
  \multicolumn{1}{c|}{0.14} &
  0.55 &
  0.34 &
  \multicolumn{1}{c|}{0.12} &
  0.54 &
  0.31 &
  \multicolumn{1}{c|}{0.15} &
  0.57 &
  0.36 &
  \multicolumn{1}{c|}{0.13} &
  {\color[HTML]{FE0000} \textbf{0.41}} &
  {\color[HTML]{FE0000} \textbf{0.18}} &
  \multicolumn{1}{c|}{{\color[HTML]{FE0000} \textbf{0.11}}} &
  0.53 &
  0.31 &
  \multicolumn{1}{c|}{0.10} &
  0.52 &
  0.33 &
  \multicolumn{1}{c|}{0.15} &
  0.53 &
  0.33 &
  \multicolumn{1}{c|}{0.13} &
  {\color[HTML]{3531FF} \textbf{0.58}} &
  {\color[HTML]{3531FF} \textbf{0.37}} &
  \multicolumn{1}{c|}{{\color[HTML]{3531FF} \textbf{0.13}}} &
  {\color[HTML]{32CB00} \textbf{0.60}} &
  {\color[HTML]{32CB00} \textbf{0.40}} &
  {\color[HTML]{32CB00} \textbf{0.16}} \\
KO &
  0.53 &
  0.30 &
  \multicolumn{1}{c|}{0.13} &
  0.51 &
  0.29 &
  \multicolumn{1}{c|}{0.11} &
  0.56 &
  0.34 &
  \multicolumn{1}{c|}{0.15} &
  0.56 &
  0.34 &
  \multicolumn{1}{c|}{0.13} &
  {\color[HTML]{FE0000} \textbf{0.43}} &
  {\color[HTML]{FE0000} \textbf{0.20}} &
  \multicolumn{1}{c|}{{\color[HTML]{FE0000} \textbf{0.11}}} &
  {\color[HTML]{3531FF} \textbf{0.57}} &
  {\color[HTML]{3531FF} \textbf{0.35}} &
  \multicolumn{1}{c|}{{\color[HTML]{3531FF} \textbf{0.14}}} &
  0.54 &
  0.34 &
  \multicolumn{1}{c|}{0.13} &
  0.55 &
  0.35 &
  \multicolumn{1}{c|}{0.13} &
  0.56 &
  0.34 &
  \multicolumn{1}{c|}{0.14} &
  {\color[HTML]{32CB00} \textbf{0.59}} &
  {\color[HTML]{32CB00} \textbf{0.39}} &
  {\color[HTML]{32CB00} \textbf{0.15}} \\
ORCL &
  0.53 &
  0.30 &
  \multicolumn{1}{c|}{0.15} &
  0.51 &
  0.30 &
  \multicolumn{1}{c|}{0.14} &
  0.52 &
  0.30 &
  \multicolumn{1}{c|}{0.15} &
  0.54 &
  0.33 &
  \multicolumn{1}{c|}{0.15} &
  {\color[HTML]{FE0000} \textbf{0.40}} &
  {\color[HTML]{FE0000} \textbf{0.17}} &
  \multicolumn{1}{c|}{{\color[HTML]{FE0000} \textbf{0.12}}} &
  0.52 &
  0.30 &
  \multicolumn{1}{c|}{0.14} &
  0.52 &
  0.29 &
  \multicolumn{1}{c|}{0.15} &
  0.51 &
  0.29 &
  \multicolumn{1}{c|}{0.15} &
  {\color[HTML]{3531FF} \textbf{0.56}} &
  {\color[HTML]{3531FF} \textbf{0.34}} &
  \multicolumn{1}{c|}{{\color[HTML]{3531FF} \textbf{0.15}}} &
  {\color[HTML]{32CB00} \textbf{0.57}} &
  {\color[HTML]{32CB00} \textbf{0.36}} &
  {\color[HTML]{32CB00} \textbf{0.16}} \\
PFE &
  0.53 &
  0.32 &
  \multicolumn{1}{c|}{0.13} &
  0.53 &
  0.32 &
  \multicolumn{1}{c|}{0.10} &
  0.54 &
  0.34 &
  \multicolumn{1}{c|}{0.12} &
  0.53 &
  0.33 &
  \multicolumn{1}{c|}{0.13} &
  {\color[HTML]{FE0000} \textbf{0.42}} &
  {\color[HTML]{FE0000} \textbf{0.20}} &
  \multicolumn{1}{c|}{{\color[HTML]{FE0000} \textbf{0.10}}} &
  0.55 &
  0.35 &
  \multicolumn{1}{c|}{0.11} &
  0.55 &
  0.34 &
  \multicolumn{1}{c|}{0.13} &
  0.56 &
  0.35 &
  \multicolumn{1}{c|}{0.13} &
  {\color[HTML]{32CB00} \textbf{0.57}} &
  {\color[HTML]{32CB00} \textbf{0.38}} &
  \multicolumn{1}{c|}{{\color[HTML]{32CB00} \textbf{0.11}}} &
  {\color[HTML]{3531FF} \textbf{0.56}} &
  {\color[HTML]{3531FF} \textbf{0.37}} &
  {\color[HTML]{3531FF} \textbf{0.12}} \\
VZ &
  0.50 &
  0.27 &
  \multicolumn{1}{c|}{0.13} &
  0.46 &
  0.24 &
  \multicolumn{1}{c|}{0.13} &
  0.47 &
  0.26 &
  \multicolumn{1}{c|}{0.13} &
  0.51 &
  0.30 &
  \multicolumn{1}{c|}{0.15} &
  {\color[HTML]{FE0000} \textbf{0.39}} &
  {\color[HTML]{FE0000} \textbf{0.14}} &
  \multicolumn{1}{c|}{{\color[HTML]{FE0000} \textbf{0.11}}} &
  {\color[HTML]{3531FF} \textbf{0.53}} &
  {\color[HTML]{3531FF} \textbf{0.30}} &
  \multicolumn{1}{c|}{{\color[HTML]{3531FF} \textbf{0.14}}} &
  0.51 &
  0.28 &
  \multicolumn{1}{c|}{0.16} &
  0.51 &
  0.28 &
  \multicolumn{1}{c|}{0.15} &
  0.50 &
  0.28 &
  \multicolumn{1}{c|}{0.12} &
  {\color[HTML]{32CB00} \textbf{0.52}} &
  {\color[HTML]{32CB00} \textbf{0.31}} &
  {\color[HTML]{32CB00} \textbf{0.15}} \\ \hline
\end{tabular}
\end{adjustbox}
\end{table}
}
Looking at Table \ref{tab:comprehensive_comparison_models_h100}, we notice that, at $\text{H}\Delta_\tau = 100$, the HLOB model outperforms  state-of-the art (SOTA) alternatives in the $33$\% of cases ($37$\% less than what happens at $\text{H}\Delta_\tau = 10$ and $27$\% less than what happens at $\text{H}\Delta_\tau = 50$). For small-tick stocks, HLOB is the best-performing model in $1/6$ scenarios (in particular for IBM stock); in the case of CHTR, it is the second-best alternative, while in all the other cases (i.e., GOOG, GS, MCD, and NVDA), it is the third-best alternative. For medium-tick stocks, it is the third-best performing model in $3/3$ scenarios (i.e., AAPL, ABBV, PM). For large-tick stocks, it is the best-performing model in $4/6$ cases (i.e., BAC, CSCO, KO, ORCL), being the second-best alternative in the case of PFE and the third-best alternative in the case of VZ. 

The \textit{HLOB average F1 score} is equal to $0.32$ for small-tick stocks (with a percentage decrease of $31.25$\% compared to the realization at $\text{H}\Delta_\tau = 10$ and a decrease of $12.50$\% compared to the realization at $\text{H}\Delta_\tau = 50$), $0.35$ for medium-tick stocks (with a percentage decrease of $17.14$\% compared to the realization at $\text{H}\Delta_\tau = 10$ and a decrease of $14.28$\% compared to the realization at $\text{H}\Delta_\tau = 50$), and to $0.52$ for large-tick stocks (with a percentage increase of $7.69$\% compared to the realization at $\text{H}\Delta_\tau = 10$ and a decrease of $11.54$\% compared to the realization at $\text{H}\Delta_\tau = 50$). 

The \textit{HLOB average MCC score} is equal to $0.05$ for small-tick stocks (with a percentage decrease of $220.00$\% compared to the realization at $\text{H}\Delta_\tau = 10$ and a decrease of $80.00$\% compared to the realization at $\text{H}\Delta_\tau = 50$), $0.06$ for medium-tick stocks (with a percentage decrease of $166.67$\% compared to the realization at $\text{H}\Delta_\tau = 10$ and a decrease of $83.33$\% compared to the realization at $\text{H}\Delta_\tau = 50$), and to $0.30$ for large-tick stocks (with a percentage decrease of $10.00$\% compared to the realization at $\text{H}\Delta_\tau = 10$ and a decrease of $26.66$\% compared to the realization at $\text{H}\Delta_\tau = 50$). 

The \textit{HLOB average} $p_{\text{T}}$ \textit{score} is equal to $0.05$ for small-tick stocks (with a percentage decrease of $120.00$\% compared to the realization at $\text{H}\Delta_\tau = 10$ and a decrease of $40.00$\% compared to the realization at $\text{H}\Delta_\tau = 50$), $0.07$ for medium-tick stocks (with a percentage decrease of $100.00$\% compared to the realization at $\text{H}\Delta_\tau = 10$ and a decrease of $42.86$\% compared to the realization at $\text{H}\Delta_\tau = 50$), and to $0.15$ for large-tick stocks (with a percentage increase of $40.00$\% compared to the realization at $\text{H}\Delta_\tau = 10$ and an increase of $6.67$\% compared to the realization at $\text{H}\Delta_\tau = 50$).Focusing on inter-models' dynamics, we observe that, consistently with what observed at $\text{H}\Delta_\tau \in \{10, 50\}$, for small- to medium-tick stocks, performances are similar for all the $3$ evaluation metrics with very minor oscillations; the two main exceptions are represented by the iTransformer and LobTransformer which present considerably lower realizations. Similarly to $\text{H}\Delta_\tau = 50$, but differently from $\text{H}\Delta_\tau = 10$, this behaviour persists also for large-tick stocks. 

Comparing HLOB performances with DeepLOB ones, we observe that (i) the average gain in F1 score is equal to $0.006$ for small-tick stocks (with a percentage decrease of $400.00$\% compared to what observed at $\text{H}\Delta_\tau = 10$ and a decrease of $733.33$\% compared to what observed at $\text{H}\Delta_\tau = 50$), $0.03$ for medium-tick stocks (with a percentage increase of $33.33$\% compared to what observed at $\text{H}\Delta_\tau = 10$ and a decrease of $66.67$\% compared to what observed at $\text{H}\Delta_\tau = 50$), and $0.04$ for large-tick stocks (with a percentage increase of $92.50$\% compared to what observed at $\text{H}\Delta_\tau = 10$ and an increase of $75.00$\% compared to what observed at $\text{H}\Delta_\tau = 50$); (ii) the average gain in MCC is equal to $0.04$ for small-tick stocks (with no percentage increase compared to what observed at $\text{H}\Delta_\tau = 10$ and a decrease of $25.00$\% compared to what observed at $\text{H}\Delta_\tau = 50$), $0.02$ for medium-tick stocks (with no percentage increase compared to what observed at $\text{H}\Delta_\tau \in \{10, 50\}$), and $0.04$ for large-tick stocks (with a percentage increase of $100.00$\% compared to what observed at $\text{H}\Delta_\tau = 10$ and an increase of $33.33$\% compared to what observed at $\text{H}\Delta_\tau = 50$); (iii) the average gain in $p_{\text{T}}$ is equal to $0.003$ for small-tick stocks (with a percentage decrease of $566.67$\% compared to what observed at $\text{H}\Delta_\tau = 10$ and a decrease of $900.00$\% compared to what observed at $\text{H}\Delta_\tau = 50$), $0.02$ for medium tick stocks (with a percentage increase of $50.00$\% compared to what observed at $\text{H}\Delta_\tau = 10$ and a decrease of $50.00$\% compared to what observed at $\text{H}\Delta_\tau = 50$), and $0.02$ for large-tick stocks (with a percentage increase of $200.00$\% compared to what observed at $\text{H}\Delta_\tau = 10$ and an increase of $100.00$\% compared to what observed at $\text{H}\Delta_\tau = 50$).

{
\centering
\renewcommand{\arraystretch}{2}
\begin{table}[h!]
\centering
\begin{adjustbox}{center,max width=\linewidth}
\caption{Models' performances at $\text{H}\Delta_\tau = 100$. For each deep learning architecture we report three key metrics: (i) the F1 score; (ii) the MCC; and (iii) the $p_{\text{T}}$. For each stock, we highlight the best performing model (green), the second-best performing model (blue) and the worst performing alternative (red); a model is considered superior to the others if the sum of the $3$ performance metrics is maximal.}
\label{tab:comprehensive_comparison_models_h100}
\tiny
\setlength{\tabcolsep}{1pt}
\begin{tabular}{c|cccccccccccccccccccccccccccccc}
\hline
 &
  \multicolumn{30}{c}{\textbf{H100}} \\ \cline{2-31} 
\multirow{-2}{*}{} &
  \multicolumn{3}{c|}{\textbf{cnn1}} &
  \multicolumn{3}{c|}{\textbf{cnn2}} &
  \multicolumn{3}{c|}{\textbf{dla}} &
  \multicolumn{3}{c|}{\textbf{transformer}} &
  \multicolumn{3}{c|}{\textbf{itransformer}} &
  \multicolumn{3}{c|}{\textbf{lobtransformer}} &
  \multicolumn{3}{c|}{\textbf{binbtabl}} &
  \multicolumn{3}{c|}{\textbf{binctabl}} &
  \multicolumn{3}{c|}{\textbf{deeplob}} &
  \multicolumn{3}{c}{\textbf{hlob}} \\ \cline{2-31} 
 &
  \textbf{F1} &
  \textbf{MCC} &
  \multicolumn{1}{c|}{\textbf{$p_T$}} &
  \textbf{F1} &
  \textbf{MCC} &
  \multicolumn{1}{c|}{\textbf{$p_T$}} &
  \textbf{F1} &
  \textbf{MCC} &
  \multicolumn{1}{c|}{\textbf{$p_T$}} &
  \textbf{F1} &
  \textbf{MCC} &
  \multicolumn{1}{c|}{\textbf{$p_T$}} &
  \textbf{F1} &
  \textbf{MCC} &
  \multicolumn{1}{c|}{\textbf{$p_T$}} &
  \textbf{F1} &
  \textbf{MCC} &
  \multicolumn{1}{c|}{\textbf{$p_T$}} &
  \textbf{F1} &
  \textbf{MCC} &
  \multicolumn{1}{c|}{\textbf{$p_T$}} &
  \textbf{F1} &
  \textbf{MCC} &
  \multicolumn{1}{c|}{\textbf{$p_T$}} &
  \textbf{F1} &
  \textbf{MCC} &
  \multicolumn{1}{c|}{\textbf{$p_T$}} &
  \textbf{F1} &
  \textbf{MCC} &
  \textbf{$p_T$} \\ \hline
CHTR &
  0.29 &
  0.04 &
  \multicolumn{1}{c|}{0.02} &
  0.30 &
  0.00 &
  \multicolumn{1}{c|}{0.02} &
  0.32 &
  0.05 &
  \multicolumn{1}{c|}{0.03} &
  0.31 &
  0.06 &
  \multicolumn{1}{c|}{0.03} &
  0.25 &
  0.01 &
  \multicolumn{1}{c|}{0.03} &
  {\color[HTML]{FE0000} \textbf{0.23}} &
  {\color[HTML]{FE0000} \textbf{0.00}} &
  \multicolumn{1}{c|}{{\color[HTML]{FE0000} \textbf{0.01}}} &
  {\color[HTML]{32CB00} \textbf{0.34}} &
  {\color[HTML]{32CB00} \textbf{0.11}} &
  \multicolumn{1}{c|}{{\color[HTML]{32CB00} \textbf{0.04}}} &
  {\color[HTML]{32CB00} \textbf{0.35}} &
  {\color[HTML]{32CB00} \textbf{0.10}} &
  \multicolumn{1}{c|}{{\color[HTML]{32CB00} \textbf{0.04}}} &
  0.34 &
  0.03 &
  \multicolumn{1}{c|}{0.04} &
  {\color[HTML]{3531FF} \textbf{0.35}} &
  {\color[HTML]{3531FF} \textbf{0.08}} &
  {\color[HTML]{3531FF} \textbf{0.04}} \\
GOOG &
  0.29 &
  0.04 &
  \multicolumn{1}{c|}{0.03} &
  0.28 &
  0.01 &
  \multicolumn{1}{c|}{0.02} &
  0.26 &
  0.03 &
  \multicolumn{1}{c|}{0.03} &
  0.32 &
  0.06 &
  \multicolumn{1}{c|}{0.04} &
  0.25 &
  0.02 &
  \multicolumn{1}{c|}{0.03} &
  {\color[HTML]{FE0000} \textbf{0.24}} &
  {\color[HTML]{FE0000} \textbf{0.02}} &
  \multicolumn{1}{c|}{{\color[HTML]{FE0000} \textbf{0.01}}} &
  {\color[HTML]{3531FF} \textbf{0.39}} &
  {\color[HTML]{3531FF} \textbf{0.13}} &
  \multicolumn{1}{c|}{{\color[HTML]{3531FF} \textbf{0.06}}} &
  {\color[HTML]{32CB00} \textbf{0.40}} &
  {\color[HTML]{32CB00} \textbf{0.14}} &
  \multicolumn{1}{c|}{{\color[HTML]{32CB00} \textbf{0.07}}} &
  0.32 &
  0.01 &
  \multicolumn{1}{c|}{0.05} &
  0.36 &
  0.10 &
  0.05 \\
GS &
  0.29 &
  0.01 &
  \multicolumn{1}{c|}{0.04} &
  0.30 &
  0.01 &
  \multicolumn{1}{c|}{0.03} &
  0.28 &
  0.01 &
  \multicolumn{1}{c|}{0.04} &
  0.28 &
  0.01 &
  \multicolumn{1}{c|}{0.04} &
  0.22 &
  0.01 &
  \multicolumn{1}{c|}{0.00} &
  {\color[HTML]{FE0000} \textbf{0.22}} &
  {\color[HTML]{FE0000} \textbf{0.00}} &
  \multicolumn{1}{c|}{{\color[HTML]{FE0000} \textbf{0.00}}} &
  {\color[HTML]{3531FF} \textbf{0.28}} &
  {\color[HTML]{3531FF} \textbf{0.06}} &
  \multicolumn{1}{c|}{{\color[HTML]{3531FF} \textbf{0.06}}} &
  {\color[HTML]{32CB00} \textbf{0.29}} &
  {\color[HTML]{32CB00} \textbf{0.06}} &
  \multicolumn{1}{c|}{{\color[HTML]{32CB00} \textbf{0.06}}} &
  0.31 &
  0.00 &
  \multicolumn{1}{c|}{0.06} &
  0.30 &
  0.02 &
  0.05 \\
IBM &
  0.29 &
  0.02 &
  \multicolumn{1}{c|}{0.02} &
  0.32 &
  0.01 &
  \multicolumn{1}{c|}{0.04} &
  0.29 &
  0.03 &
  \multicolumn{1}{c|}{0.04} &
  0.28 &
  0.03 &
  \multicolumn{1}{c|}{0.04} &
  {\color[HTML]{FE0000} \textbf{0.28}} &
  {\color[HTML]{FE0000} \textbf{0.01}} &
  \multicolumn{1}{c|}{{\color[HTML]{FE0000} \textbf{0.00}}} &
  0.26 &
  0.01 &
  \multicolumn{1}{c|}{0.02} &
  {\color[HTML]{3531FF} \textbf{0.29}} &
  {\color[HTML]{3531FF} \textbf{0.03}} &
  \multicolumn{1}{c|}{{\color[HTML]{3531FF} \textbf{0.06}}} &
  {\color[HTML]{3531FF} \textbf{0.29}} &
  {\color[HTML]{3531FF} \textbf{0.03}} &
  \multicolumn{1}{c|}{{\color[HTML]{3531FF} \textbf{0.06}}} &
  0.29 &
  0.00 &
  \multicolumn{1}{c|}{0.04} &
  {\color[HTML]{32CB00} \textbf{0.32}} &
  {\color[HTML]{32CB00} \textbf{0.03}} &
  {\color[HTML]{32CB00} \textbf{0.06}} \\
MCD &
  0.24 &
  0.02 &
  \multicolumn{1}{c|}{0.03} &
  0.28 &
  0.00 &
  \multicolumn{1}{c|}{0.03} &
  0.26 &
  0.03 &
  \multicolumn{1}{c|}{0.03} &
  0.27 &
  0.03 &
  \multicolumn{1}{c|}{0.05} &
  {\color[HTML]{FE0000} \textbf{0.22}} &
  {\color[HTML]{FE0000} \textbf{0.01}} &
  \multicolumn{1}{c|}{{\color[HTML]{FE0000} \textbf{0.03}}} &
  0.25 &
  0.01 &
  \multicolumn{1}{c|}{0.02} &
  {\color[HTML]{3531FF} \textbf{0.30}} &
  {\color[HTML]{3531FF} \textbf{0.05}} &
  \multicolumn{1}{c|}{{\color[HTML]{3531FF} \textbf{0.06}}} &
  {\color[HTML]{32CB00} \textbf{0.32}} &
  {\color[HTML]{32CB00} \textbf{0.05}} &
  \multicolumn{1}{c|}{{\color[HTML]{32CB00} \textbf{0.07}}} &
  0.29 &
  0.01 &
  \multicolumn{1}{c|}{0.06} &
  0.29 &
  0.04 &
  0.05 \\
NVDA &
  0.23 &
  0.01 &
  \multicolumn{1}{c|}{0.01} &
  {\color[HTML]{FE0000} \textbf{0.17}} &
  {\color[HTML]{FE0000} \textbf{0.00}} &
  \multicolumn{1}{c|}{{\color[HTML]{FE0000} \textbf{0.00}}} &
  0.24 &
  0.02 &
  \multicolumn{1}{c|}{0.03} &
  0.23 &
  0.01 &
  \multicolumn{1}{c|}{0.01} &
  0.18 &
  0.00 &
  \multicolumn{1}{c|}{0.01} &
  0.22 &
  0.00 &
  \multicolumn{1}{c|}{0.00} &
  {\color[HTML]{3531FF} \textbf{0.34}} &
  {\color[HTML]{3531FF} \textbf{0.09}} &
  \multicolumn{1}{c|}{{\color[HTML]{3531FF} \textbf{0.08}}} &
  {\color[HTML]{32CB00} \textbf{0.36}} &
  {\color[HTML]{32CB00} \textbf{0.09}} &
  \multicolumn{1}{c|}{{\color[HTML]{32CB00} \textbf{0.07}}} &
  0.31 &
  0.01 &
  \multicolumn{1}{c|}{0.03} &
  0.28 &
  0.05 &
  0.05 \\ \hline
AAPL &
  0.34 &
  0.06 &
  \multicolumn{1}{c|}{0.07} &
  0.30 &
  0.05 &
  \multicolumn{1}{c|}{0.03} &
  0.33 &
  0.06 &
  \multicolumn{1}{c|}{0.08} &
  0.33 &
  0.06 &
  \multicolumn{1}{c|}{0.06} &
  {\color[HTML]{FE0000} \textbf{0.24}} &
  {\color[HTML]{FE0000} \textbf{0.02}} &
  \multicolumn{1}{c|}{{\color[HTML]{FE0000} \textbf{0.01}}} &
  0.27 &
  0.03 &
  \multicolumn{1}{c|}{0.01} &
  {\color[HTML]{3531FF} \textbf{0.37}} &
  {\color[HTML]{3531FF} \textbf{0.08}} &
  \multicolumn{1}{c|}{{\color[HTML]{3531FF} \textbf{0.09}}} &
  {\color[HTML]{32CB00} \textbf{0.38}} &
  {\color[HTML]{32CB00} \textbf{0.08}} &
  \multicolumn{1}{c|}{{\color[HTML]{32CB00} \textbf{0.09}}} &
  0.34 &
  0.07 &
  \multicolumn{1}{c|}{0.05} &
  0.35 &
  0.08 &
  0.07 \\
ABBV &
  0.26 &
  0.04 &
  \multicolumn{1}{c|}{0.03} &
  0.32 &
  0.02 &
  \multicolumn{1}{c|}{0.03} &
  0.33 &
  0.05 &
  \multicolumn{1}{c|}{0.08} &
  0.31 &
  0.05 &
  \multicolumn{1}{c|}{0.06} &
  0.22 &
  0.03 &
  \multicolumn{1}{c|}{0.03} &
  {\color[HTML]{FE0000} \textbf{0.24}} &
  {\color[HTML]{FE0000} \textbf{0.01}} &
  \multicolumn{1}{c|}{{\color[HTML]{FE0000} \textbf{0.02}}} &
  {\color[HTML]{3531FF} \textbf{0.35}} &
  {\color[HTML]{3531FF} \textbf{0.07}} &
  \multicolumn{1}{c|}{{\color[HTML]{3531FF} \textbf{0.08}}} &
  {\color[HTML]{32CB00} \textbf{0.36}} &
  {\color[HTML]{32CB00} \textbf{0.08}} &
  \multicolumn{1}{c|}{{\color[HTML]{32CB00} \textbf{0.08}}} &
  0.31 &
  0.03 &
  \multicolumn{1}{c|}{0.05} &
  0.34 &
  0.06 &
  0.06 \\
PM &
  0.33 &
  0.03 &
  \multicolumn{1}{c|}{0.06} &
  0.27 &
  0.00 &
  \multicolumn{1}{c|}{0.02} &
  0.32 &
  0.04 &
  \multicolumn{1}{c|}{0.08} &
  0.33 &
  0.04 &
  \multicolumn{1}{c|}{0.06} &
  {\color[HTML]{FE0000} \textbf{0.19}} &
  {\color[HTML]{FE0000} \textbf{0.01}} &
  \multicolumn{1}{c|}{{\color[HTML]{FE0000} \textbf{0.01}}} &
  0.23 &
  0.00 &
  \multicolumn{1}{c|}{0.00} &
  {\color[HTML]{3531FF} \textbf{0.35}} &
  {\color[HTML]{3531FF} \textbf{0.05}} &
  \multicolumn{1}{c|}{{\color[HTML]{3531FF} \textbf{0.08}}} &
  {\color[HTML]{32CB00} \textbf{0.36}} &
  {\color[HTML]{32CB00} \textbf{0.05}} &
  \multicolumn{1}{c|}{{\color[HTML]{32CB00} \textbf{0.08}}} &
  0.31 &
  0.01 &
  \multicolumn{1}{c|}{0.05} &
  0.35 &
  0.04 &
  0.07 \\ \hline
BAC &
  0.56 &
  0.36 &
  \multicolumn{1}{c|}{0.10} &
  0.55 &
  0.34 &
  \multicolumn{1}{c|}{0.12} &
  0.55 &
  0.35 &
  \multicolumn{1}{c|}{0.13} &
  0.57 &
  0.38 &
  \multicolumn{1}{c|}{0.14} &
  {\color[HTML]{FE0000} \textbf{0.41}} &
  {\color[HTML]{FE0000} \textbf{0.21}} &
  \multicolumn{1}{c|}{{\color[HTML]{FE0000} \textbf{0.09}}} &
  0.52 &
  0.29 &
  \multicolumn{1}{c|}{0.14} &
  0.58 &
  0.36 &
  \multicolumn{1}{c|}{0.17} &
  {\color[HTML]{3531FF} \textbf{0.58}} &
  {\color[HTML]{3531FF} \textbf{0.37}} &
  \multicolumn{1}{c|}{{\color[HTML]{3531FF} \textbf{0.18}}} &
  0.57 &
  0.37 &
  \multicolumn{1}{c|}{0.16} &
  {\color[HTML]{32CB00} \textbf{0.63}} &
  {\color[HTML]{32CB00} \textbf{0.44}} &
  {\color[HTML]{32CB00} \textbf{0.17}} \\
CSCO &
  0.52 &
  0.28 &
  \multicolumn{1}{c|}{0.16} &
  0.52 &
  0.29 &
  \multicolumn{1}{c|}{0.15} &
  0.50 &
  0.27 &
  \multicolumn{1}{c|}{0.18} &
  0.52 &
  0.29 &
  \multicolumn{1}{c|}{0.15} &
  {\color[HTML]{FE0000} \textbf{0.38}} &
  {\color[HTML]{FE0000} \textbf{0.14}} &
  \multicolumn{1}{c|}{{\color[HTML]{FE0000} \textbf{0.12}}} &
  0.47 &
  0.23 &
  \multicolumn{1}{c|}{0.12} &
  0.49 &
  0.27 &
  \multicolumn{1}{c|}{0.16} &
  0.50 &
  0.27 &
  \multicolumn{1}{c|}{0.15} &
  {\color[HTML]{3531FF} \textbf{0.52}} &
  {\color[HTML]{3531FF} \textbf{0.28}} &
  \multicolumn{1}{c|}{{\color[HTML]{3531FF} \textbf{0.14}}} &
  {\color[HTML]{32CB00} \textbf{0.52}} &
  {\color[HTML]{32CB00} \textbf{0.30}} &
  {\color[HTML]{32CB00} \textbf{0.18}} \\
KO &
  0.50 &
  0.26 &
  \multicolumn{1}{c|}{0.14} &
  0.48 &
  0.23 &
  \multicolumn{1}{c|}{0.11} &
  {\color[HTML]{3531FF} \textbf{0.50}} &
  {\color[HTML]{3531FF} \textbf{0.26}} &
  \multicolumn{1}{c|}{{\color[HTML]{3531FF} \textbf{0.16}}} &
  0.49 &
  0.26 &
  \multicolumn{1}{c|}{0.14} &
  {\color[HTML]{FE0000} \textbf{0.35}} &
  {\color[HTML]{FE0000} \textbf{0.12}} &
  \multicolumn{1}{c|}{{\color[HTML]{FE0000} \textbf{0.09}}} &
  0.47 &
  0.23 &
  \multicolumn{1}{c|}{0.11} &
  0.48 &
  0.26 &
  \multicolumn{1}{c|}{0.13} &
  0.50 &
  0.27 &
  \multicolumn{1}{c|}{0.12} &
  0.51 &
  0.26 &
  \multicolumn{1}{c|}{0.13} &
  {\color[HTML]{32CB00} \textbf{0.53}} &
  {\color[HTML]{32CB00} \textbf{0.30}} &
  {\color[HTML]{32CB00} \textbf{0.16}} \\
ORCL &
  0.48 &
  0.23 &
  \multicolumn{1}{c|}{0.14} &
  0.46 &
  0.21 &
  \multicolumn{1}{c|}{0.12} &
  0.45 &
  0.22 &
  \multicolumn{1}{c|}{0.16} &
  0.47 &
  0.24 &
  \multicolumn{1}{c|}{0.16} &
  {\color[HTML]{FE0000} \textbf{0.30}} &
  {\color[HTML]{FE0000} \textbf{0.07}} &
  \multicolumn{1}{c|}{{\color[HTML]{FE0000} \textbf{0.08}}} &
  0.48 &
  0.24 &
  \multicolumn{1}{c|}{0.16} &
  0.48 &
  0.22 &
  \multicolumn{1}{c|}{0.17} &
  0.48 &
  0.22 &
  \multicolumn{1}{c|}{0.17} &
  {\color[HTML]{3531FF} \textbf{0.48}} &
  {\color[HTML]{3531FF} \textbf{0.25}} &
  \multicolumn{1}{c|}{{\color[HTML]{3531FF} \textbf{0.16}}} &
  {\color[HTML]{32CB00} \textbf{0.49}} &
  {\color[HTML]{32CB00} \textbf{0.26}} &
  {\color[HTML]{32CB00} \textbf{0.17}} \\
PFE &
  0.50 &
  0.26 &
  \multicolumn{1}{c|}{0.13} &
  0.53 &
  0.29 &
  \multicolumn{1}{c|}{0.12} &
  0.48 &
  0.26 &
  \multicolumn{1}{c|}{0.13} &
  0.48 &
  0.27 &
  \multicolumn{1}{c|}{0.12} &
  {\color[HTML]{FE0000} \textbf{0.38}} &
  {\color[HTML]{FE0000} \textbf{0.12}} &
  \multicolumn{1}{c|}{{\color[HTML]{FE0000} \textbf{0.09}}} &
  0.49 &
  0.27 &
  \multicolumn{1}{c|}{0.10} &
  {\color[HTML]{32CB00} \textbf{0.52}} &
  {\color[HTML]{32CB00} \textbf{0.27}} &
  \multicolumn{1}{c|}{{\color[HTML]{32CB00} \textbf{0.16}}} &
  {\color[HTML]{3531FF} \textbf{0.50}} &
  {\color[HTML]{3531FF} \textbf{0.27}} &
  \multicolumn{1}{c|}{{\color[HTML]{3531FF} \textbf{0.14}}} &
  0.50 &
  0.28 &
  \multicolumn{1}{c|}{0.11} &
  {\color[HTML]{3531FF} \textbf{0.50}} &
  {\color[HTML]{3531FF} \textbf{0.28}} &
  {\color[HTML]{3531FF} \textbf{0.13}} \\
VZ &
  0.45 &
  0.19 &
  \multicolumn{1}{c|}{0.12} &
  0.41 &
  0.17 &
  \multicolumn{1}{c|}{0.12} &
  0.42 &
  0.19 &
  \multicolumn{1}{c|}{0.13} &
  0.45 &
  0.20 &
  \multicolumn{1}{c|}{0.13} &
  {\color[HTML]{FE0000} \textbf{0.28}} &
  {\color[HTML]{FE0000} \textbf{0.07}} &
  \multicolumn{1}{c|}{{\color[HTML]{FE0000} \textbf{0.07}}} &
  0.36 &
  0.14 &
  \multicolumn{1}{c|}{0.08} &
  {\color[HTML]{3531FF} \textbf{0.47}} &
  {\color[HTML]{3531FF} \textbf{0.21}} &
  \multicolumn{1}{c|}{{\color[HTML]{3531FF} \textbf{0.16}}} &
  {\color[HTML]{32CB00} \textbf{0.47}} &
  {\color[HTML]{32CB00} \textbf{0.21}} &
  \multicolumn{1}{c|}{{\color[HTML]{32CB00} \textbf{0.17}}} &
  0.33 &
  0.12 &
  \multicolumn{1}{c|}{0.10} &
  0.47 &
  0.22 &
  0.11 \\ \hline
\end{tabular}
\end{adjustbox}
\end{table}
}

\begin{figure}[h!]
    \centering
    \begin{subfigure}{0.48\textwidth}
        \centering
        \includegraphics[width=\linewidth, height=5.5cm]{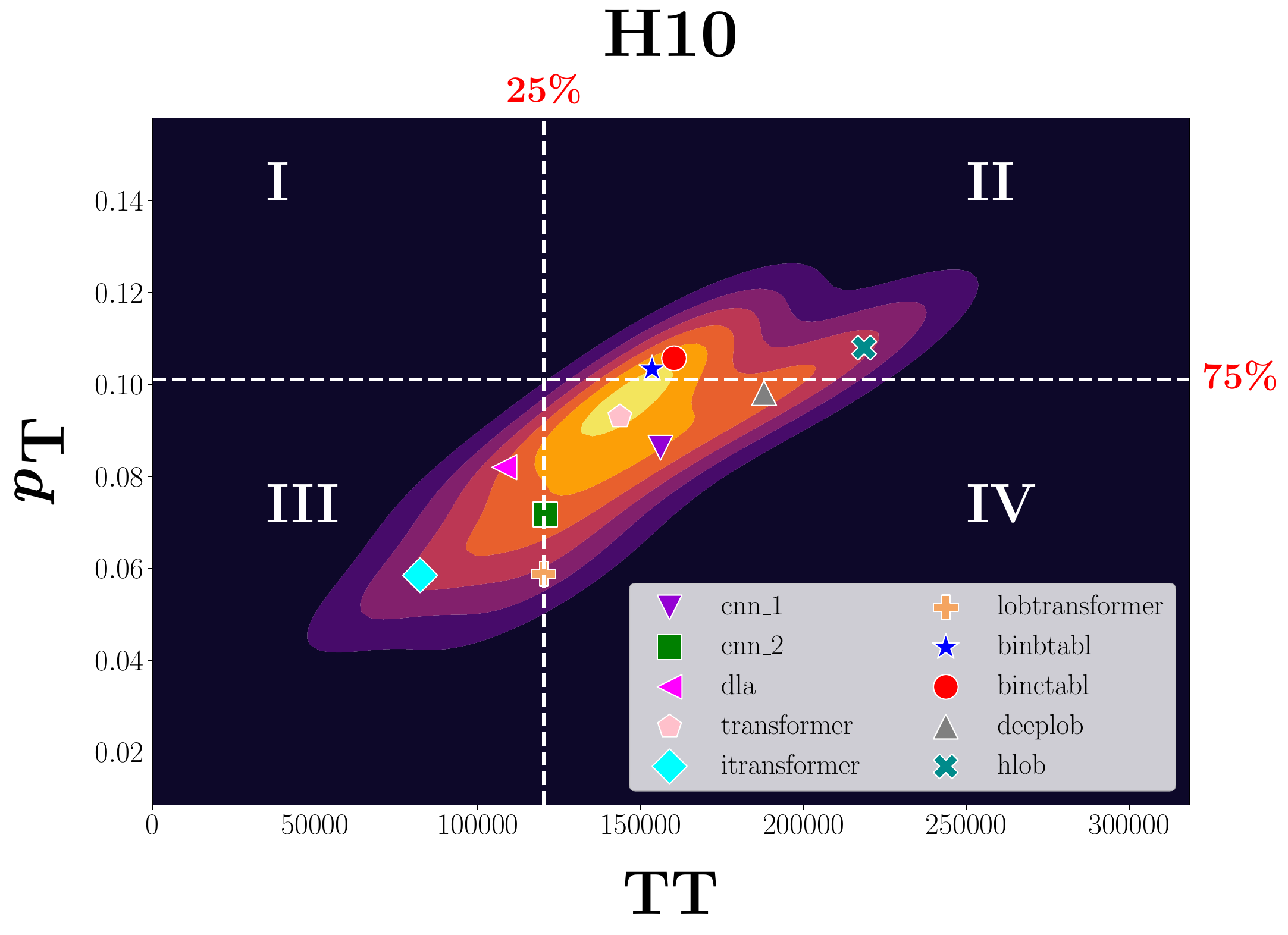}
        \captionsetup{justification=centering, margin=3.97cm}
        \caption{}
        \label{fig:h10_pT_distribution}
    \end{subfigure}
    \hfill
    \begin{subfigure}{0.48\textwidth}
        \centering
        \includegraphics[width=\linewidth, height=5.65cm]{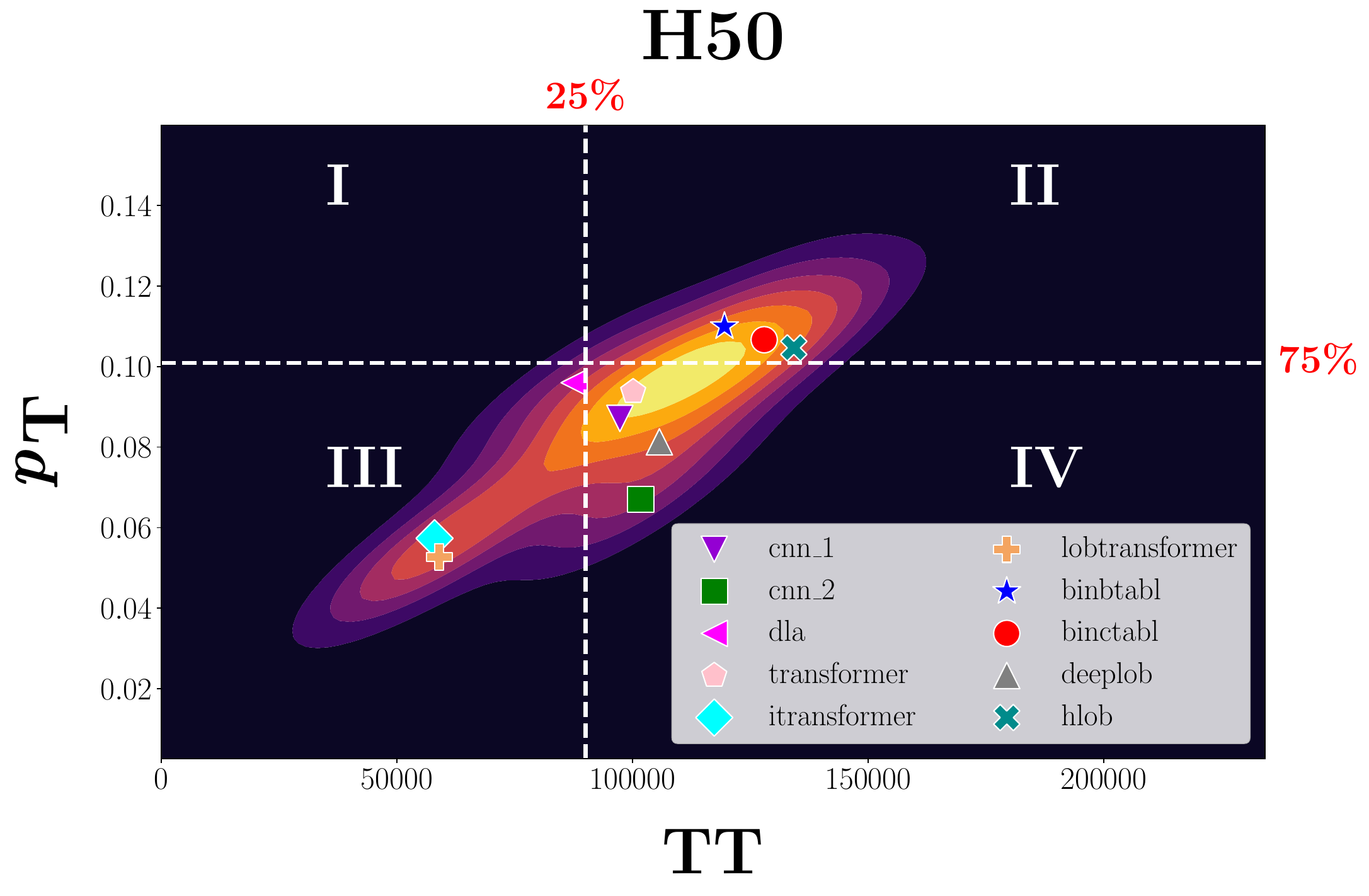}
        \captionsetup{justification=centering, margin=3.97cm}
        \caption{}
        \label{fig:h50_pT_distribution}
    \end{subfigure}
    \hfill
    \begin{subfigure}{0.48\textwidth}
        \centering
        \includegraphics[width=\linewidth, height=5.6cm]{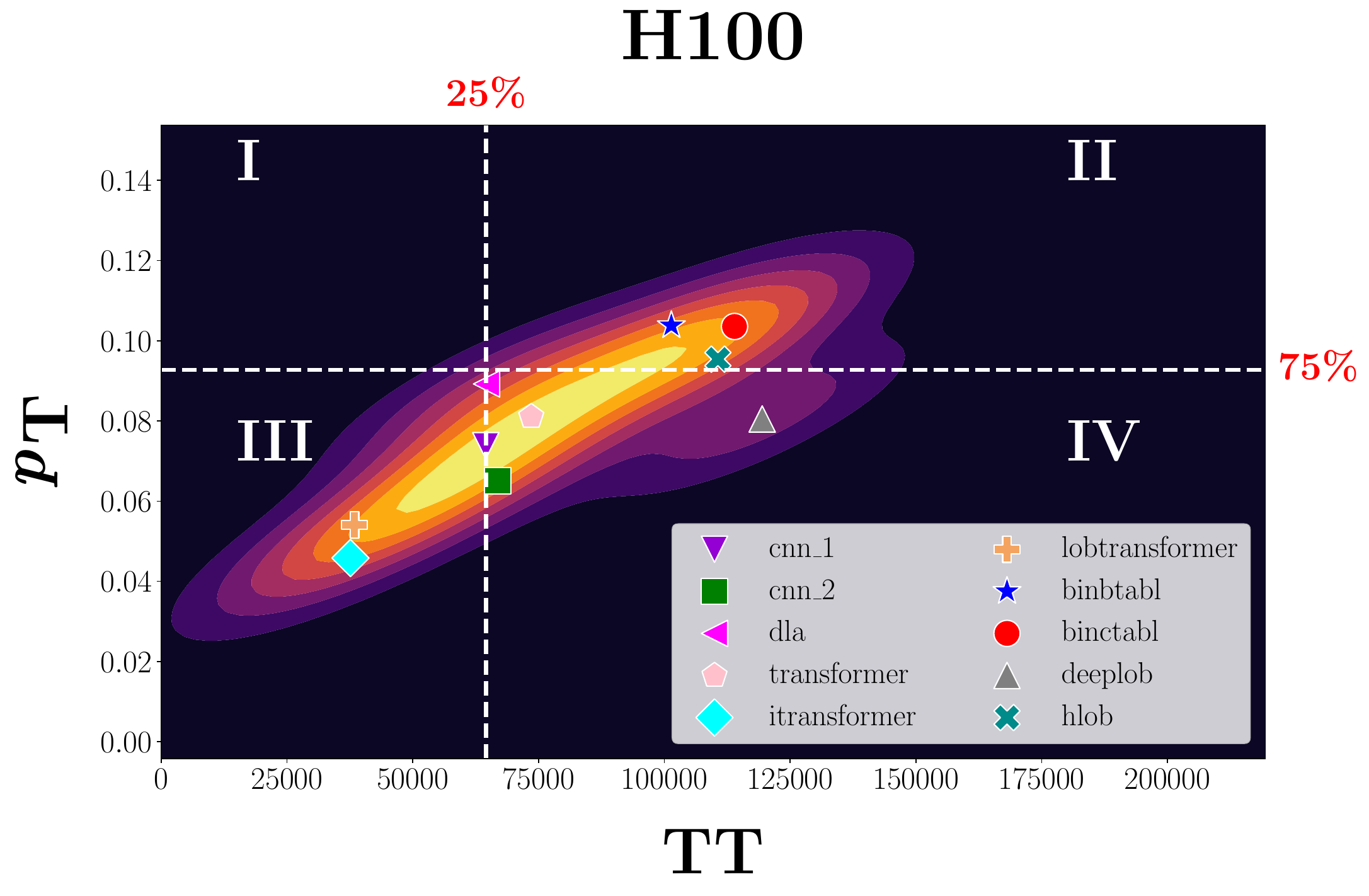}
        \captionsetup{justification=centering, margin=3.97cm}
        \caption{}
        \label{fig:h100_pT_distribution}
    \end{subfigure}
    \caption{Distribution of $p_{\text{T}}$ (see the work by \citet{briola2024deep}) as a function of the total number of executed round-trip transactions (TT) computed for each model in Table \ref{tab:models_summary} at $\text{H}\Delta_\tau \in \{10, 50, 100\}$.}
    \label{fig:pT_distributions}
\end{figure}

In Figure \ref{fig:pT_distributions}, we report, for each prediction horizon $\text{H}\Delta_\tau \in \{10, 50, 100\}$ and for each considered model, the distribution of $p_{\text{T}}$ as a function of the total number of executed round-trip transactions (TT) \cite{briola2024deep}. In this case we do not distinguish between different classes of stocks, and we use different markers to report the average models' performance values. We divide each plot into four quadrants. The \textit{upper-left quadrant} (i.e., (I)), contains the architectures with an average TT lower than the $25\%$ percentile (computed across the totality of models), and an average $p_{\text{T}}$ greater than the $75\%$ percentile (computed across the totality of models); intuitively, models located in this quadrant are the ones that achieve the best performances while remaining parsimonious in terms of executed transactions. The \textit{upper-right quadrant} (i.e., (II)) contains the architectures with an average TT higher than the $25\%$ percentile (computed across the totality of models), and an average $p_{\text{T}}$ greater than the $75\%$ percentile (computed across the totality of models); intuitively, models located in this quadrant are the ones that achieve the best performances being less parsimonious in terms of executed transactions. The \textit{lower-left} quadrant (i.e., (III)) contains the architectures with an average TT lower than the $25\%$ percentile (computed across the totality of models), and an average $p_{\text{T}}$ lower than the $75\%$ percentile (computed across the totality of models); intuitively, models located in this quadrant are the ones that achieve the worst performances being, however, parsimonious in terms of executed transactions. Finally, The \textit{lower-right} quadrant (i.e., (IV)) contains the architectures with an average TT higher than the $25\%$ percentile (computed across the totality of models), and an average $p_{\text{T}}$ higher than the $75\%$ percentile (computed across the totality of models); intuitively, models located in this quadrant are the ones that achieve the worst performances being less parsimonious in terms of executed transactions.

 From this representation, it is possible to distinguish between $3$ groups of models showing a consistent behaviour across prediction horizons. The \textit{first group} of models is made of BinBTabl, BinCTabl and HLOB. They are always placed in the upper-right quadrant, demonstrating to be the most effective models in correctly predicting round-trip transactions, even if not being particularly parsimonious in number of trading actions. At $\text{H}\Delta_\tau = 10$, HLOB is more effective than the other two benchmark alternatives, showing, however, yet the most pronounced attitude to perform active trading actions. This tendency disappears moving to $\text{H}\Delta_\tau \in \{50, 100\}$, where HLOB demonstrates to be slightly inferior to its benchmark alternatives. The \textit{second group} of models is made of iTransformer and LobTransformer. They are always placed in the lower-left quadrant, demonstrating the worst performances in terms of practicability of forecasts. Finally, the \textit{third group} is the most heterogeneous one and is made of CNN1, CNN2, DLA, Transformer, and DeepLOB architectures. Among them, DeepLOB and Transformer are the only two models which permanently remain in the same quadrant, demonstrating to be less parsimonious in terms of predicted transaction, but more accurate in terms of round-trip transactions' forecast. CNN1, CNN2, and DLA, on the other side, independently from the prediction horizon, present borderline behaviours, often placing themselves in an area between the third and the fourth quadrant.

HLOB-related results discussed previously in this Section allow us to formulate new considerations on the LOB microstructural working mechanics. As explained in Section \ref{sec:Methods}, the success of the proposed architecture mainly relies on the meaningfulness of higher-order dependency structures captured by the underlying IFN. Its effectiveness is evident and persistent across different prediction horizons since it ties SOTA performances in the case of BinCTabl and BinBTabl, constantly outperforming other benchmark alternatives and, specifically, the DeepLOB model, which represents its structure-agnostic ancestor. Specifically, compared to DeepLOB, HLOB demonstrates a broad effectiveness, capturing two microstructural aspects: (i) the LOB has an underlying spatial structure that requires the modeling of higher-order and non-trivial dependency structures among volume- (and price-) levels; (ii) the emergence of dependency structures can be modeled as a function of the asset's tick size (i.e., small-, medium-, and large-tick) and their persistence varies depending on the same factor at different prediction horizons. The \textit{first finding} can be directly derived by observing that DeepLOB, which has an architecture conceptually similar to HLOB but designed to act only on consecutive LOB's volume- and price-levels, is less effective than HLOB at all prediction horizons, remarking the necessity for modeling higher-order and deeper dependency structures. The \textit{second finding} can be derived from the observation of HLOB's performance across prediction horizons for different classes of stocks. At $\text{H}\Delta_\tau = 10$, HLOB is superior to all the other models independently of the stocks' tick size, showing that the average structure extracted through the IFN effectively models short-term mid-price change dynamics. At $\text{H}\Delta_\tau = 50$, HLOB remains effective for medium- to large-tick stocks, where the risk (expressed via the bid-ask spread) and the LOB's actual depth (see the work by \citet{briola2024deep}) are lower. This is not true for small-tick stocks, where the average structure captured by the TMFG is less robust to the LOB's changes. The same findings apply to $\text{H}\Delta_\tau = 100$, where, however, the average structure is also ineffective for medium-tick stocks. At $\text{H}\Delta_\tau \in \{50, 100\}$, HLOB is superior to DeepLOB, but slightly inferior to BinBTabl and BinCTabl. We postulate that these last two models perform better on longer horizons because they apply a dual-attention mechanism on the input's spatial and temporal dimension (the IFN behind the HLOB model only handles spatial dynamics). This means that they orchestrate a selective focus on specific elements of the input by assigning varying weights indicative of their relative significance for the given task across spatial and time LOB features, allowing the refinement of the captured non-linear relationships across time. This advantage comes with a drawback. Indeed, the high level of interpretability offered by the standard attention decreases in systems employing the dual-attention mechanism due to the inherent complexity of capturing evolving spatial dependencies over time, leading to a more dynamic and nuanced understanding of the data that might not be as easily interpreted statically. The supremacy of BinBTabl and BinCTabl models is finally annihilated in the case of large-tick stocks, which show a higher level of structure across volume levels, avoiding informational drifts that are damaging in the case of deep learning models.
 
\subsection{Spatial Distribution of Information in Limit Order Books}\label{sec:The_Informational_Content_off_LOBs}
\begin{figure}[h!]
    \centering
    \includegraphics[scale=0.25]{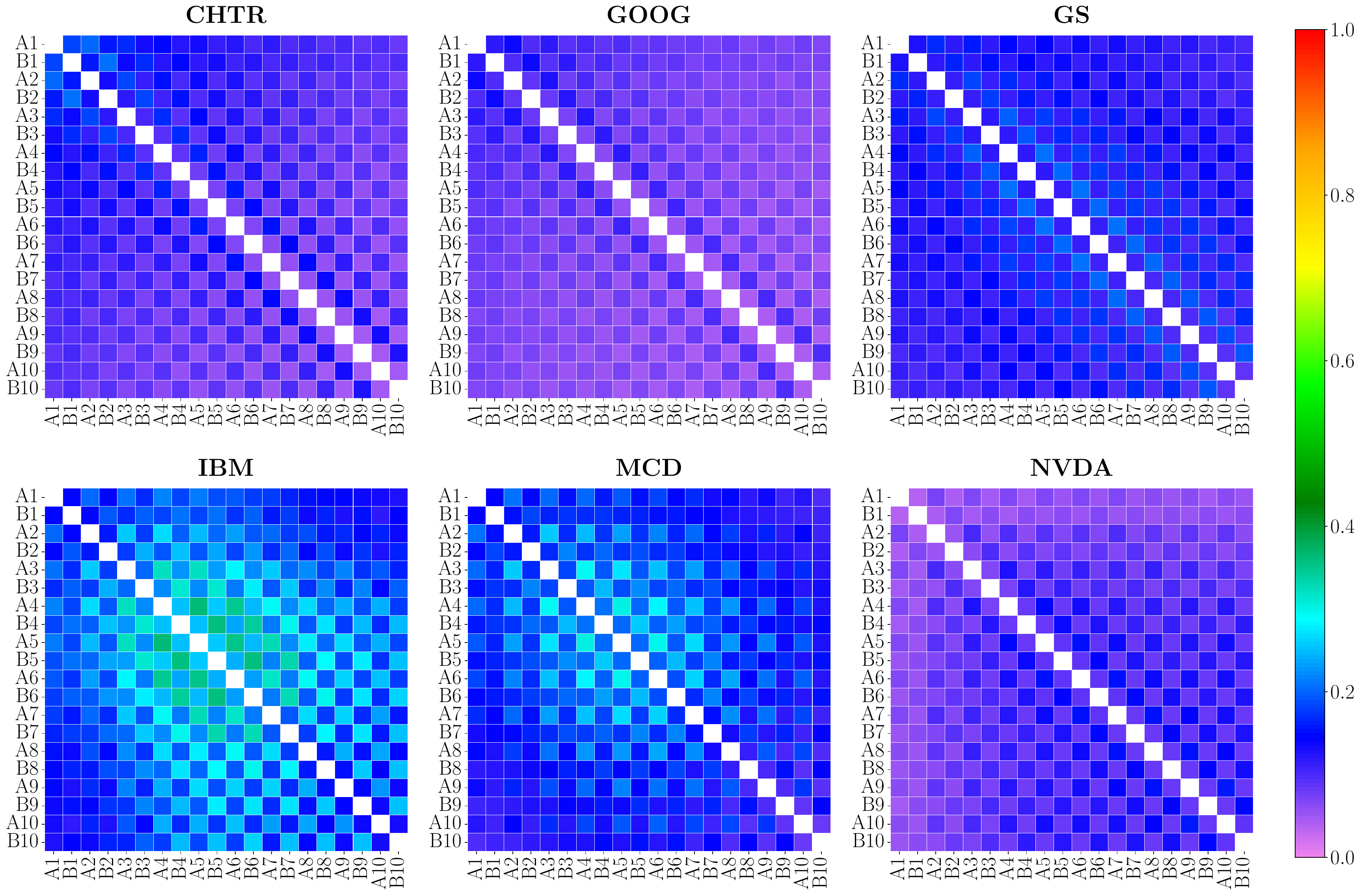}
    \caption{Normalized (over the $15$ stocks in Table \ref{tab:stocks_introduction}) version of the average (computed across the $3$-year analysis period) MI matrices computed on small-tick stocks (i.e., CHTR, GOOG, GS, IBM, MCD, NVDA).  For the sake of readability, we renamed LOB volume levels following a mapping schema that can be summarized as follows $v_\ell^{\text{ask}} \rightarrow \text{A}\ell$, $v_\ell^{\text{bid}} \rightarrow \text{B}\ell$.}
    \label{fig:mi_matrix_small_tick_stocks}
\end{figure}

\begin{figure}[H]
    \centering
    \includegraphics[scale=0.25]{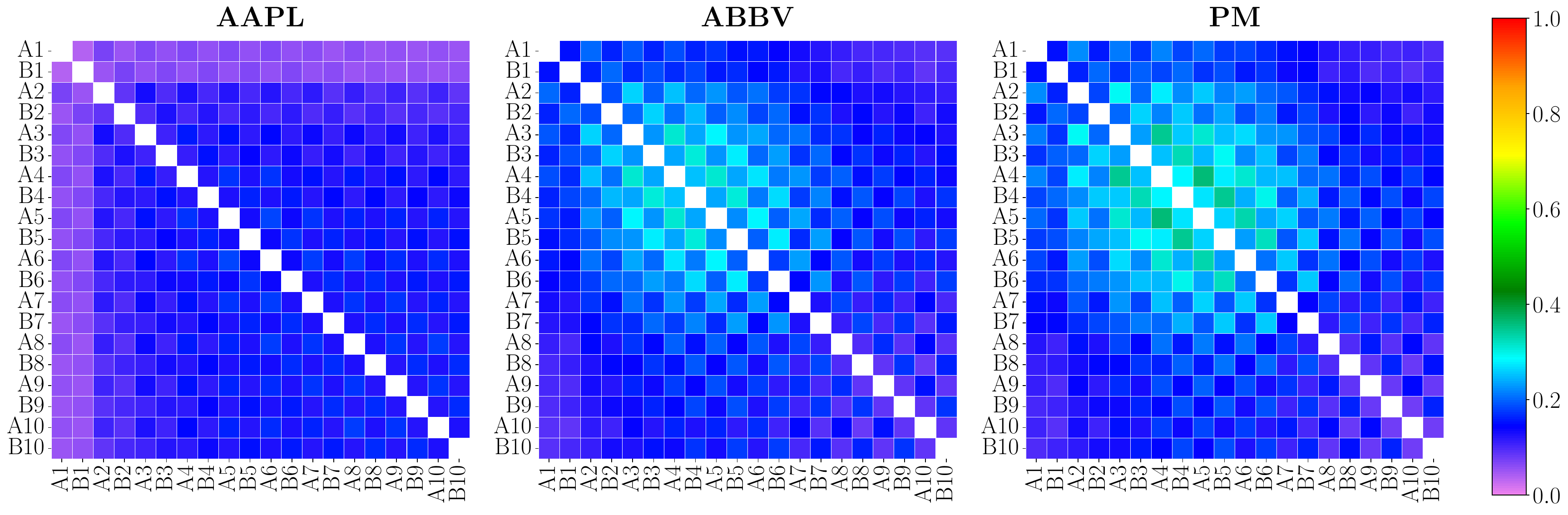}
    \caption{Normalized (over the $15$ stocks in Table \ref{tab:stocks_introduction}) version of the average (computed across the $3$-year analysis period) MI matrices computed on medium-tick stocks (i.e., AAPL, ABBV, PM).  For the seek of readability, we renamed LOB volume levels following a mapping schema that can be summarized as follows $v_\ell^{\text{ask}} \rightarrow \text{A}\ell$, $v_\ell^{\text{bid}} \rightarrow \text{B}\ell$.}
    \label{fig:mi_matrix_medium_tick_stocks}
\end{figure}

\begin{figure}[h!]
    \centering
    \includegraphics[scale=0.25]{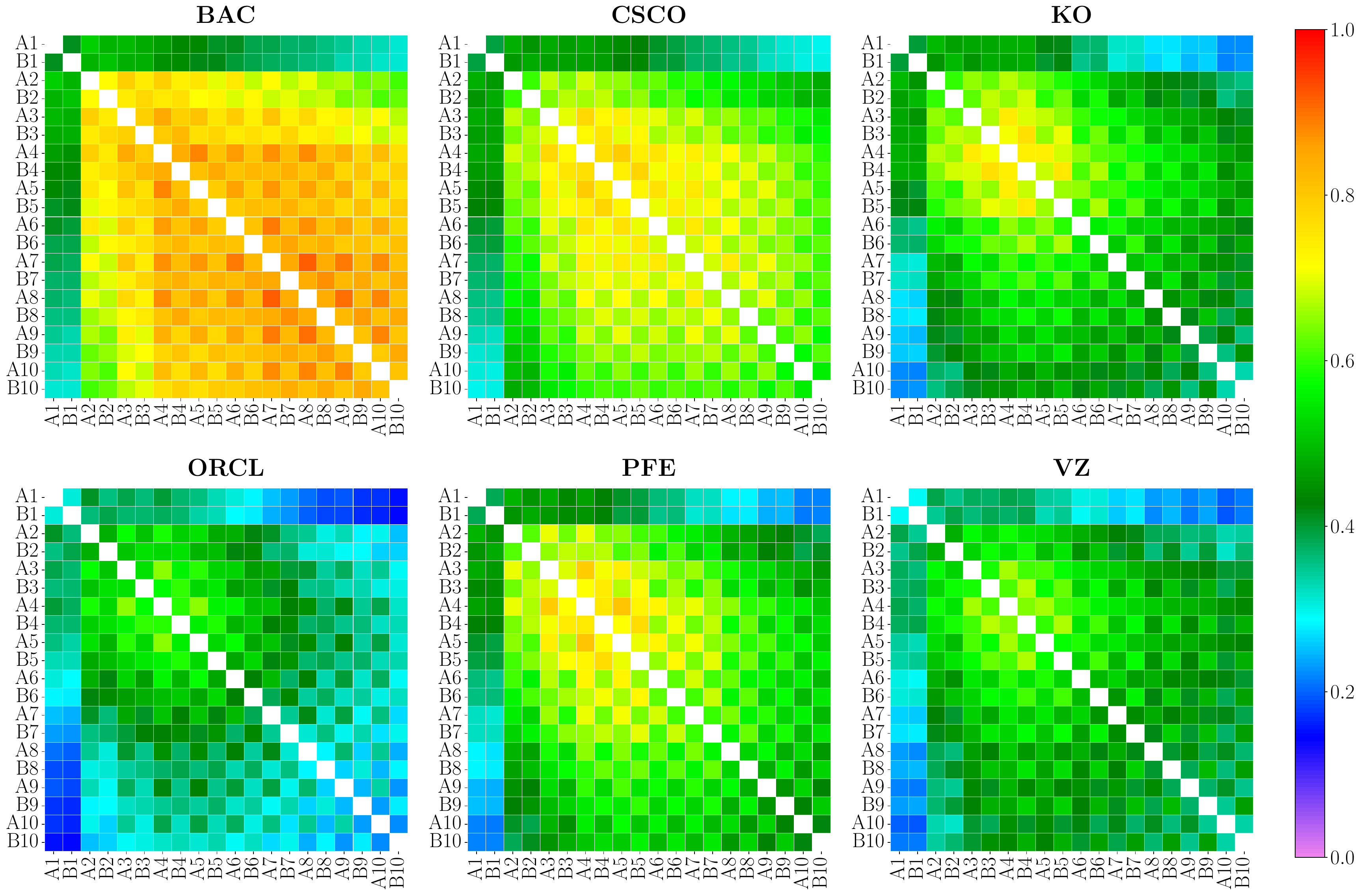}
    \caption{Normalized (over the $15$ stocks in Table \ref{tab:stocks_introduction}) version of the average (computed across the $3$-year analysis period) MI matrices computed on large-tick stocks (i.e., BAC, CSCO, KO, ORCL, PFE, VZ). For the seek of readability, we renamed LOB volume levels following a mapping schema that can be summarized as follows $v_\ell^{\text{ask}} \rightarrow \text{A}\ell$, $v_\ell^{\text{bid}} \rightarrow \text{B}\ell$.}
    \label{fig:mi_matrix_large_tick_stocks}
\end{figure}

  As a further instrument to understand the theoretical implications of empirical results obtained in Section \ref{sec:Models_Performance_Comparison}, in Figures \ref{fig:mi_matrix_small_tick_stocks}, \ref{fig:mi_matrix_medium_tick_stocks}, \ref{fig:mi_matrix_large_tick_stocks}, we report the average (computed across the $3$-year analysis period) MI matrices computed on the training set for each of the $15$ stocks under investigation (see Section \ref{sec:TMFG_building_process}). This analysis sheds light on (i) the volume levels where most of the LOB information is concentrated, and (ii) how different spatial distributions impact the model's forecasting capabilities\,\footnote{Similar attempts were performed by \citet{libman2022mutual} and \citet{cont2023cross}. However, their works differ from ours both in terms of the adopted methodology, granularity of analysis and results' interpretation.}. As average matrices, the ones presented in the following Figures are not used to build the HLOB. However, they are useful in capturing the intuition behind the scenario-dependent effectiveness of the HLOB model.

Figure \ref{fig:mi_matrix_small_tick_stocks} reports the normalized average MI matrices for \textit{small-tick stocks}. CHTR and GOOG present similar dynamics. Their not-normalized average mutual information is $0.35$ and $0.26$, respectively. In the case of CHTR, we observe a weak hierarchical organization across LOB levels. The best ask and bid volume levels (i.e., $v_{1}^{s \in \{\text{ask, bid}\}}$) present the highest cumulative mutual informational, which smoothly decreases moving to deeper levels. The highest punctual realizations of the chosen similarity measure are generally expressed among contiguous levels on the same side of the LOB. In the case of GOOG, we notice that the decrease in the cumulative mutual information across volume levels is steeper, with a clear break after $v_{4}^{s \in \{\text{ask, bid}\}}$. Also in this case, the highest punctual realizations of the chosen similarity measure are generally expressed among contiguous levels on the same side of the LOB. GS presents a not-normalized average mutual information equal to $0.45$. Compared to the previous two alternatives, this value turns out to be not only higher, but also differently spatially distributed. Indeed, looking at the volume levels' cumulative mutual information, we isolate three different groups: (i) $v_{\ell \in \{1, 3\}}^{s \in \{\text{ask, bid}\}}$, (ii) $v_{\ell \in \{4, 7\}}^{s \in \{\text{ask, bid}\}}$, and (iii) $v_{\ell \in \{8, 10\}}^{s \in \{\text{ask, bid}\}}$. The first and the last group are characterised by a similar cumulative mutual information value, which, however is lower than the one of the second group. These three clusters are clearly separated, with an absence of smoothed transition. In this sense, GS presents the first signs of a hierarchical structure where the central levels of the LOB play an increasingly central role. This behaviour is markedly evident in the case of IBM. This stock presents a not-normalized average mutual information equal to $0.74$ (i.e., the highest among small-tick stocks), with a clear concentration towards the central levels of the LOB (i.e. $v_{\ell \in \{4, 6\}}^{s \in \{\text{ask, bid}\}}$). In this case, the transition from volume levels characterised by a lower cumulative mutual information to volume levels characterised by a higher cumulative mutual information is smooth and incremental moving from top to middle volume levels and is even less evident moving from middle to deep ones (i.e. $v_{\ell \in \{7, 10\}}^{s \in \{\text{ask, bid}\}}$), which are organised in a clear cluster with a medium-to-high level of cumulative mutual information. Even if visually similar to IBM, the MI matrix characterizing MCD conveys a different message. Here, the not-normalized average mutual information is equal to $0.58$ and is mostly distributed across the top $8$ levels of the LOB (i.e., $v_{\ell \in \{1, 8\}}^{s \in \{\text{ask, bid}\}}$). In this sense, the emerging hierarchical structure is less clear compared to the one of IBM, and more similar to the one of GS. The case of NVDA, finally, is unique in the class of small-tick stocks. Here, the average mutual information is equal to $0.31$ and is mainly concentrated on the deepest $6$ levels of the LOB. Complementary to what observed for CHTR and GOOG, the top volume levels (i.e., $v_{1}^{s \in \{\text{ask, bid}\}}$) are characterised by the lowest cumulative mutual information, which incrementally increases moving to deeper levels of the LOB. However, also in this case, the highest punctual realizations of the chosen similarity measure are generally expressed among contiguous levels on the same side of the LOB. All the results discussed above directly derive from one of the findings in the work by \citet{briola2024deep}. There, the authors, following the intuition proposed by \citet{wu2021towards}, introduce $\Xi^{\text{Bid}}$ and $\Xi^{\text{Ask}}$ to measure the `actual LOB depth' (see Table \ref{tab:actual_lob_depth}) on the bid and ask side of the LOB, respectively. Indeed, as described in Section \ref{sec:Limit_Order_Book}, the LOB representation characterizing the data used in the current paper, suffers a lack of homogeneity in the spatial structure (since there is no assumption for adjacent price levels to be separated by fixed intervals). As a consequence, when the average $\Xi^{\{\text{Bid, Ask}\}} \gg 9.0$, as it happens for CHTR and GOOG, the computation of the average mutual information across levels is negatively affected due to the drifts that make the concept of `level' a pure theoretical artifact with a short-term practical feedback. On the contrary, the meaningfulness of MI matrices and, consequently, the persistence of related higher-order structures across longer time horizons, increases when $\Xi^{\{\text{Bid, Ask}\}} \simeq 9.0$, with IBM providing an example of ideal environment to challenge spatially-informed deep learning models.

Figure \ref{fig:mi_matrix_medium_tick_stocks} reports the normalized average MI matrices for \textit{medium-tick stocks}. AAPL is characterized by a not-normalized average mutual information equal to $0.41$ which is mainly concentrated across $v_{\ell \in \{2, 10\}}^{s \in \{\text{ask, bid}\}}$. The top volume levels of the LOB are markedly detached from the others, which, in contrast, show a strong interdependence. This behavior is far from that of ABBV and PM, which have a not-normalized average mutual information equal to $0.59$ and $0.63$, respectively. In both cases, the distribution of the chosen similarity metric is very similar to the one of MCD, with most of the mutual information concentrated on the top $7$ volume levels of the LOB and a clear drop for the remaining $3$ ones. Looking at Table \ref{tab:actual_lob_depth}, we notice that, in the case of AAPL, a lower average mutual information is compensated by a higher stability of the LOB, which increases the persistence of the structure extracted from the MI matrix through the IFN (described in Section \ref{sec:TMFG_building_process}). ABBV and PM, in contrast, present average $\Xi^{\text{Bid}}$ and $\Xi^{\text{Ask}}$ values that are more similar to the ones observed for small-tick stocks and are consequently exposed to the adverse consequences described previously in this Section.

Figure \ref{fig:mi_matrix_large_tick_stocks} reports the normalized average MI matrices for \textit{large-tick stocks}. The not-normalized average mutual information of BAC is equal to $1.18$. It is unevenly spatially distributed across LOB levels with an evident hierarchical organization: (i) $v_1^{s \in \{\text{ask, bid}\}}$ express the lowest pairwise mutual information realizations; (ii) $v_{\ell \in \{2, 3\}}^{s \in \{\text{ask, bid}\}}$ express an intermediate amount of pairwise mutual information; and (iii) $v_{\ell \in \{4, 10\}}^{s \in \{\text{ask, bid}\}}$ contain the highest concentration of mutual information. For each of these $3$ groups, it is possible to notice a smooth decrease of mutual information moving from top to deeper LOB levels. A similar dynamics can be observed for all the other stocks characterized by the same tick size, with the only difference of being able to clearly identify two (instead of three) groups of volume levels. In all the cases, $v_1^{s \in \{\text{ask, bid}\}}$ express the lowest pairwise mutual information, while $v_{\ell \in \{3, 5\}}^{s \in \{\text{ask, bid}\}}$ contain the highest realizations. The not-normalized average mutual information of CSCO, KO, ORCL, PFE, and VZ is equal to $1.00$, $0.83$, $0.62$, $0.90$, and $0.74$, respectively. ORCL and VZ present the lowest realizations; however, compared to small- to medium-tick stocks, they maintain a considerably higher level of structure. Looking at Table \ref{tab:actual_lob_depth}, we observe that all the stocks consistently exhibit the lowest realizations of $\Xi^{\text{Bid}}$ and $\Xi^{\text{Ask}}$. This indicates a high level of stability in LOB structures, consequently justifying the advantage of deep learning-based models in the related forecasting tasks. Additionally, the distinct emerging structure observed across LOB levels supports their sustained effectiveness over extended prediction horizons. 

To summarize, we state that (i) \textit{small- and medium-tick stocks} generally suffer from a lack of structure in the LOB informational content, which causes a faster degradation of deep learning models' forecasting capabilities moving from closer to farther prediction horizons; (ii) while \textit{large-tick stocks} present a more compact and meaningful structure of the LOB, guaranteeing a direct mapping between the theoretical concept of 'level' and its practical realization as an informational channel for deep learning models, which consequently has a positive effect on deep learning models' forecasting performances at both closer and farther prediction horizons.

{
\centering
\renewcommand{\arraystretch}{2.3}
\begin{table}[h!]
\centering
\begin{adjustbox}{center,max width=\linewidth}
\caption{Mean and median `actual depth' for bid and ask side of the LOB (i.e., $\Xi^{\text{Bid}}$ and $\Xi^{\text{Ask}}$) for the $15$ stocks of interest, in the $3$-year analysis period.}
\label{tab:actual_lob_depth}
\tiny
\setlength{\tabcolsep}{3pt}
\begin{tabular}{@{}c|cccc|cccc|cccc@{}}
\toprule
\multirow{3}{*}{\textbf{}} &
  \multicolumn{4}{c|}{\textbf{2017}} &
  \multicolumn{4}{c|}{\textbf{2018}} &
  \multicolumn{4}{c}{\textbf{2019}} \\ \cmidrule(l){2-13} 
 &
  \multicolumn{2}{c|}{\textbf{Mean}} &
  \multicolumn{2}{c|}{\textbf{Median}} &
  \multicolumn{2}{c|}{\textbf{Mean}} &
  \multicolumn{2}{c|}{\textbf{Median}} &
  \multicolumn{2}{c|}{\textbf{Mean}} &
  \multicolumn{2}{c}{\textbf{Median}} \\ \cmidrule(l){2-13} 
 &
  \textbf{Ask} &
  \multicolumn{1}{c|}{\textbf{Bid}} &
  \textbf{Ask} &
  \textbf{Bid} &
  \textbf{Ask} &
  \multicolumn{1}{c|}{\textbf{Bid}} &
  \textbf{Ask} &
  \textbf{Bid} &
  \textbf{Ask} &
  \multicolumn{1}{c|}{\textbf{Bid}} &
  \textbf{Ask} &
  \textbf{Bid} \\ \midrule
\textbf{CHTR} &
  53.83 &
  \multicolumn{1}{c|}{50.14} &
  34.00 &
  35.00 &
  71.67 &
  \multicolumn{1}{c|}{68.84} &
  55.00 &
  54.00 &
  44.56 &
  \multicolumn{1}{c|}{45.64} &
  35.00 &
  36.00 \\
\textbf{GOOG} &
  55.45 &
  \multicolumn{1}{c|}{53.57} &
  49.00 &
  47.00 &
  93.67 &
  \multicolumn{1}{c|}{91.83} &
  82.00 &
  81.00 &
  61.92 &
  \multicolumn{1}{c|}{62.14} &
  58.00 &
  58.00 \\
\textbf{GS} &
  14.97 &
  \multicolumn{1}{c|}{15.30} &
  13.00 &
  13.00 &
  19.07 &
  \multicolumn{1}{c|}{20.33} &
  15.00 &
  16.00 &
  13.98 &
  \multicolumn{1}{c|}{14.12} &
  13.00 &
  13.00 \\
\textbf{IBM} &
  10.29 &
  \multicolumn{1}{c|}{10.47} &
  9.00 &
  10.00 &
  12.47 &
  \multicolumn{1}{c|}{12.70} &
  11.00 &
  11.00 &
  9.93 &
  \multicolumn{1}{c|}{9.93} &
  9.00 &
  9.00 \\
\textbf{MCD} &
  126.97 &
  \multicolumn{1}{c|}{9.95} &
  9.00 &
  9.00 &
  12.32 &
  \multicolumn{1}{c|}{12.73} &
  10.00 &
  11.00 &
  13.35 &
  \multicolumn{1}{c|}{13.59} &
  12.00 &
  12.00 \\
\textbf{NVDA} &
  10.79 &
  \multicolumn{1}{c|}{10.73} &
  9.00 &
  9.00 &
  16.30 &
  \multicolumn{1}{c|}{16.12} &
  15.00 &
  15.00 &
  10.87 &
  \multicolumn{1}{c|}{10.88} &
  10.00 &
  10.00 \\ \midrule
\textbf{AAPL} &
  9.02 &
  \multicolumn{1}{c|}{9.02} &
  9.00 &
  9.00 &
  9.52 &
  \multicolumn{1}{c|}{9.54} &
  9.00 &
  9.00 &
  9.13 &
  \multicolumn{1}{c|}{9.14} &
  9.00 &
  9.00 \\
\textbf{ABBV} &
  10.95 &
  \multicolumn{1}{c|}{11.13} &
  9.00 &
  9.00 &
  23.27 &
  \multicolumn{1}{c|}{19.79} &
  11.00 &
  11.00 &
  9.74 &
  \multicolumn{1}{c|}{9.69} &
  9.00 &
  9.00 \\
\textbf{PM} &
  10.09 &
  \multicolumn{1}{c|}{10.07} &
  9.00 &
  9.00 &
  13.64 &
  \multicolumn{1}{c|}{13.60} &
  11.00 &
  11.00 &
  10.22 &
  \multicolumn{1}{c|}{10.20} &
  9.00 &
  9.00 \\ \midrule
\textbf{BAC} &
  9.00 &
  \multicolumn{1}{c|}{9.00} &
  9.00 &
  9.00 &
  9.00 &
  \multicolumn{1}{c|}{9.00} &
  9.00 &
  9.00 &
  9.00 &
  \multicolumn{1}{c|}{9.00} &
  9.00 &
  9.00 \\
\textbf{CSCO} &
  9.00 &
  \multicolumn{1}{c|}{9.00} &
  9.00 &
  9.00 &
  9.01 &
  \multicolumn{1}{c|}{9.01} &
  9.00 &
  9.00 &
  9.00 &
  \multicolumn{1}{c|}{9.00} &
  9.00 &
  9.00 \\
\textbf{KO} &
  9.14 &
  \multicolumn{1}{c|}{9.19} &
  9.00 &
  9.00 &
  9.04 &
  \multicolumn{1}{c|}{9.04} &
  9.00 &
  9.00 &
  9.01 &
  \multicolumn{1}{c|}{9.01} &
  9.00 &
  9.00 \\
\textbf{ORCL} &
  9.11 &
  \multicolumn{1}{c|}{9.11} &
  9.00 &
  9.00 &
  9.05 &
  \multicolumn{1}{c|}{9.06} &
  9.00 &
  9.00 &
  9.01 &
  \multicolumn{1}{c|}{9.01} &
  9.00 &
  9.00 \\
\textbf{PFE} &
  9.20 &
  \multicolumn{1}{c|}{9.21} &
  9.00 &
  9.00 &
  9.03 &
  \multicolumn{1}{c|}{9.03} &
  9.00 &
  9.00 &
  9.01 &
  \multicolumn{1}{c|}{9.01} &
  9.00 &
  9.00 \\
\textbf{VZ} &
  9.09 &
  \multicolumn{1}{c|}{9.09} &
  9.00 &
  9.00 &
  9.07 &
  \multicolumn{1}{c|}{9.09} &
  9.00 &
  9.00 &
  9.01 &
  \multicolumn{1}{c|}{9.01} &
  9.00 &
  9.00 \\ \bottomrule
\end{tabular}%
\end{adjustbox}
\end{table}
}

\section{Conclusion and Future Work}\label{sec:Conclusion}
This paper introduces HLOB, a novel large-scale deep learning architecture for high-frequency Limit Order Book (LOB) mid-price changes' direction forecasting. The novelty of the model lies in the possibility to deterministically model higher-order interactions among LOB volume (and price) levels leveraging the power of a class of Information Filtering Networks (IFNs): the Triangulated Maximally Filtered Graph (TMFG). Its computation exploits the pairwise mutual information across volume levels, and its structure only retains statistically relevant dependencies, pruning the weakest ones. The informational content of the emerging topological priors (i.e., tetrahedra, triangles, and edges) is then mapped as input to the class of Homological Convolutional Neural Networks (HCNNs), and processed to forecast the direction of high-frequency mid-price changes for $15$ stocks belonging to $3$ different classes (i.e., small-, medium-, and large-tick stocks) over a $3$-year analysis period (i.e., $2017$, $2018$, and $2019$). This class of neural networks naturally models the spatial dimension of the LOB and is modified here to handle long-term temporal dependencies through the introduction of the long short-term memory (LSTM) module. This modification sets off the transition from a simple HCNN to an HLOB model. The development of this architecture is backed by the hypothesis that a more structured architectural grasp of the LOB’s spatial dependency structures would enhance a model's predictive precision. To test this hypothesis, we test our architecture against $9$ SOTA models, demonstrating not only the supremacy of our architecture in specific scenarios but also using the empirical results to prove some theoretical conjectures on the microstructural mechanics of the LOBs. Specifically, we obtain $3$ main findings:

\begin{itemize}
    \item We demonstrate that the LOB has an underlying spatial structure that requires modeling higher-order and non-trivial dependency structures among volume (and price) levels, making the modeling of interactions among consecutive levels (used for instance in DeepLOB \cite{zhang2019deeplob}) a sub-optimal solution.
    \item We show that the emergence of dependency structures can be modeled as a function of the asset’s tick size; different structures emerge for different types of assets, and the ones characterized by a clear hierarchical structure (i.e., large-tick stocks) have more chances to be correctly forecast by deep learning models.
    \item We demonstrate that the persistence of the informational content that can be captured through a modeling exercise on the LOB spatial dependency structure varies at different prediction horizons. Indeed, when the LOB structure is sparse and subject to informational drifts (i.e., small- and medium-tick stocks), the concept of `level' becomes a purely theoretical artifact with a limited-in-time realization in practical scenarios. In this case, a deep learning model built on the average mutual information across volume levels is also exposed to the adverse impact of outliers (i.e., informational drifts) being effective only at short-term prediction horizons, where the likelihood of informational drifts is lower.
\end{itemize}

HLOB represents a step forward in building microstructurally-informed models for the prediction of the direction of high-frequency mid-price changes. The overall performance of the proposed architecture is commendable. It marks a significant advancement in the field of microstructural modeling, providing practitioners and academics with a powerful tool that combines the power of deep learning with a nuanced understanding of LOB mechanics, facilitating better decision-making in high-frequency environments. From its comparison with alternative SOTA models, some limitations emerge. Specifically, applying a dual-attention mechanism on the input data spatial and temporal dimensions guarantees, at the price of a reduced interpretability, enhanced performances, allowing for a refinement of the captured non-linear relationships over time and offering an edge over HLOB, which primarily handles spatial dynamics.

There are several avenues for further development of the HLOB model. As a future research work, (i) we should reason about more refined ways to compute the similarity matrices at the core of the proposed architecture and, as a consequence of this, (ii) we should think about the possibility of modifying the HLOB to incorporate temporally-evolving IFNs capturing the evolving complexity of the LOB. This research marks an initial step towards developing microstructurally informed models that can adapt to the complexities of high-frequency market phase transitions. Future work will build on these foundations to enhance models' adaptability and accuracy, paving the way for more robust and practical applications in financial markets forecasting.

\section*{Acknowledgments}
The author, T.A., acknowledges the financial support from ESRC (ES/K002309/1), EPSRC (EP/P031730/1) and EC (H2020-ICT-2018-2 825215). The authors declare no conflict of interest. The funders had no role in the design of the study; in the collection, analyses, or interpretation of data; in the writing of the manuscript; or in the decision to publish the results. The author, A.B., acknowledges Kashif Rasul for the valuable help in the coding of some of the SOTA deep learning models used for comparative purposes. 

\newpage

\bibliographystyle{plainnat}
\bibliography{references}

\begin{thebibliography}{63}
\providecommand{\natexlab}[1]{#1}
\providecommand{\url}[1]{\texttt{#1}}
\expandafter\ifx\csname urlstyle\endcsname\relax
  \providecommand{\doi}[1]{doi: #1}\else
  \providecommand{\doi}{doi: \begingroup \urlstyle{rm}\Url}\fi

\bibitem[Aste(2022)]{aste2022topological}
Tomaso Aste.
\newblock Topological regularization with information filtering networks.
\newblock \emph{Information Sciences}, 608:\penalty0 655--669, 2022.

\bibitem[Aste and Di~Matteo(2006)]{aste2006dynamical}
Tomaso Aste and Tiziana Di~Matteo.
\newblock Dynamical networks from correlations.
\newblock \emph{Physica A: Statistical Mechanics and its Applications}, 370\penalty0 (1):\penalty0 156--161, 2006.

\bibitem[Aste et~al.(2005)Aste, Di~Matteo, and Hyde]{aste2005complex}
Tomaso Aste, Tiziana Di~Matteo, and ST~Hyde.
\newblock Complex networks on hyperbolic surfaces.
\newblock \emph{Physica A: Statistical Mechanics and its Applications}, 346\penalty0 (1-2):\penalty0 20--26, 2005.

\bibitem[Barfuss et~al.(2016)Barfuss, Massara, Di~Matteo, and Aste]{barfuss2016parsimonious}
Wolfram Barfuss, Guido~Previde Massara, Tiziana Di~Matteo, and Tomaso Aste.
\newblock Parsimonious modeling with information filtering networks.
\newblock \emph{Physical Review E}, 94\penalty0 (6):\penalty0 062306, 2016.

\bibitem[Bouchaud et~al.(2009)Bouchaud, Farmer, and Lillo]{bouchaud2009markets}
Jean-Philippe Bouchaud, J~Doyne Farmer, and Fabrizio Lillo.
\newblock How markets slowly digest changes in supply and demand.
\newblock In \emph{Handbook of financial markets: dynamics and evolution}, pages 57--160. Elsevier, 2009.

\bibitem[Bouchaud et~al.(2018)Bouchaud, Bonart, Donier, and Gould]{bouchaud2018trades}
Jean-Philippe Bouchaud, Julius Bonart, Jonathan Donier, and Martin Gould.
\newblock \emph{Trades, quotes and prices: financial markets under the microscope}.
\newblock Cambridge University Press, 2018.

\bibitem[Briola and Aste(2022)]{briola2022dependency}
Antonio Briola and Tomaso Aste.
\newblock Dependency structures in cryptocurrency market from high to low frequency.
\newblock \emph{Entropy}, 24\penalty0 (11):\penalty0 1548, 2022.

\bibitem[Briola et~al.(2020)Briola, Turiel, and Aste]{briola2020deep}
Antonio Briola, Jeremy Turiel, and Tomaso Aste.
\newblock Deep learning modeling of limit order book: A comparative perspective.
\newblock \emph{arXiv preprint arXiv:2007.07319}, 2020.

\bibitem[Briola et~al.(2021)Briola, Turiel, Marcaccioli, Cauderan, and Aste]{briola2021deep}
Antonio Briola, Jeremy Turiel, Riccardo Marcaccioli, Alvaro Cauderan, and Tomaso Aste.
\newblock Deep reinforcement learning for active high frequency trading.
\newblock \emph{arXiv preprint arXiv:2101.07107}, 2021.

\bibitem[Briola et~al.(2023)Briola, Wang, Bartolucci, and Aste]{briola2023homological}
Antonio Briola, Yuanrong Wang, Silvia Bartolucci, and Tomaso Aste.
\newblock Homological convolutional neural networks.
\newblock \emph{arXiv preprint arXiv:2308.13816}, 2023.

\bibitem[Briola et~al.(2024)Briola, Bartolucci, and Aste]{briola2024deep}
Antonio Briola, Silvia Bartolucci, and Tomaso Aste.
\newblock Deep limit order book forecasting.
\newblock \emph{arXiv preprint arXiv:2403.09267}, 2024.

\bibitem[Brown et~al.(2020)Brown, Mann, Ryder, Subbiah, Kaplan, Dhariwal, Neelakantan, Shyam, Sastry, Askell, et~al.]{brown2020language}
Tom Brown, Benjamin Mann, Nick Ryder, Melanie Subbiah, Jared~D Kaplan, Prafulla Dhariwal, Arvind Neelakantan, Pranav Shyam, Girish Sastry, Amanda Askell, et~al.
\newblock Language models are few-shot learners.
\newblock \emph{Advances in neural information processing systems}, 33:\penalty0 1877--1901, 2020.

\bibitem[Cluster(2023)]{ucl2023cluster}
UCL CS~HPC Cluster.
\newblock Ucl cs hpc cluster.
\newblock \url{https://hpc.cs.ucl.ac.uk}, 2023.
\newblock Accessed: 2023-06-16.

\bibitem[{companiesmarketcap.com}()]{capitalization_provider}
{companiesmarketcap.com}.
\newblock Companies market cap.
\newblock \url{https://companiesmarketcap.com}.
\newblock Accessed: 24/01/2024.

\bibitem[Cont et~al.(2023)Cont, Cucuringu, and Zhang]{cont2023cross}
Rama Cont, Mihai Cucuringu, and Chao Zhang.
\newblock Cross-impact of order flow imbalance in equity markets.
\newblock \emph{Quantitative Finance}, 23\penalty0 (10):\penalty0 1373--1393, 2023.

\bibitem[Farmer and Skouras(2013)]{farmer2013ecological}
J~Doyne Farmer and Spyros Skouras.
\newblock An ecological perspective on the future of computer trading.
\newblock \emph{Quantitative Finance}, 13\penalty0 (3):\penalty0 325--346, 2013.

\bibitem[Gorodkin(2004)]{gorodkin2004comparing}
Jan Gorodkin.
\newblock Comparing two k-category assignments by a k-category correlation coefficient.
\newblock \emph{Computational biology and chemistry}, 28\penalty0 (5-6):\penalty0 367--374, 2004.

\bibitem[Guo and Chen(2023)]{guo2023forecasting}
Yanhong Guo and Xinxin Chen.
\newblock Forecasting the mid-price movements with high-frequency lob: A dual-stage temporal attention-based deep learning architecture.
\newblock \emph{Arabian Journal for Science and Engineering}, 48\penalty0 (8):\penalty0 9597--9618, 2023.

\bibitem[Hochreiter and Schmidhuber(1997)]{hochreiter1997long}
Sepp Hochreiter and J{\"u}rgen Schmidhuber.
\newblock Long short-term memory.
\newblock \emph{Neural computation}, 9\penalty0 (8):\penalty0 1735--1780, 1997.

\bibitem[Isichenko(2021)]{isichenko2021quantitative}
Michael Isichenko.
\newblock \emph{Quantitative portfolio management: The art and science of statistical arbitrage}.
\newblock John Wiley \& Sons, 2021.

\bibitem[{Karpathy}()]{nanoGPT}
{Karpathy}.
\newblock nanogpt.
\newblock \url{https://github.com/karpathy/nanoGPT/tree/master}.
\newblock Accessed: 12/01/2024.

\bibitem[Kingma and Ba(2014)]{kingma2014adam}
Diederik~P Kingma and Jimmy Ba.
\newblock Adam: A method for stochastic optimization.
\newblock \emph{arXiv preprint arXiv:1412.6980}, 2014.

\bibitem[Kisiel and Gorse(2022)]{kisiel2022axial}
Damian Kisiel and Denise Gorse.
\newblock Axial-lob: High-frequency trading with axial attention.
\newblock In \emph{2022 IEEE Symposium Series on Computational Intelligence (SSCI)}, pages 1327--1333. IEEE, 2022.

\bibitem[Kolm and Westray(2024)]{kolm2024improving}
Petter~N Kolm and Nicholas Westray.
\newblock Improving deep learning of alpha term structures from the order book.
\newblock \emph{Available at SSRN}, 2024.

\bibitem[Kolm et~al.(2023)Kolm, Turiel, and Westray]{kolm2023deep}
Petter~N Kolm, Jeremy Turiel, and Nicholas Westray.
\newblock Deep order flow imbalance: Extracting alpha at multiple horizons from the limit order book.
\newblock \emph{Mathematical Finance}, 33\penalty0 (4):\penalty0 1044--1081, 2023.

\bibitem[Kullback and Leibler(1951)]{kullback1951information}
Solomon Kullback and Richard~A Leibler.
\newblock On information and sufficiency.
\newblock \emph{The annals of mathematical statistics}, 22\penalty0 (1):\penalty0 79--86, 1951.

\bibitem[LeCun et~al.(2015)LeCun, Bengio, and Hinton]{lecun2015deep}
Yann LeCun, Yoshua Bengio, and Geoffrey Hinton.
\newblock Deep learning.
\newblock \emph{nature}, 521\penalty0 (7553):\penalty0 436--444, 2015.

\bibitem[Lehalle and Laruelle(2018)]{lehalle2018market}
Charles-Albert Lehalle and Sophie Laruelle.
\newblock \emph{Market microstructure in practice}.
\newblock World Scientific, 2018.

\bibitem[Libman et~al.(2022)Libman, Ariel, Schaps, and Haber]{libman2022mutual}
Daniel Libman, Gil Ariel, Mary Schaps, and Simi Haber.
\newblock Mutual information between order book layers.
\newblock \emph{Entropy}, 24\penalty0 (3):\penalty0 343, 2022.

\bibitem[Liu et~al.(2023)Liu, Hu, Zhang, Wu, Wang, Ma, and Long]{liu2023itransformer}
Yong Liu, Tengge Hu, Haoran Zhang, Haixu Wu, Shiyu Wang, Lintao Ma, and Mingsheng Long.
\newblock itransformer: Inverted transformers are effective for time series forecasting.
\newblock \emph{arXiv preprint arXiv:2310.06625}, 2023.

\bibitem[{LOBSTER Data}()]{lobsterdata_what_is_lobster}
{LOBSTER Data}.
\newblock What is lobster?
\newblock \url{https://lobsterdata.com/info/WhatIsLOBSTER.php}.
\newblock Accessed: 26/12/2023.

\bibitem[Loshchilov and Hutter(2017)]{loshchilov2017decoupled}
Ilya Loshchilov and Frank Hutter.
\newblock Decoupled weight decay regularization.
\newblock \emph{arXiv preprint arXiv:1711.05101}, 2017.

\bibitem[Lucchese et~al.(2022)Lucchese, Pakkanen, and Veraart]{lucchese2022short}
Lorenzo Lucchese, Mikko Pakkanen, and Almut Veraart.
\newblock The short-term predictability of returns in order book markets: a deep learning perspective.
\newblock \emph{arXiv preprint arXiv:2211.13777}, 2022.

\bibitem[Mantegna(1999)]{mantegna1999hierarchical}
Rosario~N Mantegna.
\newblock Hierarchical structure in financial markets.
\newblock \emph{The European Physical Journal B-Condensed Matter and Complex Systems}, 11\penalty0 (1):\penalty0 193--197, 1999.

\bibitem[Massara et~al.(2016)Massara, Di~Matteo, and Aste]{massara2016network}
Guido~Previde Massara, Tiziana Di~Matteo, and Tomaso Aste.
\newblock Network filtering for big data: Triangulated maximally filtered graph.
\newblock \emph{Journal of complex Networks}, 5\penalty0 (2):\penalty0 161--178, 2016.

\bibitem[Massara et~al.(2017)Massara, Di~Matteo, and Aste]{massara2017network}
Guido~Previde Massara, Tiziana Di~Matteo, and Tomaso Aste.
\newblock Network filtering for big data: Triangulated maximally filtered graph.
\newblock \emph{Journal of complex Networks}, 5\penalty0 (2):\penalty0 161--178, 2017.

\bibitem[{NASDAQ}()]{nasdaq_stock_screener}
{NASDAQ}.
\newblock Nasdaq stock screener.
\newblock \url{https://www.nasdaq.com/market-activity/stocks/screener}.
\newblock Accessed: 26/12/2023.

\bibitem[Ntakaris et~al.(2018)Ntakaris, Magris, Kanniainen, Gabbouj, and Iosifidis]{ntakaris2018benchmark}
Adamantios Ntakaris, Martin Magris, Juho Kanniainen, Moncef Gabbouj, and Alexandros Iosifidis.
\newblock Benchmark dataset for mid-price forecasting of limit order book data with machine learning methods.
\newblock \emph{Journal of Forecasting}, 37\penalty0 (8):\penalty0 852--866, 2018.

\bibitem[O'Hara and Draper(2011)]{o2011introduction}
Stephen O'Hara and Bruce~A Draper.
\newblock Introduction to the bag of features paradigm for image classification and retrieval.
\newblock \emph{arXiv preprint arXiv:1101.3354}, 2011.

\bibitem[Passalis et~al.(2017)Passalis, Tsantekidis, Tefas, Kanniainen, Gabbouj, and Iosifidis]{passalis2017time}
Nikolaos Passalis, Avraam Tsantekidis, Anastasios Tefas, Juho Kanniainen, Moncef Gabbouj, and Alexandros Iosifidis.
\newblock Time-series classification using neural bag-of-features.
\newblock In \emph{2017 25th European Signal Processing Conference (EUSIPCO)}, pages 301--305. IEEE, 2017.

\bibitem[Passalis et~al.(2020)Passalis, Tefas, Kanniainen, Gabbouj, and Iosifidis]{passalis2020temporal}
Nikolaos Passalis, Anastasios Tefas, Juho Kanniainen, Moncef Gabbouj, and Alexandros Iosifidis.
\newblock Temporal logistic neural bag-of-features for financial time series forecasting leveraging limit order book data.
\newblock \emph{Pattern Recognition Letters}, 136:\penalty0 183--189, 2020.

\bibitem[Paszke et~al.(2019)Paszke, Gross, Massa, Lerer, Bradbury, Chanan, Killeen, Lin, Gimelshein, Antiga, et~al.]{paszke2019pytorch}
Adam Paszke, Sam Gross, Francisco Massa, Adam Lerer, James Bradbury, Gregory Chanan, Trevor Killeen, Zeming Lin, Natalia Gimelshein, Luca Antiga, et~al.
\newblock Pytorch: An imperative style, high-performance deep learning library.
\newblock \emph{Advances in neural information processing systems}, 32, 2019.

\bibitem[Prata et~al.(2023)Prata, Masi, Berti, Arrigoni, Coletta, Cannistraci, Vyetrenko, Velardi, and Bartolini]{prata2023lob}
Matteo Prata, Giuseppe Masi, Leonardo Berti, Viviana Arrigoni, Andrea Coletta, Irene Cannistraci, Svitlana Vyetrenko, Paola Velardi, and Novella Bartolini.
\newblock Lob-based deep learning models for stock price trend prediction: A benchmark study.
\newblock \emph{arXiv preprint arXiv:2308.01915}, 2023.

\bibitem[Shabani et~al.(2022)Shabani, Tran, Magris, Kanniainen, and Iosifidis]{shabani2022multi}
Mostafa Shabani, Dat~Thanh Tran, Martin Magris, Juho Kanniainen, and Alexandros Iosifidis.
\newblock Multi-head temporal attention-augmented bilinear network for financial time series prediction.
\newblock In \emph{2022 30th European Signal Processing Conference (EUSIPCO)}, pages 1487--1491. IEEE, 2022.

\bibitem[Shabani et~al.(2023)Shabani, Tran, Kanniainen, and Iosifidis]{shabani2023augmented}
Mostafa Shabani, Dat~Thanh Tran, Juho Kanniainen, and Alexandros Iosifidis.
\newblock Augmented bilinear network for incremental multi-stock time-series classification.
\newblock \emph{Pattern Recognition}, 141:\penalty0 109604, 2023.

\bibitem[Sirignano and Cont(2021)]{sirignano2021universal}
Justin Sirignano and Rama Cont.
\newblock Universal features of price formation in financial markets: perspectives from deep learning.
\newblock In \emph{Machine Learning and AI in Finance}, pages 5--15. Routledge, 2021.

\bibitem[Sirignano(2019)]{sirignano2019deep}
Justin~A Sirignano.
\newblock Deep learning for limit order books.
\newblock \emph{Quantitative Finance}, 19\penalty0 (4):\penalty0 549--570, 2019.

\bibitem[Tran et~al.(2018)Tran, Iosifidis, Kanniainen, and Gabbouj]{tran2018temporal}
Dat~Thanh Tran, Alexandros Iosifidis, Juho Kanniainen, and Moncef Gabbouj.
\newblock Temporal attention-augmented bilinear network for financial time-series data analysis.
\newblock \emph{IEEE transactions on neural networks and learning systems}, 30\penalty0 (5):\penalty0 1407--1418, 2018.

\bibitem[Tran et~al.(2021)Tran, Kanniainen, Gabbouj, and Iosifidis]{tran2021data}
Dat~Thanh Tran, Juho Kanniainen, Moncef Gabbouj, and Alexandros Iosifidis.
\newblock Data normalization for bilinear structures in high-frequency financial time-series.
\newblock In \emph{2020 25th International Conference on Pattern Recognition (ICPR)}, pages 7287--7292. IEEE, 2021.

\bibitem[Tran et~al.(2022)Tran, Passalis, Tefas, Gabbouj, and Iosifidis]{tran2022attention}
Dat~Thanh Tran, Nikolaos Passalis, Anastasios Tefas, Moncef Gabbouj, and Alexandros Iosifidis.
\newblock Attention-based neural bag-of-features learning for sequence data.
\newblock \emph{IEEE Access}, 10:\penalty0 45542--45552, 2022.

\bibitem[Tsantekidis et~al.(2017{\natexlab{a}})Tsantekidis, Passalis, Tefas, Kanniainen, Gabbouj, and Iosifidis]{tsantekidis2017forecasting}
Avraam Tsantekidis, Nikolaos Passalis, Anastasios Tefas, Juho Kanniainen, Moncef Gabbouj, and Alexandros Iosifidis.
\newblock Forecasting stock prices from the limit order book using convolutional neural networks.
\newblock In \emph{2017 IEEE 19th conference on business informatics (CBI)}, volume~1, pages 7--12. IEEE, 2017{\natexlab{a}}.

\bibitem[Tsantekidis et~al.(2017{\natexlab{b}})Tsantekidis, Passalis, Tefas, Kanniainen, Gabbouj, and Iosifidis]{tsantekidis2017using}
Avraam Tsantekidis, Nikolaos Passalis, Anastasios Tefas, Juho Kanniainen, Moncef Gabbouj, and Alexandros Iosifidis.
\newblock Using deep learning to detect price change indications in financial markets.
\newblock In \emph{2017 25th European signal processing conference (EUSIPCO)}, pages 2511--2515. IEEE, 2017{\natexlab{b}}.

\bibitem[Tsantekidis et~al.(2020)Tsantekidis, Passalis, Tefas, Kanniainen, Gabbouj, and Iosifidis]{tsantekidis2020using}
Avraam Tsantekidis, Nikolaos Passalis, Anastasios Tefas, Juho Kanniainen, Moncef Gabbouj, and Alexandros Iosifidis.
\newblock Using deep learning for price prediction by exploiting stationary limit order book features.
\newblock \emph{Applied Soft Computing}, 93:\penalty0 106401, 2020.

\bibitem[Tumminello et~al.(2005)Tumminello, Aste, Di~Matteo, and Mantegna]{tumminello2005tool}
Michele Tumminello, Tomaso Aste, Tiziana Di~Matteo, and Rosario~N Mantegna.
\newblock A tool for filtering information in complex systems.
\newblock \emph{Proceedings of the National Academy of Sciences}, 102\penalty0 (30):\penalty0 10421--10426, 2005.

\bibitem[Tumminello et~al.(2007)Tumminello, Di~Matteo, Aste, and Mantegna]{tumminello2007correlation}
Michele Tumminello, Tiziana Di~Matteo, Tomaso Aste, and Rosario~N Mantegna.
\newblock Correlation based networks of equity returns sampled at different time horizons.
\newblock \emph{The European Physical Journal B}, 55:\penalty0 209--217, 2007.

\bibitem[Vaswani et~al.(2017)Vaswani, Shazeer, Parmar, Uszkoreit, Jones, Gomez, Kaiser, and Polosukhin]{vaswani2017attention}
Ashish Vaswani, Noam Shazeer, Niki Parmar, Jakob Uszkoreit, Llion Jones, Aidan~N Gomez, {\L}ukasz Kaiser, and Illia Polosukhin.
\newblock Attention is all you need.
\newblock \emph{Advances in neural information processing systems}, 30, 2017.

\bibitem[Wallbridge(2020)]{wallbridge2020transformers}
James Wallbridge.
\newblock Transformers for limit order books.
\newblock \emph{arXiv preprint arXiv:2003.00130}, 2020.

\bibitem[Wang et~al.(2023)Wang, Briola, and Aste]{wang2023homological}
Yuanrong Wang, Antonio Briola, and Tomaso Aste.
\newblock Homological neural networks: A sparse architecture for multivariate complexity, 2023.

\bibitem[West et~al.(2001)]{west2001introduction}
Douglas~Brent West et~al.
\newblock \emph{Introduction to graph theory}, volume~2.
\newblock Prentice hall Upper Saddle River, 2001.

\bibitem[Wu et~al.(2021)Wu, Mahfouz, Magazzeni, and Veloso]{wu2021towards}
Yufei Wu, Mahmoud Mahfouz, Daniele Magazzeni, and Manuela Veloso.
\newblock Towards robust representation of limit orders books for deep learning models.
\newblock \emph{arXiv preprint arXiv:2110.05479}, 2021.

\bibitem[Zhang et~al.(2019{\natexlab{a}})Zhang, Yao, Sun, and Tay]{zhang2019deep}
Shuai Zhang, Lina Yao, Aixin Sun, and Yi~Tay.
\newblock Deep learning based recommender system: A survey and new perspectives.
\newblock \emph{ACM computing surveys (CSUR)}, 52\penalty0 (1):\penalty0 1--38, 2019{\natexlab{a}}.

\bibitem[Zhang et~al.(2019{\natexlab{b}})Zhang, Zohren, and Roberts]{zhang2019deeplob}
Zihao Zhang, Stefan Zohren, and Stephen Roberts.
\newblock Deeplob: Deep convolutional neural networks for limit order books.
\newblock \emph{IEEE Transactions on Signal Processing}, 67\penalty0 (11):\penalty0 3001--3012, 2019{\natexlab{b}}.

\bibitem[Zhang et~al.(2021)Zhang, Lim, and Zohren]{zhang2021deep}
Zihao Zhang, Bryan Lim, and Stefan Zohren.
\newblock Deep learning for market by order data.
\newblock \emph{Applied Mathematical Finance}, 28\penalty0 (1):\penalty0 79--95, 2021.

\end{thebibliography}

\end{document}